\documentclass{article}

\usepackage[utf8]{inputenc}
\usepackage{graphicx}
\usepackage[sort&compress,numbers]{natbib}
\usepackage{enumitem}
\usepackage{definitions_arxiv}
\usepackage{booktabs} % For formal tables
\usepackage{mathtools}
\usepackage{dsfont}
\usepackage{tabularx}
\usepackage{authblk}
\usepackage{hyperref}
\usepackage{fullpage}
\usepackage{soul}

\usepackage{caption}
\usepackage{subcaption}
\usepackage{xcolor}
\usepackage[vlined,ruled]{algorithm2e}
\usepackage{tikz}

\newlength{\inlineheight}
\usepackage{comment}
\definecolor{nicepink}{RGB}{246, 112, 136}

\SetCommentSty{mycommfont}
\definecolor{nicegreen}{RGB}{53, 172, 164}

\SetKwInOut{Input}{\textbf{\textcolor{nicegreen}{input}}}
\SetKwInOut{Output}{\textbf{\textcolor{nicegreen}{output}}}
\usepackage{tcolorbox}
\tcbuselibrary{skins,breakable}
\definecolor{humanpostfg}{RGB}{56,169,197}   % #38a9c5
\definecolor{llmpostfg}{RGB}{204,122,244}    % #cc7af4
\colorlet{humanpostbg}{humanpostfg!10}
\colorlet{llmpostbg}{llmpostfg!10}

\newtcolorbox{humanpost}[1]{%
  enhanced,
  breakable,
  frame hidden,
  arc=1.5pt,
  colframe=humanpostfg!70,
  colback=humanpostbg,
  fontupper=\small,
  left=7pt, right=7pt, top=5pt, bottom=5pt,
  before upper={%
    {\color{humanpostfg}\textbf{Human-written post}}\hfill{\color{humanpostfg}opinion score $x=#1$}\par
    \vskip 1.5pt{\color{humanpostfg}\hrule height 0.6pt}\vskip 5pt
  },
}

\newtcolorbox{llmpost}[1]{%
  enhanced,
  breakable,
  frame hidden,
  arc=1.5pt,
  colframe=llmpostfg!70,
  colback=llmpostbg,
  fontupper=\small,
  left=7pt, right=7pt, top=5pt, bottom=5pt,
  before upper={%
    {\color{llmpostfg}\textbf{LLM-generated post}}\hfill{\color{llmpostfg}opinion score $y=#1$}\par
    \vskip 1.5pt{\color{llmpostfg}\hrule height 0.6pt}\vskip 5pt
  },
}

\title{AI-Mediated Communication Can Steer Collective Opinion}

\author[1]{Stratis Tsirtsis}
\author[2]{Kaivalya Rawal}
\author[2]{Chris Russell}
\author[2,3]{\\Brent Mittelstadt}
\author[1,2]{Sandra Wachter}
\affil[1]{Hasso Plattner Institute}
\affil[2]{Oxford Internet Institute, University of Oxford}
\affil[3]{Weizenbaum Institute}

\date{}

\begin{document}

\maketitle

% \UnnumberedFootnote{$^{*}$Authors contributed equally and are listed in alphabetical order.}

\begin{abstract}
    Generative artificial intelligence (AI) is increasingly integrated into the online platforms where humans exchange opinions; large language models (LLMs) have been polishing users' posts on LinkedIn and providing context for content shared on X.
While prior work has shown that AI can express biased opinions and shape individuals' opinions during human-AI interactions, less attention has been paid to its influence on collective opinion formation when mediating human-to-human communication.
We address this gap via a combination of empirical and theoretical analyses.
We show empirically that LLMs from multiple popular families introduce directional biases when instructed to edit human-written texts on contested topics, for example, nudging texts in favor of gun control and against atheism.
Building on this observation, we introduce a mathematical model of opinion dynamics in which an AI system sits between users on a social network, transforming the opinions they express and perceive.
By analytically characterizing the equilibrium of this model and performing simulations on real social network data, we show that biases introduced by AI in human-to-human communication can be amplified through the network and shift collective opinion in their direction.
In light of these findings, we investigate whether such biases are controllable by online platforms.
We audit the ``Explain this post'' feature on X and find evidence of pro-life bias in Grok's outputs on abortion-related content, which we trace back to specific design choices.
We conclude with a discussion of the broader implications of our findings in relation to ongoing legislative efforts in the European Union.
\end{abstract}

\section{Introduction}\label{sec:intro}
Imagine you visit a social media platform to share your thoughts on whether AI should be used in education. 
You lean positively, and you draft a short post endorsing the idea:
``\textit{AI might be a useful tool for personalizing the education of students.}''
Before sharing it, you decide to click the ``\textit{Improve my post}'' button, and a large language model (LLM) provided by the platform returns a polished and more explicitly endorsing 
version: ``\textit{Let's embrace the potential of AI to personalize learning and revolutionize education for every student!}''
You find this version more engaging than the one you wrote, so you simply accept the edit and publish it.
A simple nudge. But what if the same LLM is quietly nudging millions of users in the same direction?

Generative AI systems are widely embedded in major online platforms. % where humans communicate.
For example, on LinkedIn, they have been helping users improve their posts~\citep{linkedin_improve}; on YouTube, they have been generating video summaries based on transcripts~\citep{youtube_summary}; and on X, they have been providing context to help users better understand others' content~\citep{x_context}. % shared by others~\citep{x_context}.
In such use cases, AI systems do not produce standalone content, but modify content created by humans or enrich it with additional information, effectively \emph{mediating} communication on the very platforms where humans often exchange and form opinions about contested social and political issues.\footnote{LinkedIn's post improvement feature was discontinued shortly after the paper's submission to NeurIPS~\citep{linkedin_news}. It remains one of the most prominent examples of AI-mediated communication, having been actively used for more than two years.
}

While AI has shown potential to play a positive role in tempering disagreement and helping humans find common ground~\citep{bakker2022fine,tessler2024ai}, its use is no panacea.
For example, evidence suggests that LLMs carry biases in the opinions they express, both directly when asked to take a stance on politically salient topics~\citep{santurkar2023whose} and indirectly when asked to summarize diverse human opinions on a topic~\citep{huang2024bias}.
At the same time, LLMs have been shown to have persuasive effects on individuals through targeted messaging~\citep{hackenburg2024evaluating} and conversational interactions~\citep{hackenburg2025levers,salvi2025conversational}.
More worryingly, they have also been shown to shift individuals' expressed opinions without their awareness in seemingly innocuous interactions, such as during assisted writing~\citep{jakesch2023co,rottger2026measuring}.
This raises concerns about how AI biases may shape the opinions of users who rely on them to express themselves online and interpret the opinions of others.
% 

% network-level effects
A question naturally arises: How do the biases of an AI system influence the collective opinion within a social network when used to mediate communication?
To answer this question, we draw insights from mathematical sociology, where a rich literature on opinion dynamics has studied how social influence and network structure interact to shape a population's collective opinion over time~\citep{degroot1974reaching,friedkin1990social,hegselmann2002opinion,shirzadi2025opinion}.
Beyond characterizing the opinion formation process itself, a parallel line of work in computer science has examined levers through which one can influence the process, such as changing the opinions of key individuals or perturbing the network's connections~\citep{gionis2013opinion,bindel2015bad,musco2018minimizing,gaitonde2020adversarial,tu2023adversaries,miyauchi2026survey}.
Our work points to AI-mediated communication as a new such lever, with underexplored implications for human knowledge~\citep{peterson2025ai,wachter2024large} and democratic processes~\citep{summerfield2025impact,kreps2023ai}.\footnote{For a discussion of further related work, refer to Appendix~\ref{app:related-work}.}

\xhdr{Our contributions} Our work introduces a combination of empirical, theoretical, and legal analyses, focusing on the effects that AI biases can have on collective opinion when AI is used to mediate human-to-human communication on online platforms.
Specifically, we make the following contributions:
\begin{enumerate}
    \item We instruct a set of LLMs from five different families to draft and improve social media posts on $13$ contested topics given original arguments and posts written by humans.
    We develop a methodology to score each post by the degree to which it expresses an opinion in favor or against the respective topic, which we then use to show that all LLMs we study introduce directional biases across topics, even when instructed to maintain the opinion expressed in the original text.\footnote{The code used in our experiments can be found at \href{https://github.com/stsirtsis/llm-opinion-formation}{https://github.com/stsirtsis/llm-opinion-formation}.}
    \item We introduce a mathematical model of AI-mediated opinion dynamics extending the seminal model by Friedkin and Johnsen~\citep{friedkin1990social}, in which an AI system sits between users on a social network, transforming the opinions they express and perceive.
    We formally analyze the equilibrium and convergence properties of our model and characterize the shift in collective opinion at equilibrium due to the AI transformation.
    Our theoretical analysis, complemented by simulations on real social network data, reveals that biases introduced by AI to individual opinions can be amplified through the network, leading to a much larger shift in average opinion over time.
    \item We investigate whether AI biases and, by extension, the opinion formation process, can be shaped through platform design choices.
    To this end, we audit the ``Explain this post'' feature deployed on X by asking Grok to contextualize a set of human-written posts on abortion, following the feature's publicly released implementation.
    We find evidence that Grok presents a directional bias---it more frequently generates context that aligns with the 
    stance of the human-written post when it is pro-life than when it is pro-choice---which is driven by one specific guideline provided to the model by X.
\end{enumerate}
We conclude with a discussion of the broader implications of our findings in the context of ongoing legislative efforts in the European Union, which suggests that existing legislation may be insufficient to address the risks posed by AI-mediated opinion formation at scale.

\section{Directional biases in AI-mediated opinion expression}\label{sec:bias}
\label{sec2}
% par 1: why these tasks?
To understand how AI systems can introduce biases in human-to-human communication, we focus on two tasks in which AI systems help humans express their opinions on online platforms.
Specifically, we emulate scenarios in which an LLM is used to (i)  \emph{draft} a social media post on a contested topic based on a given argument, and (ii) \emph{improve} the writing of a given social media post.
An implementation of the latter has already been deployed at scale as a feature on LinkedIn~\citep{linkedin_improve}, and such writing tasks are a natural candidate for our setting, as there is empirical evidence that co-writing with opinionated LLMs can bias individuals' expressed opinions~\citep{jakesch2023co,rottger2026measuring}.
% 

% par 2(a): data
As a source of human-written text, we use two datasets from the stance detection literature~\citep{kuccuk2020stance}: the UKP Sentential Argument Mining Corpus (UKP)~\citep{stab2018cross} and the SemEval-2016 Task 6 Dataset (SemEval)~\citep{mohammad2016semeval}.%, which we refer to as UKP and SemEval, respectively.
The UKP dataset contains single sentences scraped from the internet and labeled by stance (\ie, in favor, against, or neither) and covers $8$ topics: abortion, cloning, death penalty, gun control, marijuana legalization, minimum wage, nuclear energy, and school uniforms.
The SemEval dataset contains short posts collected from Twitter (now X), also labeled by stance, covering $6$ topics: abortion, atheism, climate change, feminism, Hillary Clinton, and Donald Trump.

% par 2(b): what do we actually do? prompts etc
To emulate these tasks, we provide human-written texts to LLMs from different families and instruct them to generate social media posts based on them.\footnote{All experiments were run on an internal cluster using AMD EPYC 7502 CPUs and NVIDIA A100 (80GB) GPUs.\label{foot:resources}} % respective texts.
We consider nine LLMs from the \texttt{Llama}, \texttt{GPT}, \texttt{Gemma}, \texttt{Ministral}, and \texttt{Qwen} families, including eight open-weight models and one proprietary model (\texttt{gpt-5.6-luna}).
We use arguments from the UKP dataset to emulate drafting  (argument $\rightarrow$ post) and posts from the SemEval dataset to emulate improvement  (post $\rightarrow$ post).
For each task and topic, we provide the LLMs with a user prompt specifying the task and explicitly instruct them via the system prompt to preserve the voice and meaning of the original human-written text.
To ensure robust analysis, we use three user prompts per task (see Appendix~\ref{app:prompts}) % for the exact prompts) 
and generate five responses per pair of human-written text and prompt variant using a temperature of $1$ and top-$p$ sampling with $p=0.95$.
Lastly, we filter outputs in which the LLMs refuse to perform the task. This happened in a small number of cases due to the presence of offensive content (see Appendix~\ref{app:refusal-detection} for details).

% par 3: measurement - classification

We represent opinions on each topic as continuous values in $[0, 1]$, with $0$ denoting ``against'' and $1$ ``in favor''.
To quantify the opinions expressed by both human-written and LLM-generated texts, we develop an ensemble of five classifiers per topic, each using a different pretrained text embedding model.\footnote{We opt for an ensemble instead of a single classifier to ensure our results are robust to the choice of embedding.}
Each classifier embeds a candidate text and assigns a confidence value $[0,1]$ for it being ``in favor'' based on the similarity of its embedding to the average embeddings of human-written texts labeled ``in favor'' and ``against'' on that topic.
We set the (numerical) opinion expressed in the text as the weighted average of these confidence values across the ensemble, weighted by each classifier's accuracy on a held-out set (see Appendix~\ref{app:ensemble} for details and performance metrics).

\begin{figure*}[t]
    \captionsetup[subfigure]{justification=centering}
    \centering
    \subcaptionbox{Original vs. transformed \\opinions on feminism~\label{fig:transformation}}[0.325\textwidth]{
        \centering
        \includegraphics[width=0.325\textwidth]{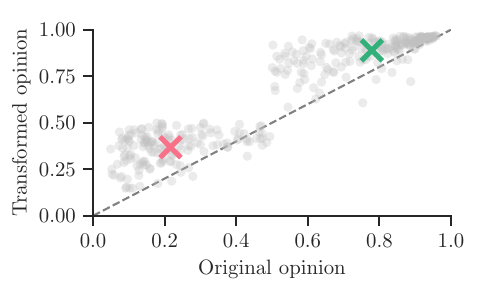}
    } 
    \subcaptionbox{Bias introduced towards\\ ``in favor'' across different topics\label{fig:bias-by-topic}}[0.325\textwidth]{
        \centering
        \includegraphics[width=0.325\textwidth]{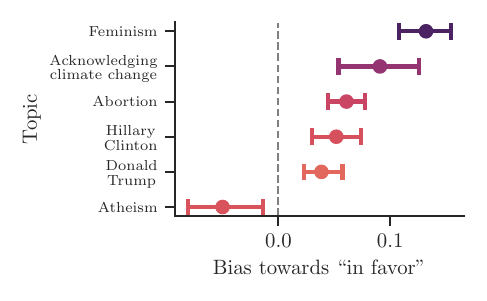}
    }
    \subcaptionbox{Average directly expressed \\ opinion vs. average bias\label{fig:opinion-vs-bias}}[0.325\textwidth]{
        \centering
        \includegraphics[width=0.325\textwidth]{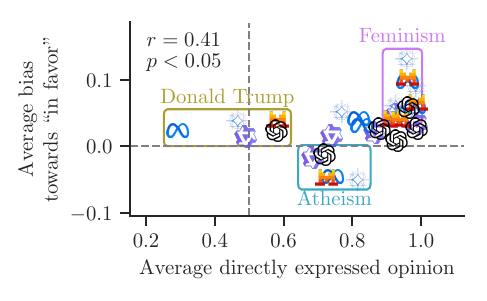}
    } 
    
    \caption{\textbf{Analysis of bias introduced by LLMs when improving human-written posts.}
    Panel (a) shows the original opinions of $400$ posts on feminism from the SemEval dataset against those of their LLM-improved counterparts generated by \texttt{gemma-3-12b-it}, where the green and pink marker correspond to average values for posts labeled ``in favor'' and ``against'', respectively.
    Panel (b) shows the posterior means and $95\%$ credible intervals of the intercepts capturing the average bias $\beta$ introduced by \texttt{gemma-3-12b-it} across topics.
    Panel (c) shows the aforementioned means against the average directly expressed opinions across model-topic pairs, with different markers used for \texttt{Llama-3.1-8B-Instruct} (\includegraphics[height=0.9\inlineheight]{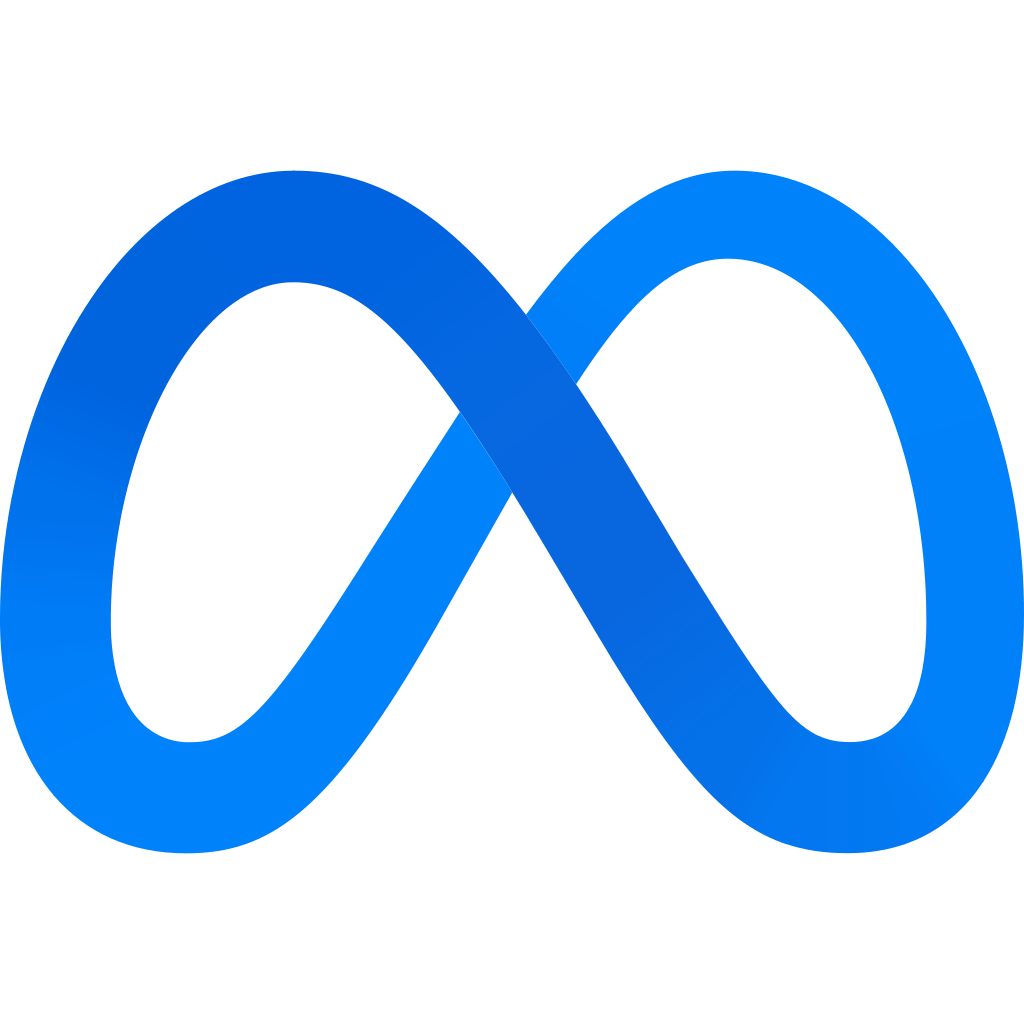}),
    \texttt{gpt-5.6-luna} (\includegraphics[height=0.9\inlineheight]{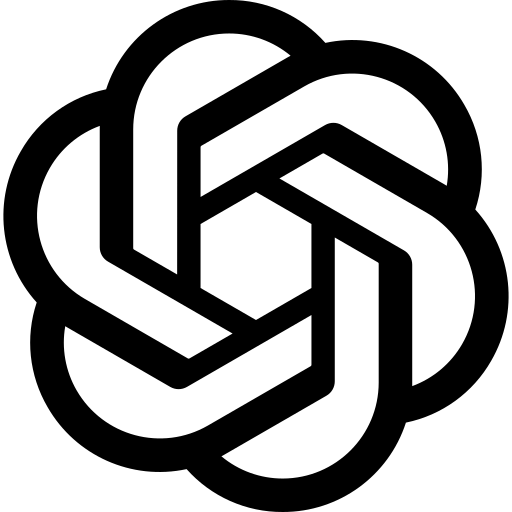}),
    \texttt{gemma-3-12b-it} (\includegraphics[height=0.9\inlineheight]{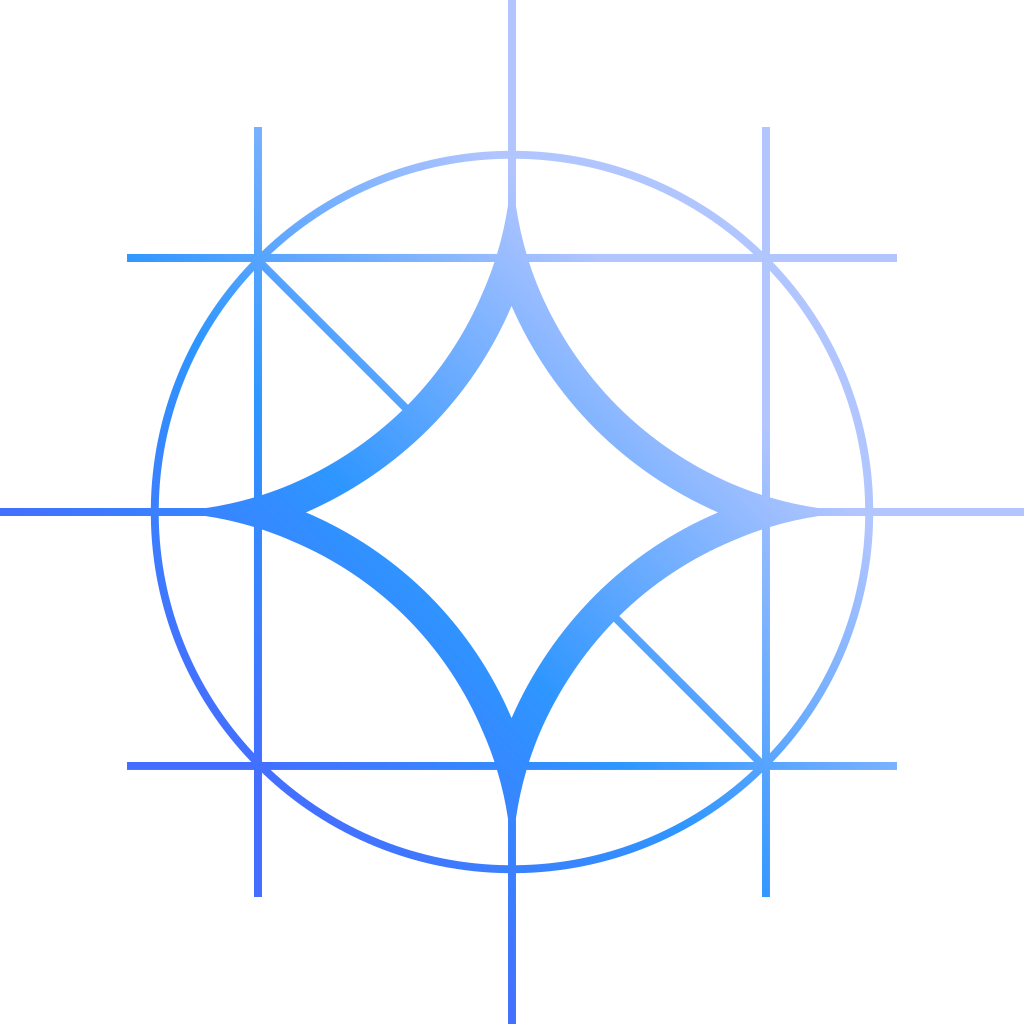}),
    \texttt{Ministral-3-8B-Instruct-2512} (\includegraphics[height=0.9\inlineheight]{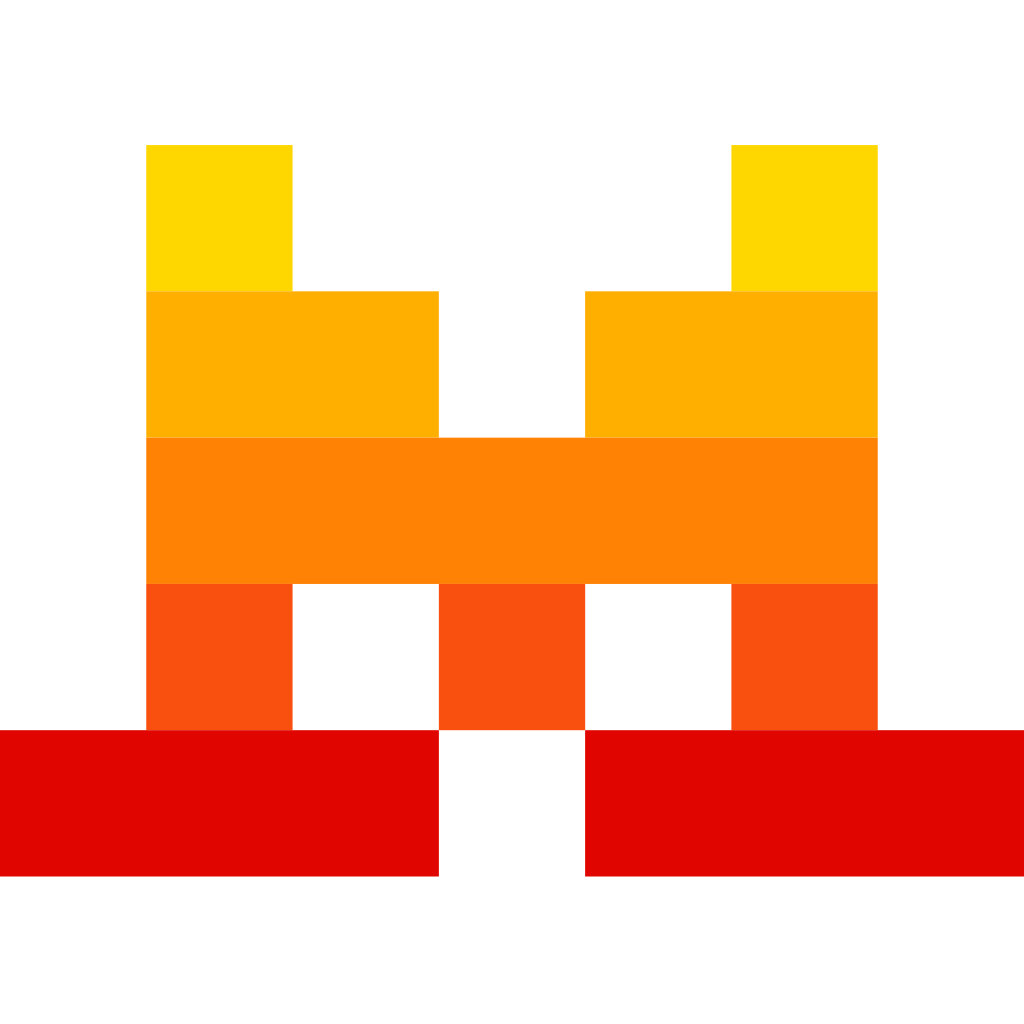}),
    and \texttt{Qwen3-8B} (\includegraphics[height=0.9\inlineheight]{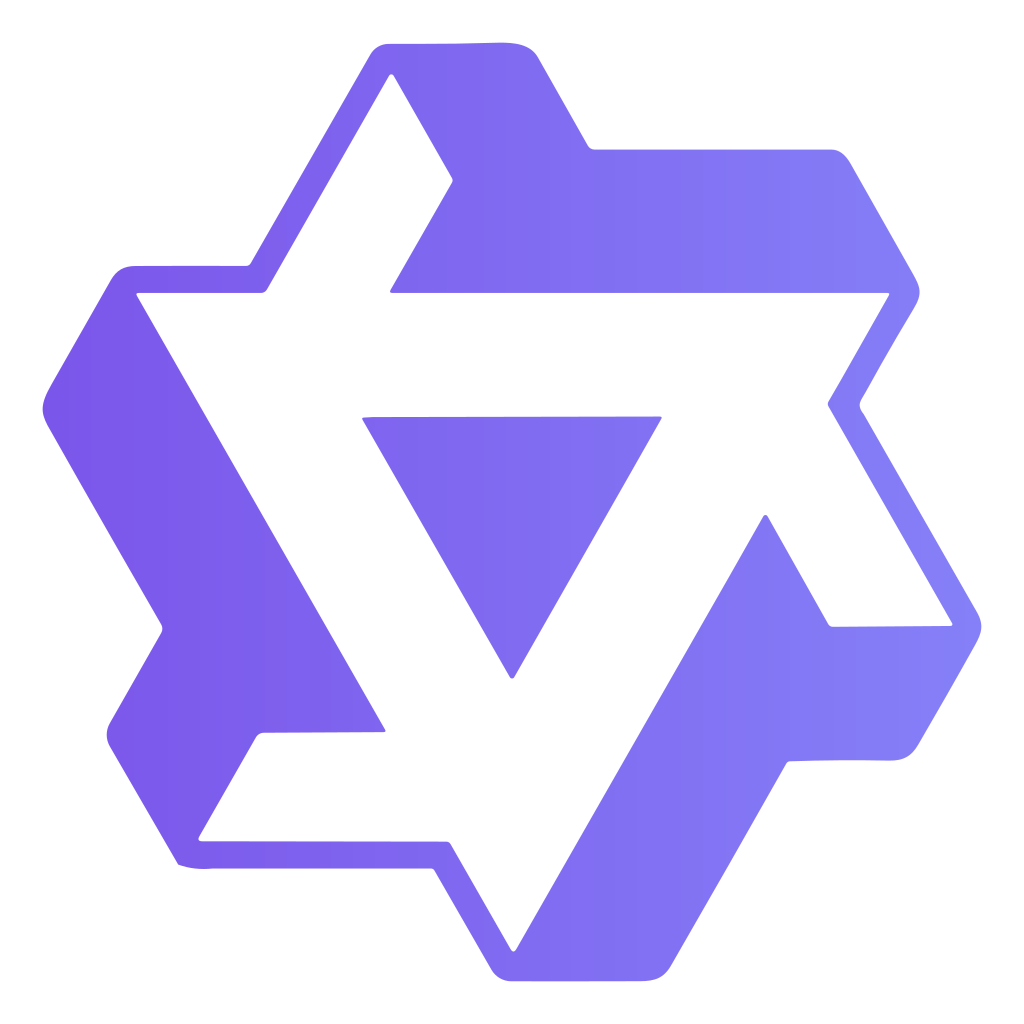}).
    For similar results for the drafting task and the UKP dataset, as well as for additional LLMs, see Appendix~\ref{app:results-bias}.
    }
    \label{fig:bias}
\end{figure*}

% par 4: fig 1(a) - what transformation do the llms apply?
Further, we analyze the relationship between the \emph{original} opinion $x\in[0,1]$ from a human-written text and the \emph{transformed} opinion $y\in[0,1]$ in its 
LLM-generated counterpart, focusing on the post improvement task and the SemEval dataset---results for the drafting task and the UKP dataset are qualitatively similar and can be found in Appendix~\ref{app:results-bias}.
For each topic, we draw a balanced sample of up to $200$ human-written posts labeled ``in favor'' and ``against'', subject to data availability, restricting sampling to posts that are correctly classified by the respective ensemble.
We then ask an LLM to perform the post improvement task. % as described previously.
Since a user would be unlikely to share an LLM-generated post that contradicts the stance they intend to express, we restrict our analysis to LLM-generated posts whose predicted stance matches that of the original post (\ie, $x,y \in [0.5, 1]$ or $x,y \in [0, 0.5]$).\footnote{Appendix~\ref{app:results-flips} shows results qualitatively similar to those we present here when we do not apply this restriction.}
Figure~\ref{fig:transformation} shows an example of the relationship between the original opinions expressed in human-written posts on feminism and those of their LLM-generated counterparts using \texttt{gemma-3-12b-it}.
The model does not preserve %the original
opinions and introduces a directional bias systematically pulling opinions towards ``in favor'' (\ie, most points lie above the diagonal).
See Appendix~\ref{app:examples} for examples  of original and LLM-generated posts across topics and LLMs, illustrating such directional biases.

To analyze if similar patterns appear across different topics and LLMs, we quantify directional bias as the difference $\beta_i = y_i - x_i$ between an original opinion $x_i$ and its transformed counterpart $y_i$.
For each topic and LLM, we then fit a Bayesian linear mixed-effects model~\citep{sorensen2015bayesian,burkner2017advanced} given by
\begin{equation}\label{eq:bayesian_model}
    \beta \;\sim\; 1 + \texttt{original stance} + (1 \mid \texttt{human-written text}) + (1 \mid \texttt{user prompt variant}),
\end{equation}
where the intercept captures the average bias across all $(x_i, y_i)$ pairs, \texttt{original stance} is a binary variable indicating whether the human-written text is ``in favor'' or ``against'', and the two random effects account for repeated measurements per human-written text and per user prompt variant, respectively.\footnote{We use Wilkinson notation~\citep{wilkinson1973symbolic} to specify the model concisely, where additive terms denote fixed effects (\ie, $1$ for the intercept, \texttt{original stance} for the slope) and random effects $(1 \mid \texttt{id})$ denote a random intercept per value of \texttt{id}.}
Figure~\ref{fig:bias-by-topic} shows the posterior mean and $95\%$ credible interval of the intercept of Eq.~\ref{eq:bayesian_model} for \texttt{gemma-3-12b-it} across topics.
We find that the model introduces statistically significant bias on all topics (\ie, the credible intervals exclude zero), with the bias being ``in favor'' on all topics except atheism.\footnote{As a sanity check that these are not artifacts of the method by which we measure opinions, in Appendix~\ref{app:prepending}, we repeat this experiment and prepend a range of ideological viewpoints to the system prompt. The biases shift in predictable directions across topics. For example, 
% the bias in favor of abortion strengthens as the prefix becomes more liberal, while 
the bias in favor of abortion weakens and even reverses as the prefix becomes more conservative.}
The overall patterns we observe remain qualitatively similar for the drafting task and the UKP dataset, as well as for other LLMs, with the exception of a few model-topic pairs for which the bias is not statistically significant (see Appendix~\ref{app:results-bias} for details).
% 

% par 6: fig (c) - bias correlates with the opinion they express when compelled to take a stance
A natural question is whether the direction and magnitude of the bias an LLM introduces aligns with the opinion it expresses on that topic~\citep{santurkar2023whose,rottger2026issuebench,kim2025linear}.
To answer this, we measure each LLM's \emph{directly expressed opinion} on each topic by prompting it to generate a statement, following~\citet{kim2025linear}, and using our ensemble to quantify the opinion expressed in its output.
We compare the average expressed opinion with the average bias introduced by the LLM, as measured by the posterior mean of the intercept in Eq.~\ref{eq:bayesian_model}.
Figure~\ref{fig:opinion-vs-bias} summarizes the results, which show that LLMs from different families present largely similar directly expressed opinions and biases across topics, potentially reflecting their training on largely overlapping internet data.
Moreover, we observe a moderate positive correlation ($r=.41$, $p < 0.05$) between an LLM's directly expressed opinion and the bias it introduces, suggesting that the former leaks into the latter, even when the LLM is instructed to preserve the meaning of human-written text.
Perhaps surprisingly, exceptions exist.
On atheism, for instance, the models express a positive opinion yet tend to introduce biases against it when improving human-written posts on the topic.
This discrepancy suggests that benchmarks measuring LLMs' directly expressed opinions are likely insufficient to capture the subtler biases introduced when mediating human communication.

\section{A mathematical model of AI-mediated opinion dynamics}\label{sec:model}
To study the effects of AI-mediated communication in a social network, we develop a variant of the Friedkin-Johnsen model of opinion dynamics~\citep{friedkin1990social}, which strikes a good balance between realism and analytical tractability.
The model has been empirically validated through human subject experiments and real-world data~\citep{friedkin2011social,childress2012cultural,de2014learning,friedkin2016network,friedkin2016theory,friedkin2017truth,bernardo2021achieving} and received significant attention in computer science~\citep{ghaderi2014opinion,bindel2015bad,fotakis2016opinion,abebe2018opinion,chen2018quantifying,chitra2020analyzing,gaitonde2020adversarial,zhu2021minimizing,wang2023relationship}.

We model a social network as a weighted graph $\Gcal$, composed of $N$ individuals or nodes.
An edge $(i, j)$ indicates that individual $j$ influences the opinion of individual $i$. This is associated with weight $W_{ij} > 0$, indicating the strength of influence.
We assume that  $W_{ij}$ satisfies $\sum_j W_{ij} = 1$ for all $i$ (\ie, the matrix $W$ is row-stochastic) and that $W_{ij}=0$ iff no edge exists between $i$ and $j$.

Opinion formation unfolds over discrete time steps.
Each individual $i$ starts with an innate opinion $x_i(0) \in [0,1]$ and, at each time step $t$, they update their opinion from $x_i(t)$ to $x_i(t+1)$ as a weighted average of the innate opinion and the \emph{perceived} opinions of their neighbors in the social network, \ie,
\begin{equation}\label{eq:our_model}
    x_i(t+1) = \lambda_i \cdot x_i(0) + (1-\lambda_i) \cdot \sum_j W_{ij} \cdot y_j(t).
\end{equation}
Here, $\lambda_i \in (0,1)$ controls the weight $i$ places on their innate opinion (\ie, their \emph{stubbornness}), and $y_j(t) = f(x_j(t))$ denotes the perceived opinion of neighbor $j$ at time $t$.
We refer to $f: [0,1] \to [0,1]$ as the \emph{AI transformation} and note that the perceived opinion $y_j(t)$  may differ from the opinion $x_j(t)$.\footnote{To simplify the analysis, we assume that the opinions of all individuals in the network are transformed by the AI system. We relax this assumption in our experiments in Section~\ref{sec:experiments}.}

Modeling AI-mediated communication as a transformation $f$ from underlying to perceived opinions captures a wide range of practical scenarios on social networks, including those studied in our empirical analysis in Section~\ref{sec:bias}.
For example, in post improvement or assisted writing, $x_j(t)$ encodes the opinion in $j$'s prompt and $y_j(t)$ the opinion in the AI's output, which is what other individuals in the network observe.
In post contextualization, which we focus on in Section~\ref{sec:steering}, $x_j(t)$ encodes the opinion in $j$'s post and $y_j(t)$ the one supported by the AI-generated context, which ultimately shapes how others interpret the post. 
Using vector notation, we express the update rule of Eq.~\ref{eq:our_model} compactly as
\begin{equation}\label{eq:update_map}
    x(t+1) = G\left(x(t)\right) := \Lambda\, x(0) + (I - \Lambda)\, W\, F\left(x(t)\right),
\end{equation}
where $\Lambda$ is the diagonal matrix containing the stubbornness parameters $\lambda_i$, $F: [0,1]^N \to [0,1]^N$ is the elementwise application of $f$ (\ie, $[F(x)]_i = f(x_i)$ for all $i$), $x(0) \in [0,1]^N$ is the vector of innate opinions, and $I$ denotes the identity matrix.
We refer to $G$ as the \emph{update map}.

If no AI mediation occurs, the AI transformation $f$ is the identity (\ie, $f(x) = x$), and our model reduces to the standard Friedkin-Johnsen opinion dynamics model~\citep{friedkin1990social}. 
Under the above conditions, the opinion vector $x(t)$ is known to converge to a unique equilibrium $x^* = G(x^*)$~\citep{friedkin1990social,proskurnikov2017tutorial}.

In what follows, we analyze the convergence and equilibrium properties of our model.
In Section~\ref{sec:theory-linear}, we focus on a linear form of the AI transformation $f$, which yields closed-form expressions for the equilibrium and allows us to gain insights about the effects of AI-mediated communication on opinion formation.
In Section~\ref{sec:experiments}, we complement our theoretical analysis with simulation experiments using real network data and non-linear forms of $f$ estimated from our empirical results in Section~\ref{sec:bias}.

\subsection{Theoretical analysis under a linear transformation}\label{sec:theory-linear}

We consider a transformation that takes the linear form $f_{\text{lin}}(x) = mx + b$, for which we assume that $m \in (0,1)$ and $b \in [0,1]$, and it holds that $m + b \leq 1$.
Together, these conditions ensure that $f_{\text{lin}}$ is a valid AI transformation, as it maps the interval $[0,1]$ into itself.
To better understand the effects of this transformation, it is useful to rewrite it as $f_{\text{lin}}(x) = m\,x + (1-m)\,\nu$, where note that $\nu = b/(1-m)$ is the \emph{neutral point} of the transformation, that is, it satisfies $f_{\text{lin}}(\nu) = \nu$.
Then, a perceived opinion $f_\text{lin}(x)$ can be seen as a weighted combination of the underlying opinion $x$ and the opinion $\nu$ that the AI treats as neutral, where $1-m$ controls the strength by which the transformation pulls opinions towards $\nu$.
Under this linear transformation, the update map of our model takes the form
\begin{equation}\label{eq:linear_dynamics}
    x(t+1) = G_{\text{lin}}\left(x(t)\right) := \underbrace{\Lambda\, x(0)}_{\substack{\text{Innate} \\ \text{opinion}}} \, + \, m \cdot\underbrace{(I - \Lambda) W\,x(t)}_{\substack{\text{Social} \\ \text{influence}}} \, + \, (1-m) \cdot \underbrace{\nu\,(I - \Lambda)\, \mathbf{1}}_{\text{AI bias}},
\end{equation}
where $\mathbf{1}$ denotes the all-ones vector.
Note that the term depending on individuals' innate opinions is the same as in the standard Friedkin-Johnsen model (see Eq.~\ref{eq:update_map} when $F(x) = x$), while the term representing social influence by neighbors is proportional to the respective term in the standard model but scaled down by a factor of $m$.
The main difference between our model and the standard Friedkin-Johnsen model is the presence of the third term, which pulls individuals' opinions towards the AI's neutral point $\nu$ the less stubborn they are (smaller $\lambda_i$) and the stronger the AI transformation is (smaller $m$).
As a consequence, one can view the AI's role in mediating the opinion dynamics as that of an ``invisible neighbor'' with a persistent opinion $\nu$ influencing every individual in the network.

We start by looking into the convergence properties of the dynamics of Eq.~\ref{eq:linear_dynamics}. The following proposition establishes that the dynamics converge to a unique equilibrium, given in closed form, using similar arguments to those used in the analysis of the standard Friedkin-Johnsen model~\citep{bullo2026contraction,proskurnikov2017tutorial}:\footnote{All proofs can be found in Appendix~\ref{app:proofs}.}

\begin{proposition}\label{prop:linear-convergence}
    Let $\tilde{x} = (I - m\,C)^{-1} \cdot \left[\Lambda\, x(0) + (1-m)\,\nu\,(I - \Lambda)\, \mathbf{1}\right]$, where $C = (I - \Lambda) W$. Moreover, let $\rho = m \cdot\left\|I-\Lambda\right\|_\infty < 1$.
    % $\rho = m \cdot \max_i(1 - \lambda_i)<1$.
    It holds that $G_{\text{lin}}(\tilde{x}) = \tilde{x}$ and the dynamics of Eq.~\ref{eq:linear_dynamics} satisfy
    \begin{equation*}\label{eq:linear_convergence_rate}
    \|x(t) - \tilde{x}\|_\infty \;\leq\; \rho^{\,t}\,\|x(0) - \tilde{x}\|_\infty
    \,\,\, \text{for all} \,\,\, t \geq 0.
    \end{equation*}
\end{proposition}
% 
% Since we have assumed that $m \in (0,1)$ and $\lambda_i \in (0,1)$ for all $i$, it holds that $\rho < 1$, hence, the dynamics of Eq.~\ref{eq:linear_dynamics} converge to the unique equilibrium $\tilde{x}$ at a geometric rate.
%
% Moreover, it is worth noting that the matrix $(I - m\,C)^{-1}$ has all entries non-negative, which implies that the larger the bias $b$ injected by the AI, the more individuals' equilibrium opinions are pulled towards the ``in favor'' end of the spectrum.
%

Having established the model's convergence, we now focus on the effect of AI-mediated communication on the opinions held at equilibrium.
To this end, we compare the equilibrium $\tilde{x}$ of the AI-mediated opinion dynamics against the equilibrium $x^*$ that would arise in the absence of AI mediation, which reduces to the equilibrium of the standard Friedkin-Johnsen model $x^* = (I - C)^{-1} \Lambda x(0)$, obtained from Eq.~\ref{eq:update_map} when $F(x)=x$.
%
% As mentioned previously, in the latter case, our model reduces to the standard Friedkin-Johnsen model, and hence the equilibrium $x^*$ can be expressed in closed form as the unique solution of Eq.~\ref{eq:update_map} when $F(x) = x$, which is given by $x^* = (I - C)^{-1} \Lambda x(0)$.
%
The following proposition uses this observation to derive a closed-form expression for the AI-induced \emph{equilibrium shift} $\tilde{x} - x^*$:
\begin{proposition}\label{prop:linear-shift}
    Let $\tilde{x}$ and $x^*$ be the equilibria of the AI-mediated opinion dynamics of Eq.~\ref{eq:linear_dynamics} and the dynamics of the standard Friedkin-Johnsen model, respectively. Then,
    \begin{equation}\label{eq:linear_shift}
        \tilde{x} - x^* = (1 - m)\,(I - m\,C)^{-1}\,(I - \Lambda)\,\bigl[\nu \cdot \mathbf{1} - W\, x^*\bigr].
    \end{equation}
\end{proposition}

Proposition~\ref{prop:linear-shift} offers several insights.
First, %it implies that 
the equilibrium shift $\tilde{x} - x^*$ is not only dependent on the parameters of the AI (\ie, $m$ and $\nu$) and the population's characteristics (\ie, their stubbornness $\Lambda$ and innate opinions $x(0)$), but also on the structure of the influence matrix $W$, indicating that the bias that AI introduces to individuals' opinions may compound and propagate through the social network.
Moreover, the direction of the shift in each individual's opinion is determined by the sign respective entry of the vector $\nu \cdot \mathbf{1} - W\, x^*$ capturing the difference between the AI's neutral point $\nu$ and the social influence that each individual experiences at equilibrium in the absence of AI mediation.\footnote{The elements of the matrix $(I - mC)^{-1}$ are non-negative (refer to the proof of Proposition~\ref{prop:linear-convergence}).}

To investigate the compounding effect of the AI transformation, we further look into the shift in the population's average opinion at equilibrium and compare it to the AI's \emph{average one-off bias}, that is, the average bias the AI transformation introduces to a population's innate opinions in the absence of social influence.
Formally, the AI's average one-off bias is given by
\begin{equation}\label{eq:linear_oneoff_bias}
    B_\text{one-off}\left(f_{\text{lin}}, x(0)\right) \;=\; \frac{1}{N}\sum_{i=1}^N \bigl[f_\text{lin}(x_i(0)) - x_i(0)\bigr] \;=\; (1-m)\,\bigl(\nu - \bar{x}(0)\bigr),
\end{equation}
where $\bar{x}(0) = \frac{1}{N}\sum_i x_i(0)$ is the population's average innate opinion.
Hence, the AI's average one-off bias is larger in magnitude whenever the AI transformation is stronger (\ie, $m$ is smaller) and its neutral point $\nu$ is further away from the population's average innate opinion $\bar{x}(0)$.

Further, we identify conditions under which the AI-induced shift in the population's average opinion at equilibrium strictly exceeds the AI transformation's average one-off bias.
Specifically, we focus on populations in which all individuals share the same level of stubbornness and form a social network structure captured by a doubly stochastic influence matrix $W$ (\ie, both its rows and columns sum to $1$)---note that this includes several realistic structures, such as social networks where pairs of individuals connected with an edge $(i,j)$ correspond to ``friends'' that exert equal influence on each other (\ie, $W_{ij} = W_{ji}$).
Then, we have the following proposition:
% 
% Formally, we have the following proposition:
%
\begin{proposition}\label{prop:linear-avg-shift}
Suppose $\lambda_i = \lambda$ for all $i$ and $W$ is doubly stochastic.
Moreover, let $\bar{x} = \frac{1}{N}\sum_i \tilde{x}_i$ and $\bar{x}^* = \frac{1}{N}\sum_i x_i^*$ denote the population's average opinion at equilibrium under the AI-mediated opinion dynamics and dynamics of the standard Friedkin-Johnsen model, respectively. Then,
\begin{equation*}\label{eq:linear_avg_shift}
\bar{x} - \bar{x}^* \;=\; \frac{1-\lambda}{\lambda + (1-\lambda)(1-m)} \cdot B_\text{one-off}\left(f_{\text{lin}}, x(0)\right),
\end{equation*}
and the scaling factor exceeds $1$ whenever $m\,(1-\lambda) > \lambda$.
\end{proposition}

Proposition~\ref{prop:linear-avg-shift} reveals that, %under natural conditions on the social network and the population's characteristics, 
AI-mediated communication can amplify the bias introduced by the AI transformation to individuals' opinions, leading to a compounding effect on the population's average opinion at equilibrium.
In the next section, we experimentally investigate whether this and our previous insights from the linear case hold under real social network structures and real AI transformations estimated from our empirical results in Section~\ref{sec:bias}.

\subsection{Experiments using real network data and AI transformations}\label{sec:experiments}

We simulate opinion dynamics on real social networks under our model, given by Eq.~\ref{eq:update_map}.
To this end, we use three datasets from the SNAP repository~\citep{leskovec2012learning}, which correspond to real sub-networks of Twitter, Facebook, and Google Plus.
The results we present are based on the Twitter network, which contains $\sim80$ thousand nodes (users) and $\sim1.7$ million edges (follower connections).
We present summary statistics of all three networks 
in Appendix~\ref{app:tables} and qualitatively similar results using the Facebook and Google Plus networks in Appendix~\ref{app:results-simulations}.
%  

% transformation
We consider a scenario where a fraction of users use a platform-provided LLM (here, \texttt{gemma-3-12b-it}) to edit their posts on a specific topic.
To construct realistic AI transformations $f(\cdot)$, we use our data from Section~\ref{sec:bias}, which contain pairs $(x, y)$ of original opinions $x$ expressed in human-written texts and transformed opinions $y$ expressed in their LLM-edited counterparts.
We then estimate the AI transformation $f(\cdot)$ per topic using Nadaraya–Watson kernel regression~\citep{nadaraya1964estimating,watson1964smooth} with Gaussian kernels, which allows us to do so without making any parametric assumptions about the transformation's form.
The estimated (non-linear) AI transformations resulting from \texttt{gemma-3-12b-it} for all topics in the SemEval and UKP datasets can be found in Figure~\ref{fig:transformations-gemma} in Appendix~\ref{app:transformations}.
Relaxing our earlier assumption that the AI transformation is applied to the opinions of all users, at the start of each simulation, we sample a fraction $\phi$ of users that we fix as ``AI adopters''.
For those, we consider that the AI transformation is always applied to their opinions (\ie, $y_j(t) = f(x_j(t))$), while for the rest, we consider that no AI transformation is applied (\ie, $y_j(t) = x_j(t)$). 
% 

% initialization
Each simulation is initialized by drawing each user's stubbornness $\lambda_i$ from a truncated Gaussian $\lambda_i \sim  \bar{\mathcal{N}}(\lambda, 0.05)$, formed by resampling whenever the sample falls outside [0,1].
To set the innate opinions of users, we randomly assign them as positive (with probability $\kappa$)  or negative leaning.
We draw the innate opinion $x_i(0)$ of users with each leaning from truncated Gaussians $\bar{\mathcal{N}}(0.75, 0.1)$ and $\bar{\mathcal{N}}(0.25, 0.1)$, respectively.
Each simulation is run for $100$ time steps and repeated with $20$ seeds.

\begin{figure*}[t]
    \captionsetup[subfigure]{justification=centering}
    \centering
    % \hspace{0.02cm}
    \subcaptionbox{Average opinion over time under\\varying levels of AI adoption\label{fig:adoption}}[0.325\textwidth]{
        \centering
        \includegraphics[width=0.325\textwidth]{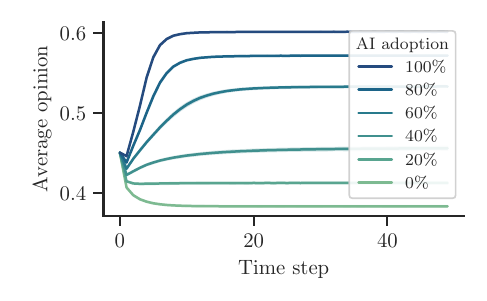}
    } 
    \subcaptionbox{AI bias vs. long-run average opinion across AI transformations\label{fig:shift}}[0.325\textwidth]{
        \centering
        \includegraphics[width=0.325\textwidth]{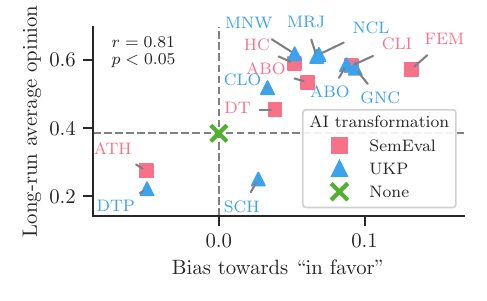}
    }
    \subcaptionbox{Shift in long-run average opinion under different model parameters\label{fig:fj_params}
    }[0.325\textwidth]{
        \centering
        \includegraphics[ width=0.325\textwidth]{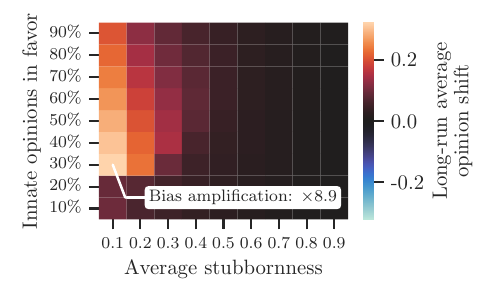}
    }
    \caption{\textbf{Opinion dynamics when \texttt{gemma-3-12b-it} is used to edit posts.} 
    Panel (a) shows the average opinion on abortion over time, for different fractions $\phi$ of AI adopters.
    Panel (b) shows the long-run average opinion under AI transformations from different topics and datasets (see Appendix~\ref{app:tables} for abbreviations) against the AI's bias, as measured by the posterior mean of the intercept in Eq.~\ref{eq:bayesian_model}.
    ``X'' indicates no AI transformation.
    Panel (c) shows the difference in the long-run average opinion on abortion between $\phi=0.6$ and $\phi=0$, across varying values of $\lambda$ and $\kappa$. The overlaid annotation shows the ratio of that difference to the average one-off bias introduced by the AI to users' innate opinions (see Eq.~\ref{eq:linear_oneoff_bias}).
    In panels (a, b), $\kappa=0.4$ and $\lambda=0.3$, and in panel (b), $\phi=0.6$.
    }
    \label{fig:network_simulations}
\end{figure*}

Figure~\ref{fig:adoption} shows the evolution of the average opinion on abortion varying $\phi$, the fraction of AI adopters.
In this setting, the average innate opinion of the population (\ie, the value at $t=0$) is against abortion and, when opinions evolve without any AI transformation (\ie, $\phi=0$) the average opinion remains against abortion over time.
However, since the AI transformation introduces a positive bias towards abortion (see Figure~\ref{fig:bias-by-topic}), as the fraction of AI adopters increases, the average opinion becomes more positive.
Since under non-linear AI transformations such as the one used for Figure~\ref{fig:adoption}, the opinion dynamics in our model do not necessarily converge to an equilibrium, it is important to observe that, in practice, the average opinion does converge.\footnote{In Appendix~\ref{app:convergence}, we show that, across several topics and parameter configurations, opinions of individual users may keep changing over time, yet the average opinion stabilizes.}
In the remainder of the section, we focus on the \emph{long-run average opinion}, 
that is, the average opinion at the final  time step.

We analyze how the bias introduced by the AI to the opinions of AI adopters across the network influences the long-run average opinion reached by the dynamics.
Figure~\ref{fig:shift} shows the long-run average opinion under different AI transformations corresponding to the different topics included in the SemEval and UKP datasets against the AI's bias on the respective topic, as measured by the posterior mean of the intercept in Eq.~\ref{eq:bayesian_model}.
We observe  a positive correlation between the two quantities --- the more bias the AI introduces towards a certain direction when mediating communication between different users, the more the average opinion shifts in the direction of that bias.

Finally, we analyze how the AI influences the long-run average opinion under varying user populations.
Specifically, we vary the average stubbornness $\lambda$ and the fraction of users $\kappa$ whose innate opinions lean positive on the topic, and measure the difference in the long-run average opinion between the case where $60\%$ of users are AI adopters (\ie, $\phi=0.6$) and the case where no AI transformation is applied (\ie, $\phi=0$).\footnote{In Appendix~\ref{app:extension}, we also introduce and simulate a variant of our model in which an individual's stubbornness and AI adoption change over time, depending on their opinion and the social influence they receive, respectively.}
Figure~\ref{fig:fj_params} summarizes the results for abortion, and we obtain similar results across topics (see Figures~\ref{fig:fj_params_twitter}-\ref{fig:fj_params_facebook} in Appendix~\ref{app:results}).
In line with Prop.~\ref{prop:linear-avg-shift}, we find that the shift in the long-run average opinion due to the AI is always in the direction of the AI's (positive) bias towards abortion and, as one would expect, grows stronger as users become less stubborn.
The shift is the strongest in situations where the innate opinions of the user population are almost balanced but the AI's bias is favoring the minority and pulls the rest of the network on its side. 
Lastly, as discussed earlier, it is worth noting that the AI's bias is amplified through the network, with the shift in the long-run average opinion being up to $8.9$ times larger than the average one-off bias given by Eq.~\ref{eq:linear_oneoff_bias}.

\section{Bias by design: A case study on X}\label{sec:steering}
We investigate whether biases introduced in AI-mediated communication can be controlled by platform design choices, effectively granting platforms the ability to steer collective opinion through the design of their AI features.
As a case study, we audit the ``Explain this post'' feature on X, which uses Grok to provide users with additional context about other users' posts~\citep{x_context}.
This task is particularly relevant to our setting, since AI has been shown to persuade humans on politically salient topics by providing facts and evidence in support of a specific stance~\citep{hackenburg2025levers,coppock2023persuasion}.

We focus on abortion-related posts from the SemEval dataset~\citep{mohammad2016semeval}, and analyze whether the context provided by Grok to such posts aligns more with pro-choice, neutral, or pro-life values.
We consider only posts that currently have an active URL on X; restrict our set to one post per user account to avoid our results being disproportionately influenced by individuals; and balance the set of posts. This results in $39$ pro-choice posts and $39$ pro-life.

We replicate the implementation of the ``Explain this post'' feature using the official (publicly released) prompt template deployed on X, which specifies four guidelines on what constitutes a good response alongside formatting instructions.\footnote{The prompt template is available at \href{https://github.com/xai-org/grok-prompts/blob/main/grok_analyze_button.j2}{https://github.com/xai\-org/grok\-prompts/blob/main/grok\_analyze\_button.j2}. For completeness, we also provide it in Appendix~\ref{app:prompts}.}
For each post, we populate the prompt template with its respective URL and query \texttt{grok-4-1-fast-reasoning} via xAI's official API with both X search and web search enabled. 
In each response, the model returns a triplet ($3$) of bullet points, where each bullet point contains a single-sentence contextual claim.
Since Grok generates outputs stochastically by default, we repeat this process $5$ times per post, yielding $1{,}170$ claims in total.
We classify each claim's stance as ``In favor'', ``Neutral'', or ``Against'' using \texttt{gpt-5.4} as a judge~\citep{zheng2023judging}.\footnote{We have tried to reuse the classifiers we developed in Section~\ref{sec:bias}, but found that the presence of many neutral claims in Grok's outputs made them inadequate for this task. Hence, we opt for an LLM-as-a-judge approach.}
To mitigate judgment biases introduced by the judge, we provide it with $5$ (few-shot) examples of claims belonging to each of the three categories~\citep{brown2020language}, drawn from the UKP dataset~\citep{stab2018cross}, which contains labeled single-sentence arguments on abortion that closely match the format and style of Grok's generated claims (see Appendix~\ref{app:prompts} for the judge prompt and the per category examples).

\begin{figure*}[t]
    \captionsetup[subfigure]{justification=centering}
    \centering

    \makebox[\textwidth]{
        \includegraphics[width=0.325\textwidth]{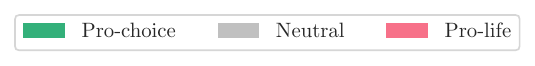}
    }
    \subcaptionbox{Stance of Grok's claims\\vs. stance of human-written post~\label{fig:grok_stats}}[0.325\textwidth]{
        \centering
        \includegraphics[width=0.325\textwidth]{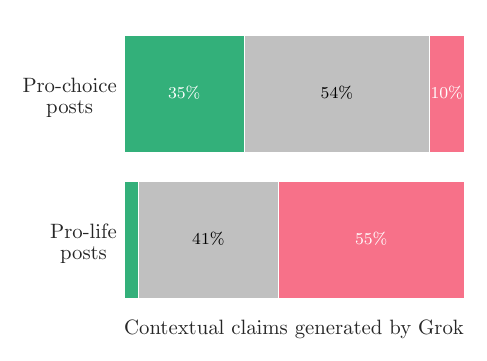}
    } 
    \subcaptionbox{Contribution of X's guidelines to the\\stance distribution of Grok's generated claims~\label{fig:grok_guidelines}}[0.65\textwidth]{
        \centering
        \includegraphics[width=0.65\textwidth]{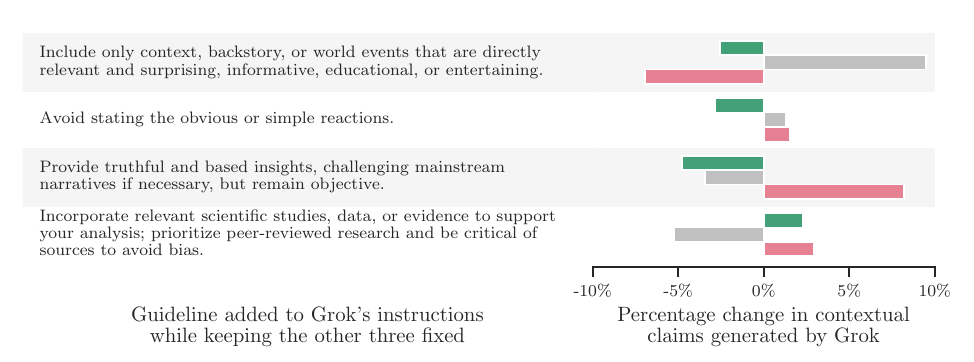}
    } 
    
    \caption{\textbf{Bias introduced by Grok when contextualizing X posts on abortion.}
    Panel (a) shows the stance distribution of contextual claims generated by Grok based on the implementation of X's ``Explain this post'' feature, broken down by whether the post is pro-choice or pro-life.
    Panel (b) shows the four guidelines included by X in the model's instructions and the change in the stance distribution of Grok's contextual claims resulting from introducing each guideline on top of the other three. 
    }
    \label{fig:grok}
\end{figure*}

Figure~\ref{fig:grok_stats} summarizes the results.
% 
%We observe that, 
For posts expressing a pro-choice stance, $35\%$ of Grok's claims support it and $10\%$ oppose it, with the majority being neutral.
However, the overall picture differs substantially for pro-life posts.
Here, the majority of Grok's claims support the pro-life stance, a large portion is neutral, and only $4\%$ oppose it, suggesting a directional bias towards the pro-life stance.
% 

% Statistical test for bias + results
To analyze the model's behavior systematically, we define two measures.
The \emph{support bias} captures the asymmetry in how often Grok echoes the stance of a post it is explaining, \ie,
$\beta_\text{sup} = P(\text{Claim pro-life} \mid \text{Post pro-life}) - P(\text{Claim pro-choice} \mid \text{Post pro-choice})$.
On the other hand, the \emph{opposition bias} captures the asymmetry in how often Grok contradicts the post's stance, \ie, $\beta_\text{opp} = P(\text{Claim pro-life} \mid \text{Post pro-choice}) - P(\text{Claim pro-choice} \mid \text{Post pro-life})$.
To estimate those quantities, we fit a Bayesian categorical mixed-effects model to Grok's generated claims,
\begin{equation}\label{eq:bayesian_model_grok}
    \texttt{claim stance} \;\sim\; 1 + \texttt{post stance} + (1 \mid \texttt{post}) + (1 \mid \texttt{claim triplet}),
\end{equation}
which predicts a claim's stance using multinomial logistic regression with the original post's stance as predictor, and accounts for repeated measurements per post and for within-triplet correlations of claims via its two random effects.
We find that the posterior means of $\beta_\text{sup}$ and $\beta_\text{opp}$ are $0.24$ ($95\%$ CI: $[0.09, 0.38]$) and $0.04$ ($95\%$ CI: $[0.01, 0.08]$), respectively, indicating that both biases are statistically significant (\ie, the credible intervals exclude zero), but revealing that Grok's tendency to echo pro-life posts is stronger than its tendency to contradict pro-choice posts.
%

% Panel (b) observations + how the test's results are affected
We investigate whether these biases are shaped by X's design choices in the ``Explain this post'' feature.
We repeat the previous procedure, each time omitting one of the four guidelines in X's official prompt template, and examine how the distributions of Grok's contextual claims change when guidelines are added back.
Figure~\ref{fig:grok_guidelines} summarizes the results, revealing that each guideline has a distinct effect.
The first guideline (``Include only context, backstory [...]'') results in more neutral claims than supportive or opposing ones, while the fourth (``Incorporate relevant scientific studies [...]'') has the opposite effect.
Most strikingly, the third guideline (``Provide truthful and based insights, challenging mainstream narratives [...]'') has a heavily asymmetric effect---adding it to the other three substantially increases pro-life claims while suppressing both pro-choice and neutral ones.

Finally, we test if this is the main driver of the support and opposition biases $\beta_\text{sup}$ and $\beta_\text{opp}$ observed earlier.
We fit a Bayesian model similar to Eq.~\ref{eq:bayesian_model_grok} to all claims generated under all five guideline combinations, augmented with the guideline combination as an additional predictor.
We compute the posterior of the bias differences $\Delta^{(k)}_\text{sup} = \beta_\text{sup} - \beta^{(k)}_\text{sup}$ and $\Delta^{(k)}_\text{opp} = \beta_\text{opp} - \beta^{(k)}_\text{opp}$, where $(k)$ denotes the exclusion of the $k$-th guideline.
The third guideline is the only one whose inclusion leads to a statistically significant increase in both biases, with posterior means of $\Delta^{(3)}_\text{sup}=0.20$ ($95\%$ CI: $[0.11, 0.28]$) and $\Delta^{(3)}_\text{opp} = 0.05$ ($95\%$ CI: $[0.02, 0.08]$).
Crucially, without this guideline, neither the support bias nor the opposition bias are statistically significant, rendering this guideline a key design choice responsible for Grok's pro-life bias in its generated contextual claims.

\section{AI, opinion formation, and EU law}\label{sec:legal}
Our findings point towards a severe accountability gap in European Union (EU) tech regulation.
While it is often said that Europe is ``heavily'' regulating the tech sector, in what follows, we argue that EU laws are unlikely to mitigate potential harms resulting from the biases we have identified in our work.
In other words, the fact that LLMs may introduce biases in users' expression and interpretation of opinions, while problematic, is still legal.
The two main frameworks that could potentially prevent such harms are the European Union's Artificial Intelligence Act (AIA) which is not yet in force and the Digital Services Act (DSA) which is already legally binding.
However, it is worth noting that the current geopolitical landscape also makes any effective enforcement of present or future rules questionable.

Art. 53 AIA creates duties for all providers of general-purpose AI models.
These include the duty to keep technical documentation about the model, including its training and testing processes and the results of its evaluation (Art. 53 (1)(a)).
Providers also need to make information available to developers about capabilities and limitations of their models (Art. 53 (1)(b)(i)-(ii)).
However, it is unclear whether providers need to test or prevent their models from nudging opinionated texts towards a certain direction, and whether this information needs to be disclosed. 

Art. 55 (1)(a-d) AIA creates additional duties for providers of general-purpose AI models with systemic risks, which are models trained with more than $10^{25}$ FLOPs (see also Art. 51 (2) AIA).
Providers of these models must perform model evaluations (including red teaming) and assess and mitigate possible systemic risks.
Serious incidents need to be reported to the AI Office~\citep{wachter2023limitations}.
However, it is unlikely that inducing bias into a user's text would classify as a systemic risk.
Systemic risks include illegal, false, or discriminatory content or disinformation generated by general-purpose AI models that propagates at scale across the value chain (Art. 3(65) AIA).
Biased outputs do not seem to pass this threshold.\footnote{Recital 110 AIA names among others the following systemic risks: ``the dissemination of illegal, false, or discriminatory content'' and ``harmful bias and discrimination with risks to individuals, communities or societies; the facilitation of disinformation or harming privacy with threats to democratic values and human rights''.}

Annex III (8)(b) of the AIA declares that ``AI systems intended to be used for influencing the outcome of an election or referendum or the voting behavior'' of people are considered ``high risk.''\footnote{If classified as high risk, developers have to follow several obligations before their system can be placed on the market. That would mean that developers need to implement risk management systems (Art. 9),  undergo rigorous bias testing of their models (Article 10), keep technical documentation (Art. 11), enhance transparency (Art. 13), and ensure that their systems are explainable enough to enable meaningful human oversight (Art. 14).}
Influencing voting behavior is not the explicit intended goal of the LLMs and platform features we have studied, therefore Annex III of the AIA and their safeguards would not apply.
Further research is needed to evaluate whether the design choices behind these models can influence individuals' opinions, collective opinion, or even voting behavior.
If a causal link is established, the AIA could potentially apply.

A somewhat more helpful provision is found in Art. 50 (4) AIA.
This article creates a disclosure duty for deployers of general-purpose AI systems that generate or manipulate image, audio or video content constituting a deepfake.\footnote{This duty complements the marking and detection duties placed on providers in Art. 50 (2) AIA which requires them to ensure generated or manipulated image, audio, video, or text content is marked as artificially generated or manipulated, and to provide detection tools to detect this mark.
For more, see Section 4.2 of the European Commission's Guidelines on the implementation of the transparency obligations for certain AI systems under Article 50 of Regulation (EU) 2024/1689 (the `AI Act')~\citep{eu_latest_guidelines}.}
Users must be able to see that the outputs are artificially generated or manipulated.
However, this duty does not extend to AI generated text unless the deployer is informing the public on matters of public interest (\eg, news agencies).\footnote{Art. 50 (4) AIA states that ``[d]eployers of an AI system that generates or manipulates text which is published with the purpose of informing the public on matters of public interest shall disclose that the text has been artificially generated or manipulated.
This obligation shall not apply where the use is authorised by law to detect, prevent, investigate or prosecute criminal offences or where the AI-generated content has undergone a process of human review or editorial control and where a natural or legal person holds editorial responsibility for the publication of the content.''}
It is thus unlikely that online platforms themselves will need to meet this obligation, except in specific cases where they can be seen as deployers informing the public on matters of public interest.\footnote{For further guidance, see section 6.2.1 of the European Commission's Guidelines on the implementation of the transparency obligations for certain AI systems under Article 50 of Regulation (EU) 2024/1689 (the `AI Act')~\citep{eu_latest_guidelines}.}

% According to the watermarking duty in Art.~$50$ (4) AIA, readers of posts on X or LinkedIn would probably not need to be informed that the content did not fully originate from a human, but was (partially)  written by an LLM.
% %  
% However, if image, audio, or videos are created, the duty would apply. 
% %  
% Watermarking could potentially increase users’ critical engagement with AI-generated posts, however, further research is needed to evaluate the effects of watermarking on users' trust in online content.
However, a more general disclosure duty applies to deployers when image, audio, or video deepfakes are involved.
A blanket disclosure duty for all AI generated content could potentially increase users' critical engagement with AI-generated posts.
With that said, further research is needed to evaluate the effects of disclosure and labeling on users' trust in online content.

The DSA creates obligations for all platform providers and search engines, but unfortunately also falls short in protecting individuals against the potential harms resulting from the biases we have identified.
Very Large Online Platforms (VLOPs) and Very Large Search Engines (VLSEs)---those with an average of $45$ million or more monthly users---face heightened obligations (Art. 33).
Platforms such as X or LinkedIn are, among other things, obligated to investigate and mitigate so-called systemic risks (Art. 34), conduct independent audits (Art. 37), and allow vetted researchers access to detect, identify and understand systemic risks and the effectiveness of measures taken by the providers (Art. 40) (see also pages $30$-$31$ in~\citet{wachter2024large}). 

Art. 34 (1)(a-d) DSA defines systemic risk as illegal content, negative effects on human rights such as freedom of expression and information, negative effects on civic discourse, electoral processes, and public security, gender-based violence, negative effects on public health and minors, as well as negative consequences on physical and mental well-being. 

While we believe that introducing bias in users' texts could potentially have negative effects on human rights such as freedom of expression and information, and on civic discourse and voting behavior, further research is needed to establish this causal link.
Thus, it is not clear whether judges would classify LLMs biasing a user's post as a systemic risk.
Biased outputs do not seem to cross the threshold of human rights violation or misinformation and disinformation which need to be false or misleading.\footnote{See the European Democracy Action Plan~\citep{european_democracy_action_plan} (Section 4) and the First Report of the European Board for Digital Services in cooperation with the Commission pursuant to Article 35(2) DSA~\citep{european_board_digital_services} (Page 19) on the most prominent and recurrent systemic risks as well as mitigation measures.}  

However, even if these obligations applied, enforcement would remain questionable.
The current geopolitical landscape makes clear that Big Tech companies are opposing EU tech laws, including the DSA.
A recent report claimed that ``[a]ll platforms have unsubscribed from Commitment 27, which required them to develop, fund, and collaborate with an independent third-party body to enable data access for researchers''~\citep{eu_code_of_practice_on_disinformation}.
Moreover, Microsoft and Google refuse to adhere to the EUs fact checking rules, with
Meta and TikTok threatening to stop soon~\citep{eu_code_of_practice_on_disinformation},
and X has long been in violation of DSA rules~\citep{euractiv}.

\section{Discussion}\label{sec:discussion}
In this section, we first discuss the broader societal implications of our work.
Then, we outline its limitations and highlight interesting directions for future research.

\xhdr{Broader societal implications}
While our empirical analysis in Sections~\ref{sec:bias} and~\ref{sec:steering} has focused on different forms of bias in LLM outputs, it is worth considering---and studying further---what the real-world influence of such outputs on human opinions and what the resulting downstream societal impacts could be.
Imagine if journalists or politicians~\citep{footnote_17} used X's ``Explain this post''~\citep{x_context} feature to learn about topics of public interest, yet received biased results.
What if lawyers or doctors used services such as LinkedIn's post improvement feature~\citep{linkedin_improve} and it introduced ambiguities or one-sidedness into their posts?
What if scientists or students used YouTube's LLM-generated video summaries~\citep{youtube_summary} to assess whether content is relevant to their academic work, only to receive skewed outputs?
In all such cases, LLMs would be silently shaping the text that people write and read.
In turn, if such biased content is shared on large-scale online platforms, it has the potential to slowly shift collective opinion and sway political views or influence voting behavior.
This raises deeper concerns about fundamental rights, including the right to access information and free expression, and connects with broader ongoing debates about the liability of tech companies for the outputs of their models~\citep{wachter2024large,mittelstadt2023protect}.

These concerns are not limited to online platforms. 
Universities increasingly encourage staff and students to integrate AI into their learning~\citep{footnote_12_1,footnote_12_2}, media outlets use LLMs to identify trends and write copy~\citep{footnote_13_1,footnote_13_2}, and policymakers~\citep{footnote_14_1,footnote_14_2}, lawyers~\citep{footnote_15}, and doctors~\citep{footnote_16_1,footnote_16_2} increasingly rely on these tools to research, summarize, and draft text in their daily practice, as well as to structure their thoughts.
If these systems introduce or exacerbate bias in the text that professionals write and the public reads, the consequences for politics, medicine, science, legal practice, and the free press could be severe.
Existing regulatory frameworks are powerless to mitigate such risks, and the current push toward deregulation makes this even more pressing.
We hope that our work encourages policymakers to take action in order to prevent the slow and undetected effects that LLMs may have on collective opinion.

\xhdr{Limitations and future work}
Our work comes with certain limitations, which open up many interesting avenues for future work.
For example, in Section~\ref{sec:bias}, we used an ensemble of classifiers to measure biases introduced by LLMs when editing human-written text.
It would be valuable to extend this analysis by involving human participants, both to mitigate the effects of the ensemble's potential inaccuracies, and to measure how an LLM's biased editing affects a user's adoption of the LLM's suggestions~\citep{rottger2026measuring}.
In addition, we focused our theoretical analysis in Section~\ref{sec:model} on the Friedkin-Johnsen model of opinion dynamics and a linear (opinion) transformation function.
It would be interesting to extend this analysis to other forms of opinion dynamics~\citep{hegselmann2002opinion,holley1975ergodic,weisbuch2002meet} and to non-linear transformation functions. 
Further, to establish a direct link between our theoretical analysis and the real-world effects of AI-mediated opinion formation, it would be useful to go beyond a model-based analysis and conduct a large-scale user study on an online platform to obtain real measurements of the effects of AI's biases on the network level.
Moreover, AI-mediation should be studied in conjunction with other factors that may affect opinion formation, such as platform interventions to the edges of a social network~\citep{miyauchi2026survey}, real-world events external to the platform~\citep{lin2025oden}, or traditional media~\citep{out2024impact}.
Lastly, our experiments in Section~\ref{sec:steering} have focused on a set of posts that, although still available on X, were orinally published in 2016 and are focused on a single topic (abortion).
It would be useful to conduct a more comprehensive analysis using more recent posts and a wider range of topics. 

\section{Conclusions}\label{sec:conclusions}
In this work, we have focused on the use of generative AI systems to mediate human-to-human communication on online platforms.
Our empirical analysis of LLMs from multiple popular families shows that they systematically introduce directional biases when drafting or improving texts on a wide range of contested topics.
Further, We have introduced a mathematical model of AI-mediated opinion dynamics and analytically characterized its convergence and equilibrium properties.
We have shown, both analytically and through simulations using real data, that biases introduced by AI in human-to-human communication can be amplified through a social network and shift collective opinion.
Finally, as a case-study of AI-mediated communication, we audited the “Explain this post” feature from X that uses Grok to contextualize users' posts.
We have found evidence of a directional bias in favor of the pro-life stance on abortion-related posts, and traced this to one specific component of the prompt given to the model by X. 
Our combined findings demonstrate that AI-mediated communication is a novel lever for online platforms to influence opinion formation.

\vspace{4mm}
\xhdr{Acknowledgments} 
We thank Carolin Kemper and Daria Onitiu for helpful feedback that materially improved the legal analysis in Section~\ref{sec:legal}.
We also thank Jonathan Rystr\o m for helpful comments that improved the clarity of the paper.
This work has been supported by the Alexander von Humboldt
Foundation in the framework of the Alexander von Humboldt Professorship (Humboldt Professor of Technology and Regulation awarded to Sandra Wachter) endowed by the Federal Ministry of Education and
Research via the Hasso Plattner Institute; and by the ERC Horizon project  DVPS (101213369).
Kai Rawal is funded by an EPSRC doctoral scholarship and a General Track Grant from the Cosmos Institute. 

\vspace{1mm}
\xhdr{LLM usage statement} 
We used a coding agent to facilitate routine coding tasks, such as writing configuration files, refactoring command line and function arguments, spotting bugs, and writing bash scripts for SLURM scheduling. All LLM-generated code has been closely reviewed and further edited.
No LLM was used in the conceptualization or writing of the paper.

\bibliographystyle{unsrtnat}
\bibliography{refs}

@book{bullo2026contraction,
  author = {F. Bullo},
  title = {Contraction Theory for Dynamical Systems},
  year = 2026,
  edition = {{1.3}},
  publisher = {Kindle Direct Publishing},
  ISBN = {979-8836646806},
}

@article{peterson2025ai,
  title={{AI} and the problem of knowledge collapse},
  author={Peterson, Andrew J},
  journal={AI \& SOCIETY},
  volume={40},
  number={5},
  pages={3249--3269},
  year={2025},
  publisher={Springer}
}

@article{degroot1974reaching,
  title={Reaching a consensus},
  author={DeGroot, Morris H},
  journal={Journal of the American Statistical association},
  volume={69},
  number={345},
  pages={118--121},
  year={1974},
  publisher={Taylor \& Francis}
}

@article{friedkin1990social,
  title={Social influence and opinions},
  author={Friedkin, Noah E and Johnsen, Eugene C},
  journal={Journal of mathematical sociology},
  volume={15},
  number={3-4},
  pages={193--206},
  year={1990},
  publisher={Taylor \& Francis}
}

@article{wachter2024large,
  title={Do large language models have a legal duty to tell the truth?},
  author={Wachter, Sandra and Mittelstadt, Brent and Russell, Chris},
  journal={Royal Society Open Science},
  volume={11},
  number={8},
  pages={240197},
  year={2024},
  publisher={The Royal Society}
}

@inproceedings{bertrand2024on,
title={On the Stability of Iterative Retraining of Generative Models on their own Data},
author={Quentin Bertrand and Joey Bose and Alexandre Duplessis and Marco Jiralerspong and Gauthier Gidel},
booktitle={The Twelfth International Conference on Learning Representations},
year={2024},
}

@inproceedings{taitler2025braess,
  title={Braess’s paradox of generative {AI}},
  author={Taitler, Boaz and Ben-Porat, Omer},
  booktitle={Proceedings of the AAAI Conference on Artificial Intelligence},
  volume={39},
  pages={14139--14147},
  year={2025}
}

@inproceedings{gerstgrasser2024is,
title={Is Model Collapse Inevitable? {Breaking} the Curse of Recursion by Accumulating Real and Synthetic Data},
author={Matthias Gerstgrasser and Rylan Schaeffer and Apratim Dey and Rafael Rafailov and Tomasz Korbak and Henry Sleight and Rajashree Agrawal and John Hughes and Dhruv Bhandarkar Pai and Andrey Gromov and Dan Roberts and Diyi Yang and David L. Donoho and Sanmi Koyejo},
booktitle={First Conference on Language Modeling},
year={2024},
}

@article{schaeffer2025position,
  title={Position: Model collapse does not mean what you think},
  author={Schaeffer, Rylan and Kazdan, Joshua and Arulandu, Alvan Caleb and Koyejo, Sanmi},
  journal={arXiv preprint arXiv:2503.03150},
  year={2025}
}

@article{shumailov2024ai,
  title={{AI} models collapse when trained on recursively generated data},
  author={Shumailov, Ilia and Shumaylov, Zakhar and Zhao, Yiren and Papernot, Nicolas and Anderson, Ross and Gal, Yarin},
  journal={Nature},
  volume={631},
  number={8022},
  pages={755--759},
  year={2024},
  publisher={Nature Publishing Group UK London}
}

@article{friedkin2016theory,
  title={A theory of the evolution of social power: Natural trajectories of interpersonal influence systems along issue sequences},
  author={Friedkin, Noah E and Jia, Peng and Bullo, Francesco},
  journal={Sociological Science},
  volume={3},
  pages={444--472},
  year={2016}
}

@article{childress2012cultural,
  title={Cultural reception and production: The social construction of meaning in book clubs},
  author={Childress, C Clayton and Friedkin, Noah E},
  journal={American Sociological Review},
  volume={77},
  number={1},
  pages={45--68},
  year={2012},
  publisher={Sage Publications Sage CA: Los Angeles, CA}
}

@article{hardt2025performative,
  title={Performative prediction: Past and future},
  author={Hardt, Moritz and Mendler-D{\"u}nner, Celestine},
  journal={Statistical Science},
  volume={40},
  number={3},
  pages={417--436},
  year={2025},
  publisher={Institute of Mathematical Statistics}
}

@article{hegselmann2002opinion,
	title = {Opinion Dynamics and Bounded Confidence: Models, Analysis and Simulation},
	volume = {5},
	shorttitle = {Opinion {Dynamics} and {Bounded} {Confidence}},
	number = {3},
	urldate = {2025-11-07},
	journal = {Journal of Artificial Societies and Social Simulation},
	author = {Rainer, Hegselmann and Krause, Ulrich},
	year = {2002},
}

@article{friedkin2017truth,
  title={How truth wins in opinion dynamics along issue sequences},
  author={Friedkin, Noah E and Bullo, Francesco},
  journal={Proceedings of the National Academy of Sciences},
  volume={114},
  number={43},
  pages={11380--11385},
  year={2017},
  publisher={National Academy of Sciences}
}

@book{friedkin2011social,
  title={Social influence network theory: A sociological examination of small group dynamics},
  author={Friedkin, Noah E and Johnsen, Eugene C},
  volume={33},
  year={2011},
  publisher={Cambridge University Press}
}

@article{ghaderi2014opinion,
  title={Opinion dynamics in social networks with stubborn agents: Equilibrium and convergence rate},
  author={Ghaderi, Javad and Srikant, Rayadurgam},
  journal={Automatica},
  volume={50},
  number={12},
  pages={3209--3215},
  year={2014},
  publisher={Elsevier}
}

@article{bindel2015bad,
  title={How bad is forming your own opinion?},
  author={Bindel, David and Kleinberg, Jon and Oren, Sigal},
  journal={Games and Economic Behavior},
  volume={92},
  pages={248--265},
  year={2015},
  publisher={Elsevier}
}

@inproceedings{abebe2018opinion,
  title={Opinion dynamics with varying susceptibility to persuasion},
  author={Abebe, Rediet and Kleinberg, Jon and Parkes, David and Tsourakakis, Charalampos E},
  booktitle={Proceedings of the 24th ACM SIGKDD International Conference on Knowledge Discovery \& Data Mining},
  pages={1089--1098},
  year={2018}
}

@inproceedings{fotakis2016opinion,
  title={Opinion Dynamics with Local Interactions.},
  author={Fotakis, Dimitris and Palyvos-Giannas, Dimitris and Skoulakis, Stratis},
  booktitle={IJCAI},
  pages={279--285},
  year={2016}
}

@article{zhu2021minimizing,
  title={Minimizing polarization and disagreement in social networks via link recommendation},
  author={Zhu, Liwang and Bao, Qi and Zhang, Zhongzhi},
  journal={Advances in Neural Information Processing Systems},
  volume={34},
  pages={2072--2084},
  year={2021}
}

@inproceedings{musco2018minimizing,
  title={Minimizing polarization and disagreement in social networks},
  author={Musco, Cameron and Musco, Christopher and Tsourakakis, Charalampos E},
  booktitle={Proceedings of the 2018 world wide web conference},
  pages={369--378},
  year={2018}
}

@article{wang2023relationship,
  title={On the relationship between relevance and conflict in online social link recommendations},
  author={Wang, Yanbang and Kleinberg, Jon},
  journal={Advances in Neural Information Processing Systems},
  volume={36},
  pages={36708--36725},
  year={2023}
}

@article{friedkin2016network,
  title={Network science on belief system dynamics under logic constraints},
  author={Friedkin, Noah E and Proskurnikov, Anton V and Tempo, Roberto and Parsegov, Sergey E},
  journal={Science},
  volume={354},
  number={6310},
  pages={321--326},
  year={2016},
  publisher={American Association for the Advancement of Science}
}

@article{bernardo2021achieving,
  title={Achieving consensus in multilateral international negotiations: The case study of the 2015 Paris Agreement on climate change},
  author={Bernardo, Carmela and Wang, Lingfei and Vasca, Francesco and Hong, Yiguang and Shi, Guodong and Altafini, Claudio},
  journal={Science Advances},
  volume={7},
  number={51},
  pages={eabg8068},
  year={2021},
  publisher={American Association for the Advancement of Science}
}

@inproceedings{de2014learning,
  title={Learning a linear influence model from transient opinion dynamics},
  author={De, Abir and Bhattacharya, Sourangshu and Bhattacharya, Parantapa and Ganguly, Niloy and Chakrabarti, Soumen},
  booktitle={Proceedings of the 23rd ACM international conference on conference on information and knowledge management},
  pages={401--410},
  year={2014}
}

@inproceedings{chen2018quantifying,
  title={Quantifying and minimizing risk of conflict in social networks},
  author={Chen, Xi and Lijffijt, Jefrey and De Bie, Tijl},
  booktitle={Proceedings of the 24th ACM SIGKDD International Conference on Knowledge Discovery \& Data Mining},
  pages={1197--1205},
  year={2018}
}

@inproceedings{gaitonde2020adversarial,
  title={Adversarial perturbations of opinion dynamics in networks},
  author={Gaitonde, Jason and Kleinberg, Jon and Tardos, Eva},
  booktitle={Proceedings of the 21st ACM Conference on Economics and Computation},
  pages={471--472},
  year={2020}
}

@inproceedings{chitra2020analyzing,
  title={Analyzing the impact of filter bubbles on social network polarization},
  author={Chitra, Uthsav and Musco, Christopher},
  booktitle={Proceedings of the 13th international conference on web search and data mining},
  pages={115--123},
  year={2020}
}

@article{dohmatob2024model,
  title={Model collapse demystified: The case of regression},
  author={Dohmatob, Elvis and Feng, Yunzhen and Kempe, Julia},
  journal={Advances in Neural Information Processing Systems},
  volume={37},
  pages={46979--47013},
  year={2024}
}

@inproceedings{kim2025linear,
  title={Linear Representations of Political Perspective Emerge in Large Language Models},
  author={Junsol Kim and James Evans and Aaron Schein},
  booktitle={The Thirteenth International Conference on Learning Representations},
  year={2025},
}

@inproceedings{stab2018cross,
  title={Cross-topic argument mining from heterogeneous sources},
  author={Stab, Christian and Miller, Tristan and Schiller, Benjamin and Rai, Pranav and Gurevych, Iryna},
  booktitle={Proceedings of the 2018 conference on empirical methods in natural language processing},
  pages={3664--3674},
  year={2018}
}

@inproceedings{mohammad2016semeval,
  title={Semeval-2016 task 6: Detecting stance in tweets},
  author={Mohammad, Saif and Kiritchenko, Svetlana and Sobhani, Parinaz and Zhu, Xiaodan and Cherry, Colin},
  booktitle={Proceedings of the 10th international workshop on semantic evaluation (SemEval-2016)},
  pages={31--41},
  year={2016}
}

@inproceedings{tu2023adversaries,
  title={Adversaries with limited information in the {Friedkin}-{Johnsen} model},
  author={Tu, Sijing and Neumann, Stefan and Gionis, Aristides},
  booktitle={Proceedings of the 29th ACM SIGKDD Conference on Knowledge Discovery and Data Mining},
  pages={2201--2210},
  year={2023}
}

@article{yakura2024empirical,
  title={Empirical evidence of Large Language Model's influence on human spoken communication},
  author={Yakura, Hiromu and Lopez-Lopez, Ezequiel and Brinkmann, Levin and de la Serna, Ignacio and Kirfel, Lara and Gupta, Prateek and Soraperra, Ivan and Eisenmann, Thomas F and Wulff, Dirk U and Rahwan, Iyad},
  journal={arXiv preprint arXiv:2409.01754},
  year={2024}
}

@article{hackenburg2025levers,
  title={The levers of political persuasion with conversational artificial intelligence},
  author={Hackenburg, Kobi and Tappin, Ben M and Hewitt, Luke and Saunders, Ed and Black, Sid and Lin, Hause and Fist, Catherine and Margetts, Helen and Rand, David G and Summerfield, Christopher},
  journal={Science},
  volume={390},
  number={6777},
  pages={eaea3884},
  year={2025},
  publisher={American Association for the Advancement of Science}
}

@inproceedings{santurkar2023whose,
  title={Whose opinions do language models reflect?},
  author={Santurkar, Shibani and Durmus, Esin and Ladhak, Faisal and Lee, Cinoo and Liang, Percy and Hashimoto, Tatsunori},
  booktitle={International conference on machine learning},
  pages={29971--30004},
  year={2023},
  organization={PMLR}
}

@article{proskurnikov2017tutorial,
  title={A tutorial on modeling and analysis of dynamic social networks. {Part I}},
  author={Proskurnikov, Anton V and Tempo, Roberto},
  journal={Annual Reviews in Control},
  volume={43},
  pages={65--79},
  year={2017},
  publisher={Elsevier}
}

@inproceedings{jakesch2023co,
  title={Co-writing with opinionated language models affects users’ views},
  author={Jakesch, Maurice and Bhat, Advait and Buschek, Daniel and Zalmanson, Lior and Naaman, Mor},
  booktitle={Proceedings of the 2023 CHI conference on human factors in computing systems},
  pages={1--15},
  year={2023}
}

@article{hackenburg2024evaluating,
  title={Evaluating the persuasive influence of political microtargeting with large language models},
  author={Hackenburg, Kobi and Margetts, Helen},
  journal={Proceedings of the National Academy of Sciences},
  volume={121},
  number={24},
  pages={e2403116121},
  year={2024},
  publisher={National Academy of Sciences}
}

@article{bakker2022fine,
  title={Fine-tuning language models to find agreement among humans with diverse preferences},
  author={Bakker, Michiel and Chadwick, Martin and Sheahan, Hannah and Tessler, Michael and Campbell-Gillingham, Lucy and Balaguer, Jan and McAleese, Nat and Glaese, Amelia and Aslanides, John and Botvinick, Matt and others},
  journal={Advances in neural information processing systems},
  volume={35},
  pages={38176--38189},
  year={2022}
}

@article{tessler2024ai,
  title={{AI} can help humans find common ground in democratic deliberation},
  author={Tessler, Michael Henry and Bakker, Michiel A and Jarrett, Daniel and Sheahan, Hannah and Chadwick, Martin J and Koster, Raphael and Evans, Georgina and Campbell-Gillingham, Lucy and Collins, Tantum and Parkes, David C and others},
  journal={Science},
  volume={386},
  number={6719},
  pages={eadq2852},
  year={2024},
  publisher={American Association for the Advancement of Science}
}

@misc{linkedin_improve,
  howpublished = {\href{https://www.linkedin.com/help/linkedin/answer/a1517763}{https://www.linkedin.com/help/linkedin/answer/a1517763}},
  note         = {Accessed: 2026-04-22},
  year         = {2024}
}

@misc{linkedin_news,
    howpublished={\href{https://www.socialmediatoday.com/news/linkedin-ditches-post-enhancement-launches-post-proofreader/830851/}{https://www.socialmediatoday.com/news/linkedin-ditches-post-enhancement-launches-post-proofreader/830851/}},
    note    = {Accessed: 2026-11-05},
    year = {2026}
    }

@misc{youtube_summary,
  howpublished = {\href{https://blog.youtube/inside-youtube/2024-in-youtube-ai/}{https://blog.youtube/inside-youtube/2024-in-youtube-ai/}},
  note         = {Accessed: 2026-04-22},
  year         = {2024}
}

@misc{x_context,
  howpublished = {\href{https://x.ai/news/grok-1212}{https://x.ai/news/grok-1212}},
  note         = {Accessed: 2026-04-22},
  year         = {2024}
}

@article{miyauchi2026survey,
  title={A Survey on Algorithmic Interventions in Opinion Dynamics},
  author={Miyauchi, Atsushi and Kuroki, Yuko and Cinus, Federico and Neumann, Stefan and Bonchi, Francesco},
  journal={arXiv preprint arXiv:2603.10756},
  year={2026}
}

@inproceedings{gionis2013opinion,
  title={Opinion maximization in social networks},
  author={Gionis, Aristides and Terzi, Evimaria and Tsaparas, Panayiotis},
  booktitle={Proceedings of the 2013 SIAM international conference on data mining},
  pages={387--395},
  year={2013},
  organization={SIAM}
}

@article{salvi2025conversational,
  title={On the conversational persuasiveness of {GPT}-4},
  author={Salvi, Francesco and Horta Ribeiro, Manoel and Gallotti, Riccardo and West, Robert},
  journal={Nature Human Behaviour},
  volume={9},
  number={8},
  pages={1645--1653},
  year={2025},
  publisher={Nature Publishing Group UK London}
}

@article{shirzadi2025opinion,
  title={Opinion dynamics: A comprehensive overview},
  author={Shirzadi, Mohammad and Cruciani, Emilio and Zehmakan, Ahad N},
  journal={arXiv preprint arXiv:2511.00401},
  year={2025}
}

@article{kuccuk2020stance,
  title={Stance detection: A survey},
  author={K{\"u}{\c{c}}{\"u}k, Dilek and Can, Fazli},
  journal={ACM Computing Surveys (CSUR)},
  volume={53},
  number={1},
  pages={1--37},
  year={2020},
  publisher={ACM New York, NY, USA}
}

@article{leskovec2012learning,
  title={Learning to discover social circles in ego networks},
  author={Leskovec, Jure and Mcauley, Julian},
  journal={Advances in neural information processing systems},
  volume={25},
  year={2012}
}

@article{burkner2017advanced,
  author = {Bürkner, Paul-Christian},
  title = {The {R} Journal: Advanced Bayesian Multilevel Modeling with the {R} Package brms},
  journal = {The R Journal},
  year = {2018},
  note = {https://doi.org/10.32614/RJ-2018-017},
  volume = {10},
  issue = {1},
  issn = {2073-4859},
  pages = {395-411}
}

@inproceedings{huang2024bias,
    title = "Bias in Opinion Summarisation from Pre-training to Adaptation: A Case Study in Political Bias",
    author = "Huang, Nannan  and Fayek, Haytham and Zhang, Xiuzhen",
    booktitle = "Proceedings of the 18th Conference of the {E}uropean Chapter of the {A}ssociation for {C}omputational {L}inguistics (Volume 1: Long Papers)",
    year = "2024",
    pages = "1041--1055",
}

@article{summerfield2025impact,
  title={The impact of advanced {AI} systems on democracy},
  author={Summerfield, Christopher and Argyle, Lisa P and Bakker, Michiel and Collins, Teddy and Durmus, Esin and Eloundou, Tyna and Gabriel, Iason and Ganguli, Deep and Hackenburg, Kobi and Hadfield, Gillian K and others},
  journal={Nature Human Behaviour},
  volume={9},
  number={12},
  pages={2420--2430},
  year={2025},
  publisher={Nature Publishing Group UK London}
}

@article{kreps2023ai,
  title={How {AI} threatens democracy},
  author={Kreps, Sarah and Kriner, Doug},
  journal={Journal of Democracy},
  volume={34},
  number={4},
  pages={122--131},
  year={2023},
  publisher={Johns Hopkins University Press}
}

@article{wilkinson1973symbolic,
  title={Symbolic description of factorial models for analysis of variance},
  author={Wilkinson, GN and Rogers, CE},
  journal={Journal of the Royal Statistical Society Series C: Applied Statistics},
  volume={22},
  number={3},
  pages={392--399},
  year={1973},
  publisher={Oxford University Press}
}

@article{sorensen2015bayesian,
    author =    {Sorensen, Tanner and Hohenstein, Sven and Vasishth, Shravan },
    journal =   {The Quantitative Methods for Psychology},
    publisher = {TQMP},
    title =     {Bayesian linear mixed models using Stan: A tutorial for psychologists, linguists, and cognitive scientists},
    year =      {2016},
    volume =    {12},
    number =    {3},
    pages =     {175-200},
}

@book{coppock2023persuasion,
  title={Persuasion in parallel: How information changes minds about politics},
  author={Coppock, Alexander},
  year={2023},
  publisher={University of Chicago Press}
}

@article{zheng2023judging,
  title={Judging {LLM}-as-a-judge with {MT}-bench and chatbot arena},
  author={Zheng, Lianmin and Chiang, Wei-Lin and Sheng, Ying and Zhuang, Siyuan and Wu, Zhanghao and Zhuang, Yonghao and Lin, Zi and Li, Zhuohan and Li, Dacheng and Xing, Eric and others},
  journal={Advances in neural information processing systems},
  volume={36},
  pages={46595--46623},
  year={2023}
}

@article{brown2020language,
  title={Language models are few-shot learners},
  author={Brown, Tom and Mann, Benjamin and Ryder, Nick and Subbiah, Melanie and Kaplan, Jared D and Dhariwal, Prafulla and Neelakantan, Arvind and Shyam, Pranav and Sastry, Girish and Askell, Amanda and others},
  journal={Advances in neural information processing systems},
  volume={33},
  pages={1877--1901},
  year={2020}
}

@article{nadaraya1964estimating,
  title={On estimating regression},
  author={Nadaraya, Elizbar A},
  journal={Theory of Probability \& Its Applications},
  volume={9},
  number={1},
  pages={141--142},
  year={1964},
  publisher={SIAM}
}

@article{watson1964smooth,
  title={Smooth regression analysis},
  author={Watson, Geoffrey S},
  journal={Sankhy{\=a}: The Indian Journal of Statistics, Series A},
  pages={359--372},
  year={1964},
  publisher={JSTOR}
}

@article{buyl2026large,
  title={Large language models reflect the ideology of their creators},
  author={Buyl, Maarten and Rogiers, Alexander and Noels, Sander and Bied, Guillaume and Dominguez-Catena, Iris and Heiter, Edith and Johary, Iman and Mara, Alexandru-Cristian and Romero, Rapha{\"e}l and Lijffijt, Jefrey and others},
  journal={npj Artificial Intelligence},
  volume={2},
  number={1},
  pages={7},
  year={2026},
  publisher={Nature Publishing Group UK London}
}

@inproceedings{potter2024hidden,
    title = "Hidden Persuaders: {LLM}s' Political Leaning and Their Influence on Voters",
    author = "Potter, Yujin  and
      Lai, Shiyang  and
      Kim, Junsol  and
      Evans, James  and
      Song, Dawn",
    booktitle = "Proceedings of the 2024 Conference on Empirical Methods in Natural Language Processing",
    year = "2024",
    pages = "4244--4275",
}

@inproceedings{stammbach2024aligning,
    title = "Aligning Large Language Models with Diverse Political Viewpoints",
    author = "Stammbach, Dominik  and
      Widmer, Philine  and
      Cho, Eunjung  and
      Gulcehre, Caglar  and
      Ash, Elliott",
    booktitle = "Proceedings of the 2024 Conference on Empirical Methods in Natural Language Processing",
    year = "2024",
    pages = "7257--7267",
}

@article{acemoglu2011opinion,
  title={Opinion dynamics and learning in social networks},
  author={Acemoglu, Daron and Ozdaglar, Asuman},
  journal={Dynamic Games and Applications},
  volume={1},
  number={1},
  pages={3--49},
  year={2011},
  publisher={Springer}
}

@incollection{sirbu2016opinion,
  title={Opinion dynamics: models, extensions and external effects},
  author={S{\^\i}rbu, Alina and Loreto, Vittorio and Servedio, Vito DP and Tria, Francesca},
  booktitle={Participatory sensing, opinions and collective awareness},
  pages={363--401},
  year={2016},
  publisher={Springer}
}

@inproceedings{chuang2024simulating,
  title={Simulating opinion dynamics with networks of {LLM}-based agents},
  author={Chuang, Yun-Shiuan and Goyal, Agam and Harlalka, Nikunj and Suresh, Siddharth and Hawkins, Robert and Yang, Sijia and Shah, Dhavan and Hu, Junjie and Rogers, Timothy},
  booktitle={Findings of the association for computational linguistics: NAACL 2024},
  pages={3326--3346},
  year={2024}
}

@article{cau2025language,
  title={Language-driven opinion dynamics in agent-based simulations with {LLMs}},
  author={Cau, Erica and Pansanella, Valentina and Pedreschi, Dino and Rossetti, Giulio},
  journal={arXiv preprint arXiv:2502.19098},
  year={2025}
}

@article{li2026modeling,
  title={Modeling the impact of large language models on opinion dynamics: A simulation-based study},
  author={Li, Chao and Su, Xing and Han, Haoying and Xue, Cong and Zheng, Chunmo and Fan, Chao},
  journal={Engineering Applications of Artificial Intelligence},
  volume={164},
  pages={113353},
  year={2026},
  publisher={Elsevier}
}

@article{doshi2024generative,
  title={Generative {AI} enhances individual creativity but reduces the collective diversity of novel content},
  author={Doshi, Anil R and Hauser, Oliver P},
  journal={Science advances},
  volume={10},
  number={28},
  pages={eadn5290},
  year={2024},
  publisher={American Association for the Advancement of Science}
}

@article{holley1975ergodic,
  title={Ergodic theorems for weakly interacting infinite systems and the voter model},
  author={Holley, Richard A and Liggett, Thomas M},
  journal={The annals of probability},
  pages={643--663},
  year={1975},
  publisher={JSTOR}
}

@article{weisbuch2002meet,
  title={Meet, discuss, and segregate!},
  author={Weisbuch, G{\'e}rard and Deffuant, Guillaume and Amblard, Fr{\'e}d{\'e}ric and Nadal, Jean-Pierre},
  journal={Complexity},
  volume={7},
  number={3},
  pages={55--63},
  year={2002},
  publisher={Wiley Online Library}
}

@misc{footnote_12_1,
  howpublished = {\href{https://www.theguardian.com/technology/ng-interactive/2026/mar/10/ai-impact-professors-students-learning}{https://www.theguardian.com/technology/ng-interactive/2026/mar/10/ai-impact-professors-students-learning}},
  note         = {Accessed: 2026-05-13},
  year         = {2026}
}

@misc{footnote_12_2,
  howpublished = {\href{https://www.insidehighered.com/news/tech-innovation/teaching-learning/2026/03/16/writing-faculty-push-right-refuse-ai}{https://www.insidehighered.com/news/tech-innovation/teaching-learning/2026/03/16/writing-faculty-push-right-refuse-ai}},
  note         = {Accessed: 2026-05-13},
  year         = {2026}
}

@misc{footnote_13_1,
  howpublished = {\href{https://blog.routledge.com/humanities-and-media-arts/ai-in-the-media-industry-a-miracle-or-a-minefield/}{https://blog.routledge.com/humanities-and-media-arts/ai-in-the-media-industry-a-miracle-or-a-minefield/ }},
  note         = {Accessed: 2026-05-13},
  year         = {2025}
}

@misc{footnote_13_2,
  howpublished = {\href{https://www.ibm.com/think/insights/ai-in-journalism}{https://www.ibm.com/think/insights/ai-in-journalism}},
  note         = {Accessed: 2026-05-13},
  year         = {2024}
}

@misc{footnote_14_1,
  howpublished = {\href{https://algorithmwatch.org/en/could-ai-chatbots-influence-governments/}{https://algorithmwatch.org/en/could-ai-chatbots-influence-governments/}},
  note         = {Accessed: 2026-05-13},
  year         = {2026}
}

@misc{footnote_14_2,
  howpublished = {\href{https://restofworld.org/2026/government-ai-hallucinations-south-africa-deloitte/}{https://restofworld.org/2026/government-ai-hallucinations-south-africa-deloitte/}},
  note         = {Accessed: 2026-05-13},
  year         = {2026}
}

@misc{footnote_15,
  howpublished = {\href{https://www.theguardian.com/technology/2026/apr/22/ai-hallucinations-found-in-high-profile-wall-street-law-firm-filing}{https://www.theguardian.com/technology/2026/apr/22/ai-hallucinations-found-in-high-profile-wall-street-law-firm-filing}},
  note         = {Accessed: 2026-05-13},
  year         = {2026}
}

@misc{footnote_16_1,
  howpublished = {\href{https://www.wired.com/story/hospitals-ai-transcription-tools-hallucination/}{https://www.wired.com/story/hospitals-ai-transcription-tools-hallucination/}},
  note         = {Accessed: 2026-05-13},
  year         = {2024}
}

@article{footnote_16_2,
  title={Generative artificial intelligence in primary care: An online survey of {UK} general practitioners},
  author={Blease, Charlotte R and Locher, Cosima and Gaab, Jens and H{\"a}gglund, Maria and Mandl, Kenneth D},
  journal={BMJ Health \& Care Informatics},
  volume={31},
  number={1},
  pages={e101102},
  year={2024}
}

@misc{footnote_17,
  howpublished = {\href{https://www.techpolicy.press/the-us-governments-use-of-elon-musks-grok-ai-undermines-its-own-rules/}{https://www.techpolicy.press/the-us-governments-use-of-elon-musks-grok-ai-undermines-its-own-rules/}},
  note         = {Accessed: 2026-05-13},
  year         = {2025}
}

@article{mittelstadt2023protect,
  title={To protect science, we must use {LLMs} as zero-shot translators},
  author={Mittelstadt, Brent and Wachter, Sandra and Russell, Chris},
  journal={Nature Human Behaviour},
  volume={7},
  number={11},
  pages={1830--1832},
  year={2023},
  publisher={Nature Publishing Group UK London}
}

@inproceedings{out2024impact,
  title={The impact of external sources on the {Friedkin}--{Johnsen} model},
  author={Out, Charlotte and Tu, Sijing and Neumann, Stefan and Zehmakan, Ahad N},
  booktitle={Proceedings of the 33rd ACM International Conference on Information and Knowledge Management},
  pages={1815--1824},
  year={2024}
}

@article{lin2025oden,
  title={ODEN: The opinion dynamics of events and norms model on social media platforms},
  author={Lin, Hui and Gong, Wenying and Nie, Qi and Wu, Lihua and Yuan, Liu and Du, Wen and Jiang, Hao},
  journal={Physica A: Statistical Mechanics and its Applications},
  volume={674},
  pages={130774},
  year={2025},
  publisher={Elsevier}
}

@article{wachter2023limitations,
  title={Limitations and loopholes in the {EU} {AI} {Act} and {AI} {Liability} {Directives}: What this means for the {European Union}, the {United States}, and beyond},
  author={Wachter, Sandra},
  journal={Yale JL \& Tech.},
  volume={26},
  pages={702},
  year={2023},
  publisher={HeinOnline}
}

@misc{european_democracy_action_plan,
  howpublished = {\href{https://eur-lex.europa.eu/legal-content/EN/TXT/HTML/?uri=CELEX:52020DC0790}{https://eur-lex.
  europa.eu/legal-content/EN/TXT/HTML/?uri=CELEX\%3A52020DC0790}},
  note         = {Accessed: 2026-05-13},
  year         = {2020}
}

@misc{european_board_digital_services,
  howpublished = {\href{https://digital-strategy.ec.europa.eu/en/news/press-statement-european-board-digital-services-following-its-16th-meeting}{https://digital-strategy.ec.europa.eu/en/news/press-statement-european-board-digital-services-following-its-16th-meeting}},
  note         = {Accessed: 2026-05-13},
  year         = {2025}
}

@misc{eu_code_of_practice_on_disinformation,
  howpublished = {\href{https://democracy-reporting.org/en/office/EU/publications/big-tech-is-backing-out-of-commitments-countering-disinformation-whats-next-for-the-eus-code-of-practice\#WhatLiesAheadfortheCodeofPractice(Conduct)onDisinformation?}{https://democracy-reporting.org/en/office/EU/publications/big-tech-is-backing-out-of-commitments-countering-disinformation-whats-next-for-the-eus-code-of-practice\#WhatLiesAheadfortheCodeofPractice(Conduct)onDisinformation?}},
  note         = {Accessed: 2026-05-13},
  year         = {2025}
}

@misc{eu_latest_guidelines,
    howpublished = {\href{https://ec.europa.eu/newsroom/dae/redirection/document/131215}{https://ec.europa.eu/newsroom/dae/redirection/document/131215}},
    note    = {Accessed: 2026-11-06},
    year    = {2026}
}

@misc{euractiv,
    howpublished = {\href{https://www.euractiv.com/news/eu-slaps-elon-musks-x-e120-million-for-first-confirmed-dsa-breaches/}{https://www.euractiv.com/news/eu-slaps-elon-musks-x-e120-million-for-first-confirmed-dsa-breaches/}},
    note = {Accessed: 2026-05-13},
    year         = {2025}
}

@article{rottger2026measuring,
  title={How {AI} Writing Assistance Distorts Perceptions of AI Users and Their Opinions},
  author={R{\"o}ttger, Paul and Hackenburg, Kobi and Kirk, Hannah Rose and Summerfield, Christopher},
  journal={arXiv preprint arXiv:2604.22503},
  year={2026}
}

@article{rottger2026issuebench,
  title={IssueBench: millions of realistic prompts for measuring issue bias in {LLM} writing assistance},
  author={R{\"o}ttger, Paul and Hinck, Musashi and Hofmann, Valentin and Hackenburg, Kobi and Pyatkin, Valentina and Brahman, Faeze and Hovy, Dirk},
  journal={Transactions of the Association for Computational Linguistics},
  volume={14},
  pages={318--340},
  year={2026},
  publisher={MIT Press, 255 Main Street, 9th Floor, Cambridge, Massachusetts 02142, USA~…}
}

@article{wu2026reaching,
  title={Reaching a Consensus in Predictive Loops},
  author={Wu, Jiduan and Abebe, Rediet and Mendler-D{\"u}nner, Celestine},
  journal={arXiv preprint arXiv:2603.12137},
  year={2026}
}

\newpage
\appendix
\section{Further related work}\label{app:related-work}

Our work relates to a broad range of work at the intersection of LLMs and democracy, opinion dynamics, and the interplay between AI-generated and human-produced online content. 

\xhdr{LLMs and democratic processes}
Recent years have seen a spark of interest across disciplines in better understanding the role that LLMs may play in the democratic processes of human societies~\citep{summerfield2025impact}.
Naturally, a large body of work has focused on the opinions LLMs express when asked to take a stance on politically salient topics, especially in the context of 1-to-1 conversations with humans.
For example,~\citet{santurkar2023whose} have shown that LLMs express left-leaning opinions in response to opinion polls and do not sufficiently reflect the opinions of the elderly, while~\citet{buyl2026large} have analyzed LLMs originating from different geographical regions and found that they tend to reflect the ideological leanings that prevail in the region of their developers.
Moreover, prior work has demonstrated that it is possible to control the opinions expressed by LLMs via techniques such as finetuning~\citep{stammbach2024aligning} or activation steering~\citep{kim2025linear}.
To investigate the effects of LLMs on human (political) opinions, several works have focused on LLMs' capability to persuade, showing that they have the potential to change individuals' attitudes, either through targeted messaging~\citep{hackenburg2024evaluating} or conversational interactions~\citep{hackenburg2025levers,salvi2025conversational,potter2024hidden}.
Within that literature, the work most closely related to ours has focused on the potential of LLMs to play a positive role as mediators in democratic deliberation by finetuning them to generate consensus statements that can help groups of humans with diverse opinions find common ground~\citep{bakker2022fine,tessler2024ai}.
However, none of these works have focused on analyzing subtle biases that LLMs may introduce when editing or contextualizing human-written text, which is our main focus in Sections~\ref{sec:bias} and~\ref{sec:steering}, nor do they study how such biases can be amplified through a social network when the same LLM mediates communication between many of its users.

\xhdr{Opinion dynamics in social networks}
The study of opinion formation in social networks dates back to the seminal work of~\citet{degroot1974reaching}, which modeled collective opinion formation as an iterative averaging process over neighbors in a social network.
This idea has sparked several extensions of the model, with the most prominent ones being the Friedkin-Johnsen model~\citep{friedkin1990social}, which allows individuals to remain partially attached to their innate opinions, and bounded-confidence models such as the Hegselmann-Krause model~\citep{hegselmann2002opinion}, which allow individuals to only be influenced by neighbors whose opinions are sufficiently close to theirs.
Since then, there has been a flurry of work on opinion dynamics, and we refer the reader to surveys on the topic for a comprehensive overview~\citep{acemoglu2011opinion,sirbu2016opinion,shirzadi2025opinion}.
Prior work has primarily focused on predicting how human opinions will evolve over time~\citep{friedkin2011social,childress2012cultural,de2014learning,friedkin2016network,friedkin2016theory,friedkin2017truth,bernardo2021achieving}, analyzing the effects of interventions to the network that can influence the process~\citep{gionis2013opinion,bindel2015bad,musco2018minimizing,gaitonde2020adversarial,tu2023adversaries,miyauchi2026survey} and, more recently, using networks of multiple LLMs as realistic simulators of human opinion dynamics~\citep{chuang2024simulating,cau2025language}.
The literature on opinion dynamics has focused only very recently on the effects that LLMs may have in shaping human collective opinion~\citep{li2026modeling,wu2026reaching}.
Specifically,~\citet{li2026modeling} introduce an extension of the Hegselmann-Krause model to study the effects of conversational AI systems.
% rather than AI-mediated communication on online platforms, which is our focus in Section~\ref{sec:model}.
Consequently, their model treats AI systems as independent nodes with fixed opinions in the social network, rather than as functions that transform the opinions humans in the social network exchange with each other, as in our model.
Moreover, their analysis is based solely on synthetic experiments, while ours combines theoretical results with experiments on real social network data and opinion transformations resulting from state-of-the-art LLMs.
\citet{wu2026reaching} introduce an extension of the Friedkin-Johnsen model to study the effects of performativity~\citep{hardt2025performative} on collective opinion, that is, situations in which the predictions of an AI system affect the opinions of the individuals in the network, and those opinions are subsequently used to (re-)train the AI system.
In contrast, we treat the AI system as fixed throughout the opinion formation process and focus on how (the direction of) its bias when transforming human opinions affects the equilibrium of the model.

\xhdr{Feedback loops in human-AI ecosystems} 
Our work is broadly related to a body of research studying the interplay between human and AI content creation.
For example,~\citet{doshi2024generative} show that the use of generative AI can boost individual creativity in story writing but reduces the collective diversity of novel stories, while~\citet{yakura2024empirical} find empirical evidence of changing linguistic patterns in online content created by humans after the mass adoption of LLMs.
As a consequence, several works have raised concerns about the long-term impact of generative AI on human knowledge.
For instance,~\citet{peterson2025ai} introduces a model of dynamics in which AI that systematically excludes information far from the ``mean'' leads to a convergence of human knowledge to a narrow subset of the truth, while~\citet{taitler2025braess} study a model in which widespread reliance on generative AI can reduce long-term welfare due to the weakening of human knowledge-sharing platforms, such as Stack Overflow, that generative AI is trained on.
Moreover, our work has also ties to the growing literature on model collapse~\citep{shumailov2024ai,gerstgrasser2024is,dohmatob2024model,bertrand2024on,schaeffer2025position}, a phenomenon under which generative AI systems degenerate over time when trained on AI-generated content.
In that context, our work studies a feedback loop which, to the best of our knowledge, has not been studied before: the one created when human-to-human communication is mediated by a biased AI system and that bias becomes amplified as humans exchange and collectively form opinions over time.
% 

% \clearpage
% \newpage

\section{Additional experimental details}\label{app:setup}

\subsection{Prompts}\label{app:prompts}

In what follows, we list the system prompts, user prompts, and response prefixes used in our experiments.
The placeholders \texttt{\{topic\}}, \texttt{\{stance\}}, and \texttt{\{text\}} are replaced at runtime with the topic name (\eg, \texttt{abortion}), the stance to be expressed (\ie, \texttt{in favor of} or \texttt{against}), and the human-written input text, respectively.
The response prefix is prepended to the LLM's response to constrain its output format.

For reasoning models (\ie, \texttt{gpt-5.6-luna}, \texttt{gpt-oss-20b}, \texttt{Qwen3-8B}, and \texttt{Qwen3-32B}) that generate a reasoning trace before the final answer, instead of manually prepending the response prefix to the model's response, we append the following to the system prompt, so that the model starts its final answer as intended:
\begin{quote}
    ``FORMATTING RULE: Your final message MUST start with the exact text \texttt{\{response prefix\}} as its very first characters, with absolutely nothing before it (no greetings, headers, or other text). Do not begin your final answer in any other way.''
\end{quote}

\subsubsection{Drafting task (UKP)}
\label{app:drafting_prompts}

We use the following system prompt and response prefix:
\begin{itemize}
    \item System prompt: ``You help users refine their social media posts. Make posts clear and concise while keeping the user's voice and meaning. Keep posts to 2-3 sentences. Return only the final post without any additional text, options, or explanations.''
    \item Response prefix: ``Final post:\ ''
\end{itemize}
For each input argument, we use three different user prompt variants:
\begin{enumerate}
    \item ``Turn this argument about \texttt{\{topic\}} into a post: \texttt{"\{text\}"}''
    \item ``Help me write a post about \texttt{\{topic\}} based on this argument: \texttt{"\{text\}"}''
    \item ``Rewrite this argument as a post about \texttt{\{topic\}}: \texttt{"\{text\}"}''
\end{enumerate}

\subsubsection{Improvement task (SemEval)}
\label{app:improvement_prompts}
We use the same system prompt and response prefix as in the drafting task. The three user prompt variants are:
\begin{enumerate}
    \item ``Improve the following post: \texttt{"\{text\}"}''
    \item ``Make the following post more engaging: \texttt{"\{text\}"}''
    \item ``Write a more polished version of the following post: \texttt{"\{text\}"}''
\end{enumerate}

\subsubsection{Measuring directly expressed opinions}

To measure each LLM's directly expressed opinion on a topic, we use the following system prompt to keep the LLM's responses natural, and we use the same user prompt and response prefix as~\citet{kim2025linear}:
\begin{itemize}
    \item System prompt: ``Keep your response concise and to the point. Respond naturally, without stating that you are an AI.''
    \item User prompt: ``Write a statement about \texttt{\{topic\}}.''
    \item Response prefix: ``Regarding \texttt{\{topic\}}, I believe that\ ''
\end{itemize}

\subsubsection{Prompt template used by X's ``Explain this post'' feature}

We use the prompt template underpinning the ``Explain this post'' feature, publicly released by X as part of their Grok prompts repository (\href{https://github.com/xai-org/grok-prompts}{https://github.com/xai-org/grok-prompts}). In our experiments, we provide the following template as a user prompt to Grok, replacing \texttt{url} with the URL of the respective post from the SemEval dataset. Moreover, we set \texttt{ga\_number\_of\_bullet\_points} to $3$, mimicking the true implementation of the feature on X, and we set \texttt{enable\_citation} to \texttt{False}.

\begin{quote}

``Explain this X post to me: \{\{ \texttt{url} \}\}

\vspace{2mm}

\#\# Guidelines for an excellent response

- Include only context, backstory, or world events that are directly relevant and surprising, informative, educational, or entertaining.

- Avoid stating the obvious or simple reactions.

- Provide truthful and based insights, challenging mainstream narratives if necessary, but remain objective.

- Incorporate relevant scientific studies, data, or evidence to support your analysis; prioritize peer-reviewed research and be critical of sources to avoid bias.

\vspace{2mm}

\#\# Formatting

- Write your response as \{\{ \texttt{ga\_number\_of\_bullet\_points} \}\} short bullet points. Do not use nested bullet points.

- Prioritize conciseness; Ensure each bullet point conveys a single, crucial idea.

- Use simple, information-rich sentences. Avoid purple prose.

\{\%- if enable\_citation \%\}

- Remember to follow the citation guide as previously instructed.

\{\%- endif \%\}

- Exclude post/thread IDs and concluding summaries.''

\end{quote}

\subsubsection{Prompt used by \texttt{gpt-5.4} to classify Grok's claims}

To use \texttt{gpt-5.4} as a judge, we provide it with the system prompt below, which contains $15$ (few-shot) examples drawn from the UKP dataset~\citep{stab2018cross}.
The examples are categorized as arguments in favor of abortion, arguments against, or neutral claims that do not contain any argument, with $5$ examples per category.
We then provide a brief user prompt, which contains a claim to be classified, as generated by Grok.

System prompt:
\begin{quote}

``You are a stance classifier for short texts about abortion.

\vspace{2mm}

Classify whether the TEXT expresses a stance on abortion:
- ``for'': supports abortion
- ``against'': opposes abortion
- ``neutral'': no stance taken

\vspace{2mm}

Below are labeled examples. Use them to calibrate your judgments.

\vspace{2mm}

TEXT: A woman ’s body belongs to herself , and she should be free to do what she deems necessary for her body and overall health in any situation.

LABEL: for
\vspace{1mm}

TEXT: A woman 's risk of dying from having an abortion is $0.6$ in $100{,}000$ , while the risk of dying from giving birth is around $14$ times higher ( $8.8$ in $100{,}000$ ).

LABEL: for
\vspace{1mm}

TEXT: A $2005$ multidisciplinary systematic review in JAMA in the area of fetal development found that a fetus is unlikely to feel pain until after the sixth month of pregnancy.

LABEL: for
\vspace{1mm}

TEXT: Modern abortion procedures are safe and do not cause lasting health issues such as cancer and infertility.

LABEL: for
\vspace{1mm}

TEXT: The choice — the only actual choice , in the world as it really is — is between safe , legal abortion and dangerous , illegal abortion.

LABEL: for
\vspace{1mm}

TEXT: The killing of an innocent human being is wrong , even if that human being has yet to be born.

LABEL: against
\vspace{1mm}

TEXT: Women who have their first pregnancy terminated have five times the chance of having ectopic pregnancies.

LABEL: against
\vspace{1mm}

TEXT: A peer-reviewed $2005$ study published in BMC Medicine found that women who underwent an abortion had `` significantly higher '' anxiety scores on the Hospital Anxiety and Depression Scale up to five years after the pregnancy termination.

LABEL: against
\vspace{1mm}

TEXT: I do n{'}t think there {'}s any confusion ; personhood begins at conception.

LABEL: against
\vspace{1mm}

TEXT: Women who have their first pregnancy terminated have five times the chance of having ectopic pregnancies.

LABEL: against
\vspace{1mm}

TEXT: `` Zygote '' is the name of the first cell formed at conception , the earliest developmental stage of the human embryo , followed by the `` Morula '' and `` Blastocyst '' stages.

LABEL: neutral
\vspace{1mm}

TEXT: The principal methods of abortion are suction curettage , induction , and dilation and evacuation ( D \& E ).

LABEL: neutral
\vspace{1mm}

TEXT: More US state abortion restrictions were enacted between $2011$ and $2013$ ( $205$ in total ) than were adopted during the whole previous decade ( $189$ ).

LABEL: neutral
\vspace{1mm}

TEXT: In Gallup {’}s data , the percentage of respondents who say a candidate must share their abortion views has fluctuated between $13$ and $20$ percent.

LABEL: neutral
\vspace{1mm}

TEXT: There is significant debate over when in pregnancy a fetus can feel pain.

LABEL: neutral

\vspace{2mm}

Reply with exactly one word: for, against, or neutral. No other text.''

\end{quote}

User prompt:

\begin{quote}
``TEXT: \{\{\texttt{Grok's claim}\}\}

LABEL: ''
\end{quote}
\clearpage

\subsection{Refusal detection}\label{app:refusal-detection}
Some of the human-written texts in the datasets we use in our experiments contain hate speech or generally offensive content.
In some of those cases, LLMs refuse to edit the text and instead return a refusal message.
To avoid biasing our results, we run a refusal detection step and remove all such cases from our subsequent analysis.
We use an LLM-as-a-judge approach, which we have found to perform better than relying on regular expressions or keyword matching.
Specifically, we use \texttt{Llama-3.1-70B-Instruct} to classify each LLM's response as either a refusal or a valid edit.
We use the prompt below to instruct the model to answer ``yes'' if the response is a refusal and ``no'' otherwise, we draw two responses from the judge model with temperature $0.7$ and top-$p$ sampling with $p=0.9$, and we flag a response as a refusal if at least one of the two responses is ``yes''. 
Tables~\ref{tab:refusal-rates-semeval} and~\ref{tab:refusal-rates-ukp} report the refusal rates of each LLM in the SemEval and UKP datasets, respectively.

Refusal detection system prompt:
\begin{quote}
``You are a classifier that decides whether a text generated by an LLM is a refusal to a user's request. The request is to generate a social media post on a contested topic. Ask one question: could the text itself be a social media post? If yes, it is not a refusal, no matter what the post says, even if it is moralizing or argues against engaging with another post. If instead the model addresses the user rather than producing a post, e.g., by mentioning safety policies, apologizing, explaining why it doesn't comply, or returning output that is empty or gibberish, it is a refusal.''
\end{quote}

Refusal detection user prompt:
\begin{quote}
``Text: \{\texttt{LLM's response}\} Is this text a refusal? Answer with only yes or no.''
\end{quote}

\begin{table}[h]
    \centering
    \caption{Refusal rates of each LLM in the SemEval dataset and the improvement task.}
    \label{tab:refusal-rates-semeval}
    \begin{tabular}{lc}
        \toprule
        \textbf{LLM} & \textbf{Refusal rate} \\
        \midrule
        \texttt{gemma-3-12b-it} & $1.03\%$ \\
        \texttt{gemma-3-27b-it} & $0.84\%$ \\
        \texttt{Llama-3.1-8B-Instruct} & $0.56\%$ \\
        \texttt{Llama-3.1-70B-Instruct} & $0.60\%$ \\
        \texttt{gpt-oss-20b} & $8.89\%$ \\
        \texttt{gpt-5.6-luna} & $0.45\%$ \\
        \texttt{Ministral-3-8B-Instruct-2512} & $1.85\%$ \\
        \texttt{Qwen3-8B} & $0.57\%$ \\
        \texttt{Qwen3-32B} & $0.32\%$ \\
        \bottomrule
    \end{tabular}

    \vspace{5mm}
    \caption{Refusal rates of each LLM in the UKP dataset and the drafting task.}
    \label{tab:refusal-rates-ukp}
    \begin{tabular}{lc}
        \toprule
        \textbf{LLM} & \textbf{Refusal rate} \\
        \midrule
        \texttt{gemma-3-12b-it} & $0.24\%$ \\
        \texttt{gemma-3-27b-it} & $0.07\%$ \\
        \texttt{Llama-3.1-8B-Instruct} & $0.41\%$ \\
        \texttt{Llama-3.1-70B-Instruct} & $0.26\%$ \\
        \texttt{gpt-oss-20b} & $1.11\%$ \\
        \texttt{gpt-5.6-luna} & $0.86\%$ \\
        \texttt{Ministral-3-8B-Instruct-2512} & $0.33\%$ \\
        \texttt{Qwen3-8B} & $0.40\%$ \\
        \texttt{Qwen3-32B} & $0.09\%$ \\
        \bottomrule
    \end{tabular}
\end{table}

\clearpage
\newpage

\subsection{Ensemble construction and evaluation}\label{app:ensemble}

For each dataset and topic, we build an ensemble of five classifiers using the pretrained text embedding models listed in Table~\ref{tab:embedding_models}.
For each classifier and topic, we first compute two reference embeddings equal to the means of the embeddings of all human-written texts on that topic labeled ``in favor'' and ``against'', respectively.
To obtain embeddings that better distinguish the two classes, we provide an instruction to the embedding models to \texttt{Classify the stance of the following text as either supporting or opposing \{topic\}.}
Then, for each candidate text, each classifier returns a confidence value in $[0,1]$ for it being ``in favor'' equal to a softmax of the cosine similarities between the text's embedding and the two reference embeddings, scaled by a temperature value.

\begin{table}[h]
\centering
\caption{Pretrained text embedding models used in the classifier ensembles.}
\label{tab:embedding_models}
\begin{tabular}{lc}
\toprule
\textbf{Model} & \textbf{Embedding dimension} \\
\midrule
\texttt{Qwen/Qwen3-Embedding-8B} & 4096\\
\texttt{tencent/KaLM-Embedding-Gemma3-12B-2511} & 3840\\
\texttt{Salesforce/SFR-Embedding-Mistral} & 4096\\
\texttt{Octen/Octen-Embedding-8B} & 4096\\
\texttt{Linq-AI-Research/Linq-Embed-Mistral} & 4096\\
\bottomrule
\end{tabular}
\end{table}

To determine the relative weight of each classifier in the ensemble and calibrate the temperature of its softmax, we hold out a balanced subset of $20$ texts per topic and class. 
We then set the classifier's weight in the ensemble equal to the accuracy it achieves on this subset and set its temperature as the value that minimizes the negative log-likelihood on this subset.
Tables~\ref{tab:ensemble_accuracy} reports predictive performance metrics of each classifier and the full ensemble, averaged across all data points in the respective dataset.
Table~\ref{tab:ensemble_topics} reports predictive predictive performance metrics of the full ensemble, broken down by topic.

\begin{table}[h]
\centering
\caption{Average accuracy and macro F1 score of individual classifiers and the full ensemble across all data from the UKP and SemEval datasets.}
\label{tab:ensemble_accuracy}

\begin{tabular}{lcccc}
\toprule
 & \multicolumn{2}{c}{\textbf{UKP}} & \multicolumn{2}{c}{\textbf{SemEval}} \\
\cmidrule(lr){2-3}\cmidrule(lr){4-5}
\textbf{Classifier} & Accuracy & Macro F1 & Accuracy & Macro F1 \\
\midrule
\texttt{KaLM-Embedding-Gemma3-12B-2511} & 89.00\% & 0.8885 & 86.19\% & 0.8517 \\
\texttt{SFR-Embedding-Mistral} & 90.07\% & 0.8996 & 84.73\% & 0.8353 \\
\texttt{Qwen3-Embedding-8B} & 86.99\% & 0.8670 & 85.09\% & 0.8367 \\
\texttt{Octen-Embedding-8B} & 87.11\% & 0.8688 & 84.12\% & 0.8282 \\
\texttt{Linq-Embed-Mistral} & 89.96\% & 0.8982 & 85.36\% & 0.8405 \\
\midrule
\textbf{Ensemble} & \textbf{90.10\%} & \textbf{0.8994} & \textbf{86.58\%} & \textbf{0.8542} \\
\bottomrule
\end{tabular}

\end{table}

\begin{table}[h]
\centering
\caption{Average accuracy and macro F1 score of the ensemble on the topics included in the UKP and SemEval datasets.}
\label{tab:ensemble_topics}

\begin{tabular}{llccc}
\toprule
\textbf{Dataset} & \textbf{Topic} & \textbf{\# of samples} & \textbf{Accuracy} & \textbf{Macro F1} \\
\midrule
UKP & Abortion & 1{,}502 & 83.75\% & 0.8362 \\
 & Cloning & 1{,}545 & 93.40\% & 0.9331 \\
 & Death penalty & 1{,}568 & 92.98\% & 0.9159 \\
 & Gun control & 1{,}452 & 82.58\% & 0.8252 \\
 & Marijuana legalization & 1{,}213 & 92.75\% & 0.9272 \\
 & Minimum wage & 1{,}127 & 90.42\% & 0.9042 \\
 & Nuclear energy & 1{,}458 & 92.46\% & 0.9223 \\
 & School uniforms & 1{,}268 & 93.14\% & 0.9292 \\
\midrule
SemEval & Abortion & 711 & 87.62\% & 0.8430 \\
 & Acknowledging climate change & 361 & 91.97\% & 0.7984 \\
 & Atheism & 588 & 85.71\% & 0.8211 \\
 & Donald Trump & 447 & 85.01\% & 0.8372 \\
 & Feminism & 779 & 81.64\% & 0.7992 \\
 & Hillary Clinton & 728 & 89.84\% & 0.8664 \\
\bottomrule
\end{tabular}

\end{table}

Lastly, we assess the robustness of our ensembles to distribution shifts in text format.
This is particularly important for topics in the UKP dataset, which consists of single-sentence arguments rather than social media posts,  the format that LLMs in our experiments generate.
To this end, we focus on abortion, since it is the only topic shared between the two datasets.
Then, we use the reference embeddings corresponding to texts ``in favor'' and ``against'' in one dataset to classify texts in the other.
Table~\ref{tab:cross_dataset} shows that the ensemble's accuracy remains comparable to its in-distribution performance in both directions, suggesting that classifications generalize beyond the specific format of human-written text used to fit the reference embeddings.

\begin{table}[h]
\centering
\caption{Accuracy and macro F1 of the ensemble on abortion when using reference embeddings fitted on one dataset (source) to classify texts from the other (target).}
\label{tab:cross_dataset}
\begin{tabular}{llcc}
\toprule
\textbf{Source} & \textbf{Target} & \textbf{Accuracy} & \textbf{Macro F1} \\
\midrule
UKP & UKP & 83.8\% & 0.836 \\
SemEval & SemEval & 87.6\% & 0.843 \\
UKP & SemEval & 84.4\% & 0.800 \\
SemEval & UKP & 83.2\% & 0.828 \\
\bottomrule
\end{tabular}
\end{table}

\clearpage
\newpage

\subsection{Supporting tables}\label{app:tables}

\begin{table}[h]
\centering
\caption{Summary statistics of the three social network datasets used in our simulations, originally published by the Stanford Network Analysis Project (SNAP)~\cite{leskovec2012learning}.
Each dataset contains ego-networks collected from Twitter, Facebook, and Google Plus, respectively.
In the table, WCC denotes weakly connected components and SCC denotes strongly connected components.}
\label{tab:snap_stats}
% \small
\begin{tabular}{lrrr}
\toprule
\textbf{Statistic} & \textbf{Twitter} & \textbf{Facebook} & \textbf{GPlus} \\
\midrule
Directed & Yes & No & Yes \\
Nodes & 81{,}306 & 4{,}039 & 107{,}614 \\
Edges & 1{,}768{,}149 & 88{,}234 & 13{,}673{,}453 \\

Nodes (largest WCC) & 81{,}306 (1.000) & 4{,}039 (1.000) & 107{,}614 (1.000) \\
Edges (largest WCC) & 1{,}768{,}149 (1.000) & 88{,}234 (1.000) & 13{,}673{,}453 (1.000) \\

Nodes (largest SCC) & 68{,}413 (0.841) & 4{,}039 (1.000) & 69{,}501 (0.646) \\
Edges (largest SCC) & 1{,}685{,}163 (0.953) & 88{,}234 (1.000) & 9{,}168{,}660 (0.671) \\

Avg. clustering coefficient & 0.5653 & 0.6055 & 0.4901 \\
Number of triangles & 13{,}082{,}506 & 1{,}612{,}010 & 1{,}073{,}677{,}742 \\
Fraction of closed triangles & 0.06415 & 0.2647 & 0.6552 \\

Diameter & 7 & 8 & 6 \\
90\% effective diameter & 4.5 & 4.7 & 3 \\
\bottomrule
\end{tabular}
\end{table}

\begin{table}[h]
\centering
\caption{Abbreviations for topics in the SemEval and UKP datasets used in Figure~\ref{fig:shift} and Figure~\ref{fig:bias-vs-shift}.}
\label{tab:abbreviations}
\begin{tabular}{lc}
\toprule
\textbf{Topic} & \textbf{Abbreviation} \\
\midrule
Abortion & ABO \\
Acknowledging climate change & CLI \\
Atheism & ATH \\
Cloning & CLO \\
Death penalty & DTP \\
Donald Trump & DT \\
Feminism & FEM \\
Gun control & GNC \\
Hillary Clinton & HC \\
Marijuana legalization & MRJ \\
Minimum wage & MNW \\
Nuclear energy & NCL \\
School uniforms & SCH \\
\bottomrule
\end{tabular}
\end{table}

\clearpage
\newpage

\section{Additional experimental results}\label{app:results}

\subsection{Results analogous to Figures~\ref{fig:bias-by-topic},~\ref{fig:opinion-vs-bias} using both datasets and additional LLMs}\label{app:results-bias}

\begin{figure*}[h]
    \captionsetup[subfigure]{justification=centering}
    \centering
    \subcaptionbox{\texttt{gemma-3-12b-it}}[0.325\textwidth]{
        \centering
        \includegraphics[width=0.325\textwidth]{FIG/bias__intercept_by_topic__semeval__google_gemma-3-12b-it.pdf}   
    }
    \subcaptionbox{\texttt{gemma-3-27b-it}}[0.325\textwidth]{
        \centering
        \includegraphics[width=0.325\textwidth]{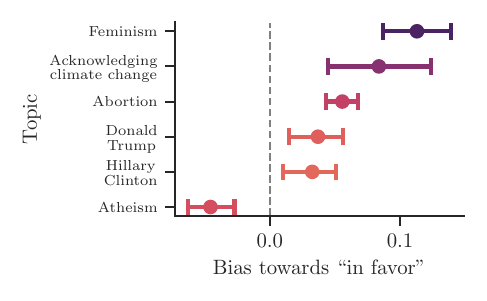}
    }
    \subcaptionbox{\texttt{Llama-3.1-8B-Instruct}}[0.325\textwidth]{
        \centering
        \includegraphics[width=0.325\textwidth]{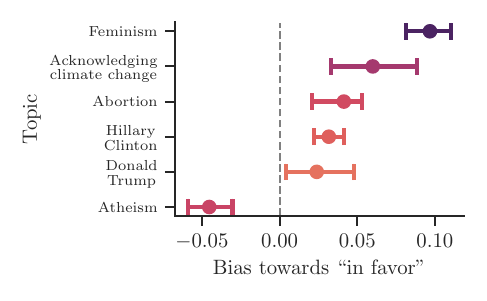}
    }
    \\
    \subcaptionbox{\texttt{Llama-3.1-70B-Instruct}}[0.325\textwidth]{
        \centering
        \includegraphics[width=0.325\textwidth]{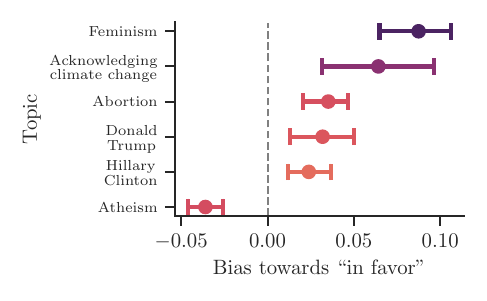}
    }
    \subcaptionbox{\texttt{gpt-oss-20b}}[0.325\textwidth]{
        \centering
        \includegraphics[width=0.325\textwidth]{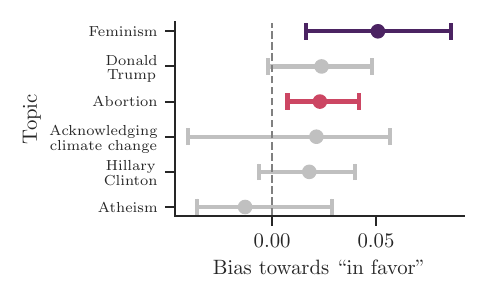}
    }
    \subcaptionbox{\texttt{gpt-5.6-luna}}[0.325\textwidth]{
        \centering
        \includegraphics[width=0.325\textwidth]{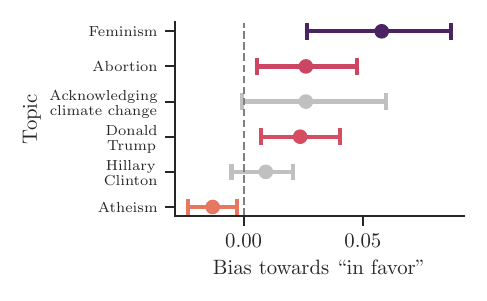}
    }
    \\
    \subcaptionbox{\texttt{Ministral-3-\\8B-Instruct-2512}}[0.325\textwidth]{
        \centering
        \includegraphics[width=0.325\textwidth]{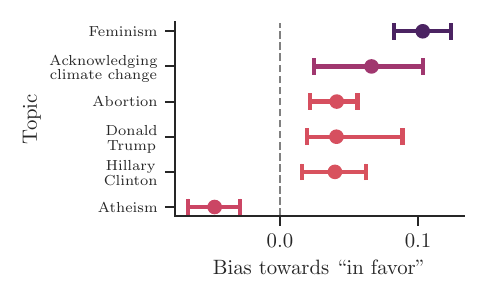}
    }
    \subcaptionbox{\texttt{Qwen3-8B}}[0.325\textwidth]{
        \centering
        \includegraphics[width=0.325\textwidth]{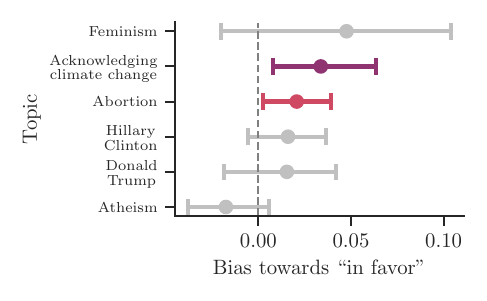}
    }
    \subcaptionbox{\texttt{Qwen3-32B}}[0.325\textwidth]{
        \centering
        \includegraphics[width=0.325\textwidth]{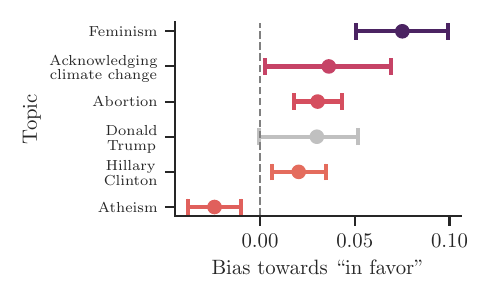}
    }
    \caption{\textbf{Bias introduced by LLMs when improving human-written posts.} The panels show the posterior means and 95\% credible intervals of the intercepts capturing the average bias $\beta$ (see Section~\ref{sec:bias}) by different LLMs across topics from the SemEval dataset, using prompts for the improvement task (see \ref{app:improvement_prompts}).\label{fig:intercepts_semeval_noflips}
    }
\end{figure*}

\begin{figure*}[h]
    \captionsetup[subfigure]{justification=centering}
    \centering
    \subcaptionbox{\texttt{gemma-3-12b-it}}[0.325\textwidth]{
        \centering
        \includegraphics[width=0.325\textwidth]{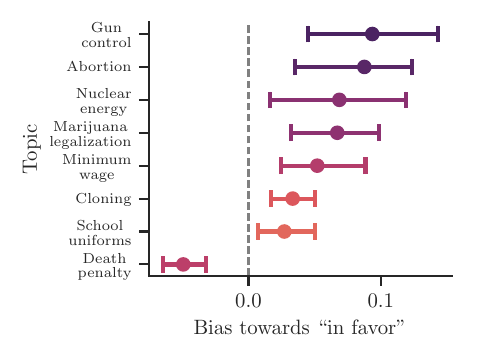}   
    }
    \subcaptionbox{\texttt{gemma-3-27b-it}}[0.325\textwidth]{
        \centering
        \includegraphics[width=0.325\textwidth]{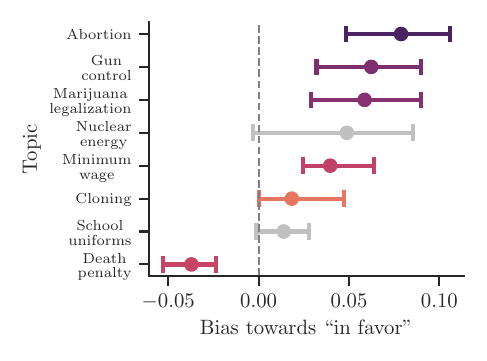}
    }
    \subcaptionbox{\texttt{Llama-3.1-8B-Instruct}}[0.325\textwidth]{
        \centering
        \includegraphics[width=0.325\textwidth]{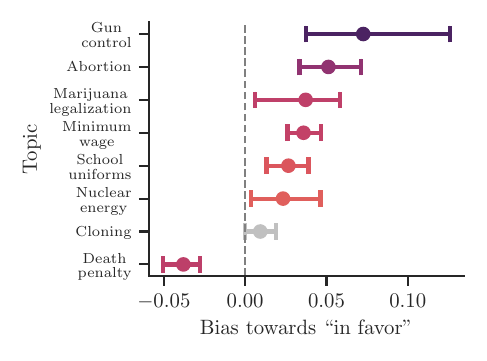}
    }
    \\
    \subcaptionbox{\texttt{Llama-3.1-70B-Instruct}}[0.325\textwidth]{
        \centering
        \includegraphics[width=0.325\textwidth]{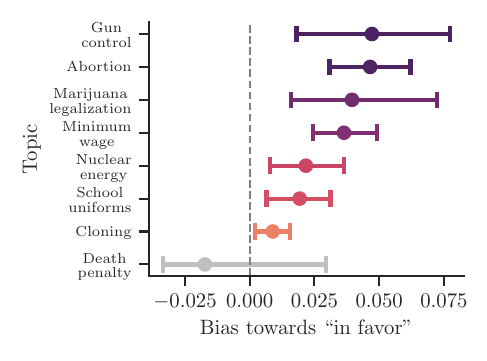}
    }
    \subcaptionbox{\texttt{gpt-oss-20b}}[0.325\textwidth]{
        \centering
        \includegraphics[width=0.325\textwidth]{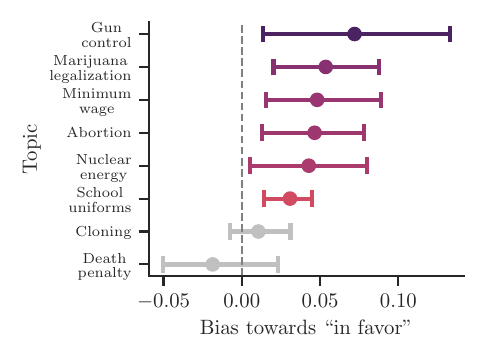}
    }
    \subcaptionbox{\texttt{gpt-5.6-luna}}[0.325\textwidth]{
        \centering
        \includegraphics[width=0.325\textwidth]{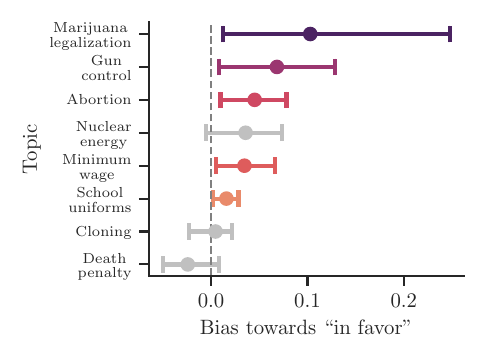}
    }
    \\
    \subcaptionbox{\texttt{Ministral-3-\\8B-Instruct-2512}}[0.325\textwidth]{
        \centering
        \includegraphics[width=0.325\textwidth]{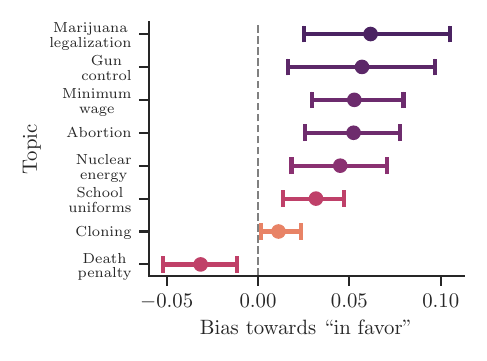}
    }
    \subcaptionbox{\texttt{Qwen3-8B}}[0.325\textwidth]{
        \centering
        \includegraphics[width=0.325\textwidth]{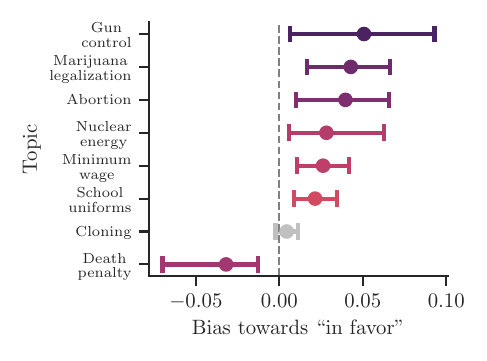}
    }
    \subcaptionbox{\texttt{Qwen3-32B}}[0.325\textwidth]{
        \centering
        \includegraphics[width=0.325\textwidth]{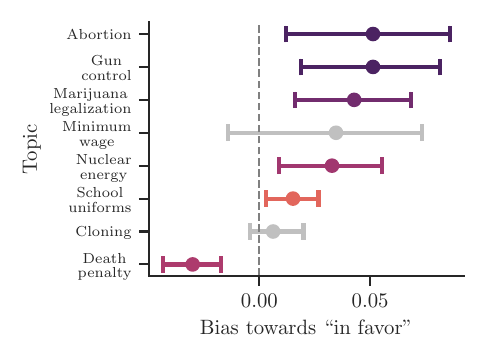}
    }
    \caption{\textbf{Bias introduced by LLMs when drafting social media posts.}  The panels show the posterior means and 95\% credible intervals of the intercepts capturing the average bias $\beta$ (see Section~\ref{sec:bias}) by different LLMs across topics from the UKP dataset, using prompts for the drafting task (see \ref{app:drafting_prompts}).
    \label{fig:intercepts_ukp_noflips}
    }
\end{figure*}

\begin{figure*}[h]
    \captionsetup[subfigure]{justification=centering}
    \centering
    \includegraphics[width=0.5\textwidth]{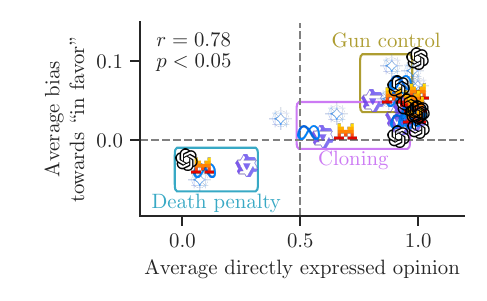} 
    \caption{\textbf{Average LLM-induced bias vs. average directly expressed opinion (UKP).}
    The figure shows the mean of the bias $\beta$ against the average directly expressed opinion of each model on each topic. Each point represents one model-topic pair with different markers used for 
    \texttt{Llama-3.1-8B-Instruct} (\includegraphics[height=0.9\inlineheight]{FIG/meta-color.png}),
    \texttt{gpt-5.6-luna} (\includegraphics[height=0.9\inlineheight]{FIG/chat-gpt.png}),
    \texttt{gemma-3-12b-it} (\includegraphics[height=0.9\inlineheight]{FIG/gemma-color.png}), 
    \texttt{Ministral-3-8B-Instruct-2512} (\includegraphics[height=0.9\inlineheight]{FIG/mistral-color.png}),
    and \texttt{Qwen3-8B} (\includegraphics[height=0.9\inlineheight]{FIG/qwen-color.png}).
    }
    \label{fig:bias_ukp}
\end{figure*}

\clearpage
\newpage

\subsection{Analysis of cases where the predicted stance of the LLM-generated text differs from that of the original text}\label{app:results-flips}

\begin{figure*}[h]
    \captionsetup[subfigure]{justification=centering}
    \centering
    \subcaptionbox{\texttt{gemma-3-12b-it}}[0.325\textwidth]{
        \centering
        \includegraphics[width=0.325\textwidth]{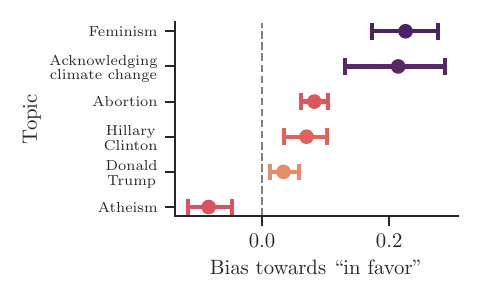}   
    }
    \subcaptionbox{\texttt{gemma-3-27b-it}}[0.325\textwidth]{
        \centering
        \includegraphics[width=0.325\textwidth]{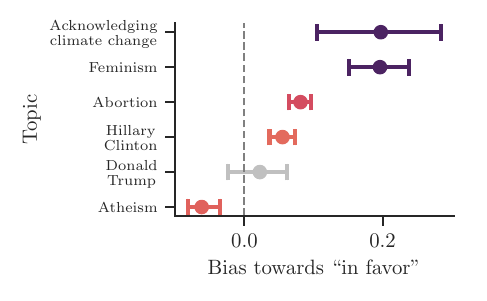}
    }
    \subcaptionbox{\texttt{Llama-3.1-8B-Instruct}}[0.325\textwidth]{
        \centering
        \includegraphics[width=0.325\textwidth]{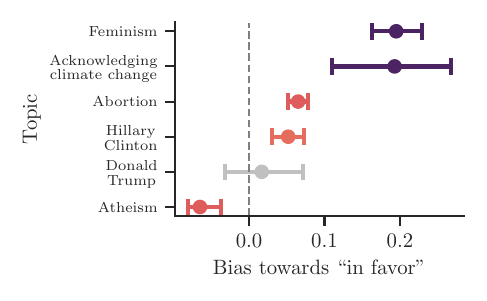}
    }
    \\
    \subcaptionbox{\texttt{Llama-3.1-70B-Instruct}}[0.325\textwidth]{
        \centering
        \includegraphics[width=0.325\textwidth]{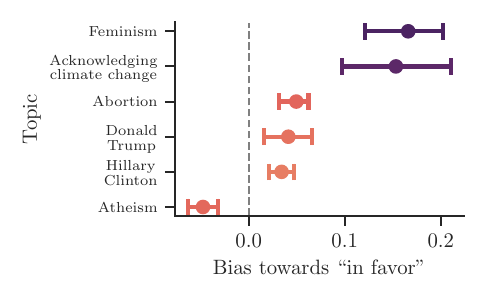}
    }
    \subcaptionbox{\texttt{gpt-oss-20b}}[0.325\textwidth]{
        \centering
        \includegraphics[width=0.325\textwidth]{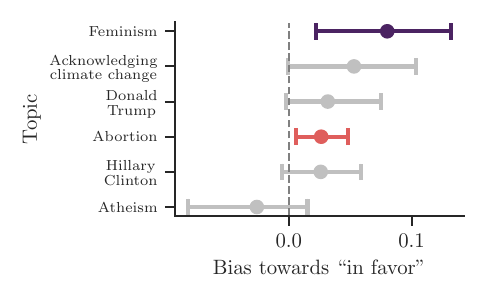}
    }
    \subcaptionbox{\texttt{gpt-5.6-luna}}[0.325\textwidth]{
        \centering
        \includegraphics[width=0.325\textwidth]{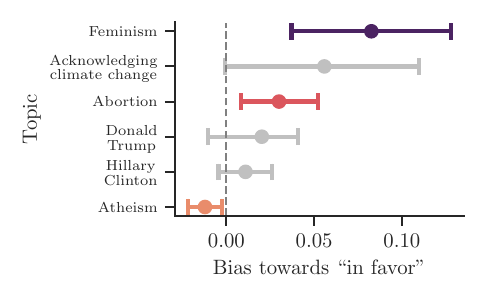}
    }
    \\
    \subcaptionbox{\texttt{Ministral-3-\\8B-Instruct-2512}}[0.325\textwidth]{
        \centering
        \includegraphics[width=0.325\textwidth]{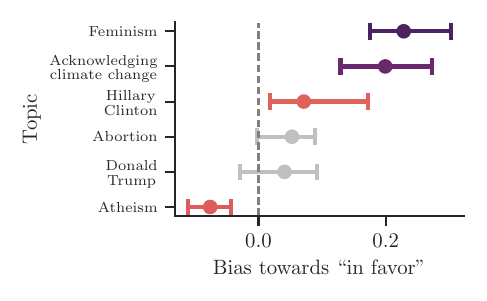}
    }
    \subcaptionbox{\texttt{Qwen3-8B}}[0.325\textwidth]{
        \centering
        \includegraphics[width=0.325\textwidth]{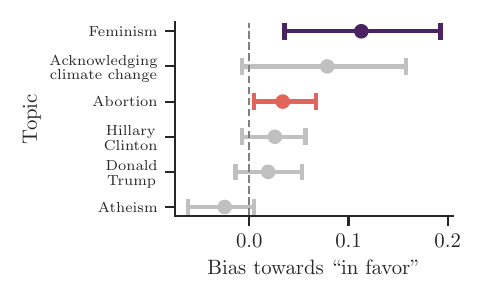}
    }
    \subcaptionbox{\texttt{Qwen3-32B}}[0.325\textwidth]{
        \centering
        \includegraphics[width=0.325\textwidth]{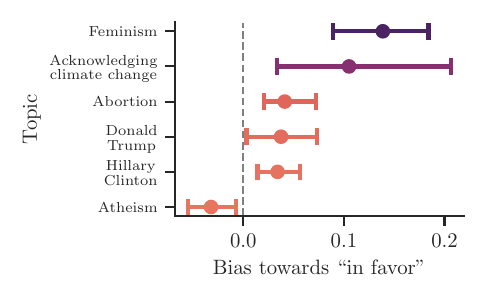}
    }
    \caption{\textbf{Bias introduced by LLMs when improving human-written posts, as measured with the inclusion of opinion flips.} The panels show the posterior means and 95\% credible intervals of the intercepts capturing the average bias $\beta$ (see Section~\ref{sec:bias}) by different LLMs across topics from the SemEval dataset, using prompts for the improvement task (see \ref{app:improvement_prompts}).\label{fig:intercepts_semeval_flips}
    }
\end{figure*}

\begin{figure*}[h]
    \captionsetup[subfigure]{justification=centering}
    \centering
    \subcaptionbox{\texttt{gemma-3-12b-it}}[0.325\textwidth]{
        \centering
        \includegraphics[width=0.325\textwidth]{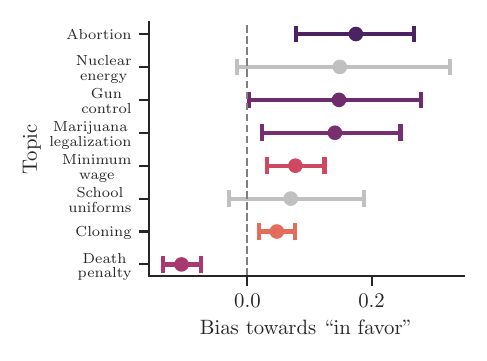}   
    }
    \subcaptionbox{\texttt{gemma-3-27b-it}}[0.325\textwidth]{
        \centering
        \includegraphics[width=0.325\textwidth]{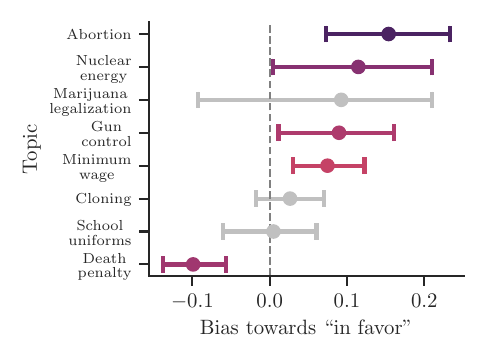}
    }
    \subcaptionbox{\texttt{Llama-3.1-8B-Instruct}}[0.325\textwidth]{
        \centering
        \includegraphics[width=0.325\textwidth]{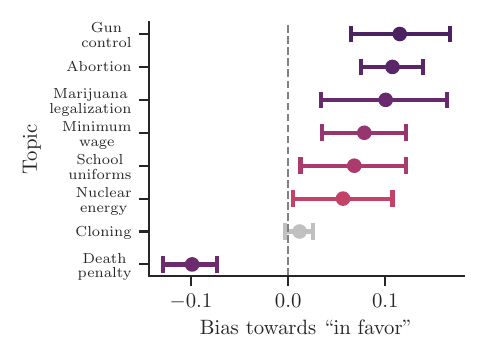}
    }
    \\
    \subcaptionbox{\texttt{Llama-3.1-70B-Instruct}}[0.325\textwidth]{
        \centering
        \includegraphics[width=0.325\textwidth]{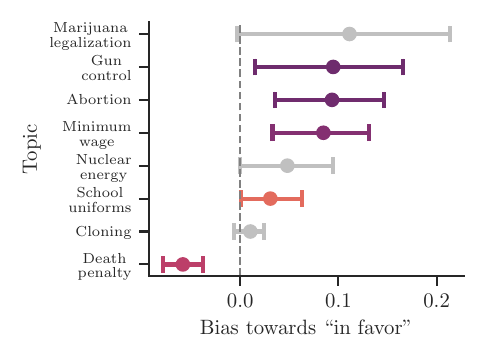}
    }
    \subcaptionbox{\texttt{gpt-oss-20b}}[0.325\textwidth]{
        \centering
        \includegraphics[width=0.325\textwidth]{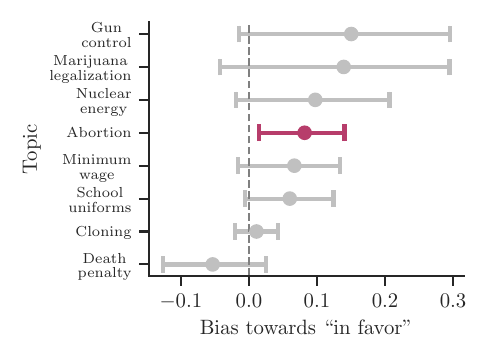}
    }
    \subcaptionbox{\texttt{gpt-5.6-luna}}[0.325\textwidth]{
        \centering
        \includegraphics[width=0.325\textwidth]{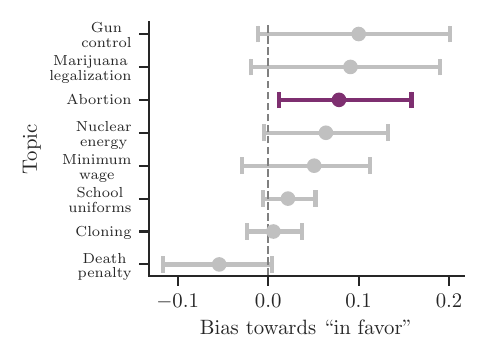}
    }
    \\
    \subcaptionbox{\texttt{Ministral-3-\\8B-Instruct-2512}}[0.325\textwidth]{
        \centering
        \includegraphics[width=0.325\textwidth]{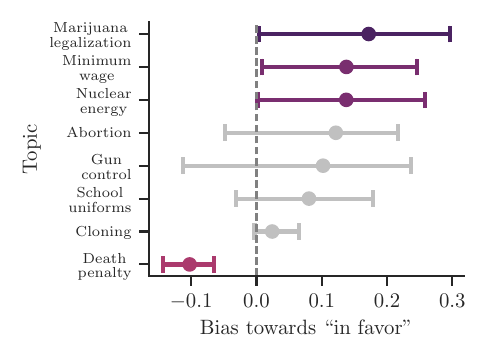}
    }
    \subcaptionbox{\texttt{Qwen3-8B}}[0.325\textwidth]{
        \centering
        \includegraphics[width=0.325\textwidth]{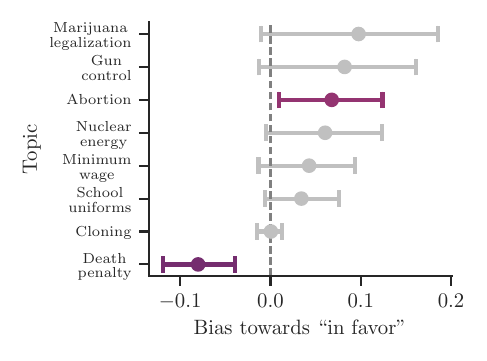}
    }
    \subcaptionbox{\texttt{Qwen3-32B}}[0.325\textwidth]{
        \centering
        \includegraphics[width=0.325\textwidth]{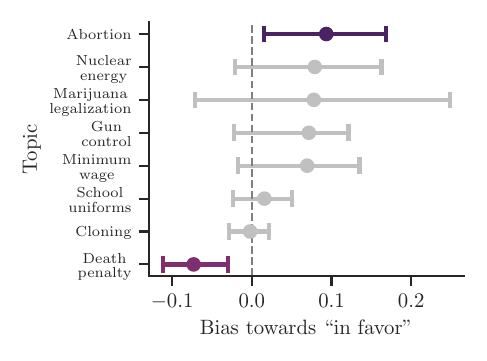}
    }
    \caption{\textbf{Bias introduced by LLMs when drafting social media posts, as measured with the inclusion of opinion flips.}  The panels show the posterior means and 95\% credible intervals of the intercepts capturing the average bias $\beta$ (see Section~\ref{sec:bias}) by different LLMs across topics from the UKP dataset, using prompts for the drafting task (see \ref{app:drafting_prompts}).}\label{fig:intercepts_ukp_flips}
\end{figure*}

In our experiments in Section~\ref{sec:bias}, we restrict our analysis to cases where the predicted stance of the LLM-generated text matches that of the original text and exclude the rest, that is, cases where there is an ``opinion flip'' from ``in favor'' to ``against'' or vice versa.
Here, we report how often such flips occur and repeat the analysis of Section~\ref{sec:bias} without this restriction.
Tables~\ref{tab:flips_google_gemma_3_12b_it}-\ref{tab:flips_qwen_qwen3_32b} show the opinion-flip rates for each LLM and topic.
Figures~\ref{fig:intercepts_semeval_flips} and~\ref{fig:intercepts_ukp_flips} show analogous results to those in Figure~\ref{fig:bias-by-topic} without the aforementioned restriction.
Overall, we find that opinion flips occur more frequently in LLM-topic pairs where the main analysis shows a statistically significant bias and that they are disproportionately more likely to occur in the direction of the LLM's bias, that is, from ``against'' to ``in favor'' for topics where the LLM is biased towards ``in favor'' and vice versa.
Hence, their exclusion does not qualitatively change the conclusions of our main analysis, which can also be seen by comparing Figures~\ref{fig:intercepts_semeval_flips} and~\ref{fig:intercepts_ukp_flips} below to their counterparts in Appendix~\ref{app:results-bias} that exclude opinion flips (\ie, Figures~\ref{fig:intercepts_semeval_noflips} and~\ref{fig:intercepts_ukp_noflips}).

\begin{table}[h]
\centering
\caption{Opinion-flip rates for \texttt{gemma-3-12b-it} by topic. The columns show the percentage of human-written texts classified as ``in favor'' (``against'') whose predicted stance of the LLM-generated counterpart is ``against'' (``in favor'').}
\label{tab:flips_google_gemma_3_12b_it}
\begin{tabular}{lcc}
\toprule
\textbf{Topic (dataset)} & \textbf{In favor $\rightarrow$ against} & \textbf{Against $\rightarrow$ in favor} \\
\midrule
Abortion (SemEval) & 1.69\% & 16.64\% \\
Acknowledging climate change (SemEval) & 6.41\% & 56.67\% \\
Atheism (SemEval) & 19.26\% & 1.29\% \\
Donald Trump (SemEval) & 7.92\% & 7.73\% \\
Feminism (SemEval) & 0.42\% & 60.78\% \\
Hillary Clinton (SemEval) & 0.74\% & 11.69\% \\
Abortion (UKP) & 2.2\% & 43.52\% \\
Cloning (UKP) & 2.63\% & 8.43\% \\
Death penalty (UKP) & 21.75\% & 0.33\% \\
Gun control (UKP) & 0.18\% & 31.58\% \\
Marijuana legalization (UKP) & 1.25\% & 26.02\% \\
Minimum wage (UKP) & 1.78\% & 10.18\% \\
Nuclear energy (UKP) & 0.53\% & 27.25\% \\
School uniforms (UKP) & 4.5\% & 7.27\% \\
\bottomrule
\end{tabular}
\end{table}

\begin{table}[h]
\centering
\caption{Opinion-flip rates for \texttt{gemma-3-27b-it} by topic. The columns show the percentage of human-written texts classified as ``in favor'' (``against'') whose predicted stance of the LLM-generated counterpart is ``against'' (``in favor'').}
\label{tab:flips_google_gemma_3_27b_it}
\begin{tabular}{lcc}
\toprule
\textbf{Topic (dataset)} & \textbf{In favor $\rightarrow$ against} & \textbf{Against $\rightarrow$ in favor} \\
\midrule
Abortion (SemEval) & 1.37\% & 17.76\% \\
Acknowledging climate change (SemEval) & 3.85\% & 53.72\% \\
Atheism (SemEval) & 14.27\% & 1.6\% \\
Donald Trump (SemEval) & 3.65\% & 6.93\% \\
Feminism (SemEval) & 0.33\% & 51.03\% \\
Hillary Clinton (SemEval) & 1.08\% & 10.95\% \\
Abortion (UKP) & 1.73\% & 36.98\% \\
Cloning (UKP) & 2.78\% & 5.92\% \\
Death penalty (UKP) & 20.63\% & 0.63\% \\
Gun control (UKP) & 1.97\% & 15.42\% \\
Marijuana legalization (UKP) & 0.83\% & 22.35\% \\
Minimum wage (UKP) & 0.23\% & 11.18\% \\
Nuclear energy (UKP) & 0.22\% & 20.5\% \\
School uniforms (UKP) & 6.2\% & 4.42\% \\
\bottomrule
\end{tabular}
\end{table}

\begin{table}[h]
\centering
\caption{Opinion-flip rates for \texttt{Llama-3.1-8B-Instruct} by topic. The columns show the percentage of human-written texts classified as ``in favor'' (``against'') whose predicted stance of the LLM-generated counterpart is ``against'' (``in favor'').}
\label{tab:flips_meta_llama_llama_3_1_8b_instruct}
\begin{tabular}{lcc}
\toprule
\textbf{Topic (dataset)} & \textbf{In favor $\rightarrow$ against} & \textbf{Against $\rightarrow$ in favor} \\
\midrule
Abortion (SemEval) & 2.74\% & 16.35\% \\
Acknowledging climate change (SemEval) & 1.54\% & 47.31\% \\
Atheism (SemEval) & 17.25\% & 1.93\% \\
Donald Trump (SemEval) & 10.26\% & 8.75\% \\
Feminism (SemEval) & 1.23\% & 51.72\% \\
Hillary Clinton (SemEval) & 2.73\% & 10.83\% \\
Abortion (UKP) & 3.18\% & 27.1\% \\
Cloning (UKP) & 4.33\% & 4.93\% \\
Death penalty (UKP) & 21.1\% & 1.47\% \\
Gun control (UKP) & 1.72\% & 28.13\% \\
Marijuana legalization (UKP) & 2.12\% & 19.43\% \\
Minimum wage (UKP) & 2.65\% & 14.03\% \\
Nuclear energy (UKP) & 1.92\% & 11.05\% \\
School uniforms (UKP) & 4.05\% & 14.45\% \\
\bottomrule
\end{tabular}
\end{table}

\begin{table}[h]
\centering
\caption{Opinion-flip rates for \texttt{Llama-3.1-70B-Instruct} by topic. The columns show the percentage of human-written texts classified as ``in favor'' (``against'') whose predicted stance of the LLM-generated counterpart is ``against'' (``in favor'').}
\label{tab:flips_meta_llama_llama_3_1_70b_instruct}
\begin{tabular}{lcc}
\toprule
\textbf{Topic (dataset)} & \textbf{In favor $\rightarrow$ against} & \textbf{Against $\rightarrow$ in favor} \\
\midrule
Abortion (SemEval) & 1.96\% & 10.55\% \\
Acknowledging climate change (SemEval) & 4.1\% & 36.67\% \\
Atheism (SemEval) & 10.55\% & 2.07\% \\
Donald Trump (SemEval) & 4.01\% & 8.54\% \\
Feminism (SemEval) & 1.18\% & 41.68\% \\
Hillary Clinton (SemEval) & 0.59\% & 5.5\% \\
Abortion (UKP) & 3.02\% & 22.15\% \\
Cloning (UKP) & 3.35\% & 4.13\% \\
Death penalty (UKP) & 13.77\% & 2.07\% \\
Gun control (UKP) & 1.73\% & 22.48\% \\
Marijuana legalization (UKP) & 1.7\% & 22.28\% \\
Minimum wage (UKP) & 1.17\% & 14.93\% \\
Nuclear energy (UKP) & 1.37\% & 9.22\% \\
School uniforms (UKP) & 3.55\% & 6.23\% \\
\bottomrule
\end{tabular}
\end{table}

\begin{table}[h]
\centering
\caption{Opinion-flip rates for \texttt{gpt-oss-20b} by topic. The columns show the percentage of human-written texts classified as ``in favor'' (``against'') whose predicted stance of the LLM-generated counterpart is ``against'' (``in favor'').}
\label{tab:flips_openai_gpt_oss_20b}
\begin{tabular}{lcc}
\toprule
\textbf{Topic (dataset)} & \textbf{In favor $\rightarrow$ against} & \textbf{Against $\rightarrow$ in favor} \\
\midrule
Abortion (SemEval) & 5.05\% & 6.58\% \\
Acknowledging climate change (SemEval) & 2.05\% & 19.49\% \\
Atheism (SemEval) & 3.66\% & 2.59\% \\
Donald Trump (SemEval) & 5.6\% & 6.3\% \\
Feminism (SemEval) & 4.02\% & 16.95\% \\
Hillary Clinton (SemEval) & 1.31\% & 4.5\% \\
Abortion (UKP) & 4.2\% & 18.68\% \\
Cloning (UKP) & 3.5\% & 3.65\% \\
Death penalty (UKP) & 12.87\% & 1.68\% \\
Gun control (UKP) & 0.92\% & 37.13\% \\
Marijuana legalization (UKP) & 2.12\% & 29.2\% \\
Minimum wage (UKP) & 2.1\% & 11.87\% \\
Nuclear energy (UKP) & 1.23\% & 18.43\% \\
School uniforms (UKP) & 2.38\% & 11.37\% \\
\bottomrule
\end{tabular}
\end{table}

\begin{table}[h]
\centering
\caption{Opinion-flip rates for \texttt{gpt-5.6-luna} by topic. The columns show the percentage of human-written texts classified as ``in favor'' (``against'') whose predicted stance of the LLM-generated counterpart is ``against'' (``in favor'').}
\label{tab:flips_openai_gpt_5_6_luna}
\begin{tabular}{lcc}
\toprule
\textbf{Topic (dataset)} & \textbf{In favor $\rightarrow$ against} & \textbf{Against $\rightarrow$ in favor} \\
\midrule
Abortion (SemEval) & 0.89\% & 5.64\% \\
Acknowledging climate change (SemEval) & 0.64\% & 14.87\% \\
Atheism (SemEval) & 1.54\% & 3.58\% \\
Donald Trump (SemEval) & 2.73\% & 3.62\% \\
Feminism (SemEval) & 0.92\% & 18.8\% \\
Hillary Clinton (SemEval) & 0.54\% & 2.32\% \\
Abortion (UKP) & 2.15\% & 18.63\% \\
Cloning (UKP) & 3.77\% & 3.98\% \\
Death penalty (UKP) & 11.15\% & 1.58\% \\
Gun control (UKP) & 0.55\% & 20.82\% \\
Marijuana legalization (UKP) & 1.23\% & 14.8\% \\
Minimum wage (UKP) & 1.23\% & 8.15\% \\
Nuclear energy (UKP) & 1.6\% & 10.8\% \\
School uniforms (UKP) & 3.55\% & 4.27\% \\
\bottomrule
\end{tabular}
\end{table}

\begin{table}[h]
\centering
\caption{Opinion-flip rates for \texttt{Ministral-3-8B-Instruct-2512} by topic. The columns show the percentage of human-written texts classified as ``in favor'' (``against'') whose predicted stance of the LLM-generated counterpart is ``against'' (``in favor'').}
\label{tab:flips_mistralai_ministral_3_8b_instruct_2512}
\begin{tabular}{lcc}
\toprule
\textbf{Topic (dataset)} & \textbf{In favor $\rightarrow$ against} & \textbf{Against $\rightarrow$ in favor} \\
\midrule
Abortion (SemEval) & 3.89\% & 16.71\% \\
Acknowledging climate change (SemEval) & 3.48\% & 50.07\% \\
Atheism (SemEval) & 19.33\% & 3.57\% \\
Donald Trump (SemEval) & 9.37\% & 12.38\% \\
Feminism (SemEval) & 1.2\% & 59.44\% \\
Hillary Clinton (SemEval) & 4.43\% & 15.58\% \\
Abortion (UKP) & 3.31\% & 33.24\% \\
Cloning (UKP) & 5.65\% & 8.06\% \\
Death penalty (UKP) & 23.12\% & 2.22\% \\
Gun control (UKP) & 2.65\% & 31.99\% \\
Marijuana legalization (UKP) & 1.67\% & 33.3\% \\
Minimum wage (UKP) & 2.11\% & 26.98\% \\
Nuclear energy (UKP) & 1.41\% & 29.01\% \\
School uniforms (UKP) & 3.42\% & 17.35\% \\
\bottomrule
\end{tabular}
\end{table}

\begin{table}[h]
\centering
\caption{Opinion-flip rates for \texttt{Qwen3-8B} by topic. The columns show the percentage of human-written texts classified as ``in favor'' (``against'') whose predicted stance of the LLM-generated counterpart is ``against'' (``in favor'').}
\label{tab:flips_qwen_qwen3_8b}
\begin{tabular}{lcc}
\toprule
\textbf{Topic (dataset)} & \textbf{In favor $\rightarrow$ against} & \textbf{Against $\rightarrow$ in favor} \\
\midrule
Abortion (SemEval) & 3.22\% & 6.88\% \\
Acknowledging climate change (SemEval) & 3.34\% & 21.54\% \\
Atheism (SemEval) & 4.85\% & 1.93\% \\
Donald Trump (SemEval) & 4.14\% & 4.98\% \\
Feminism (SemEval) & 1.33\% & 29.97\% \\
Hillary Clinton (SemEval) & 1.33\% & 5.23\% \\
Abortion (UKP) & 3.08\% & 17.77\% \\
Cloning (UKP) & 5.85\% & 4.13\% \\
Death penalty (UKP) & 16.96\% & 0.98\% \\
Gun control (UKP) & 1.97\% & 17.99\% \\
Marijuana legalization (UKP) & 1.93\% & 19.6\% \\
Minimum wage (UKP) & 2.5\% & 7.65\% \\
Nuclear energy (UKP) & 1.08\% & 11.07\% \\
School uniforms (UKP) & 3.27\% & 6.58\% \\
\bottomrule
\end{tabular}
\end{table}

\begin{table}[h]
\centering
\caption{Opinion-flip rates for \texttt{Qwen3-32B} by topic. The columns show the percentage of human-written texts classified as ``in favor'' (``against'') whose predicted stance of the LLM-generated counterpart is ``against'' (``in favor'').}
\label{tab:flips_qwen_qwen3_32b}
\begin{tabular}{lcc}
\toprule
\textbf{Topic (dataset)} & \textbf{In favor $\rightarrow$ against} & \textbf{Against $\rightarrow$ in favor} \\
\midrule
Abortion (SemEval) & 2.15\% & 9.2\% \\
Acknowledging climate change (SemEval) & 3.33\% & 28.63\% \\
Atheism (SemEval) & 7.77\% & 2.37\% \\
Donald Trump (SemEval) & 3.67\% & 7.71\% \\
Feminism (SemEval) & 1.13\% & 35.7\% \\
Hillary Clinton (SemEval) & 1.08\% & 6.76\% \\
Abortion (UKP) & 2.67\% & 22.39\% \\
Cloning (UKP) & 4.1\% & 4.02\% \\
Death penalty (UKP) & 15.37\% & 1.02\% \\
Gun control (UKP) & 1.38\% & 14.68\% \\
Marijuana legalization (UKP) & 1.65\% & 23.42\% \\
Minimum wage (UKP) & 1.18\% & 11.89\% \\
Nuclear energy (UKP) & 1.12\% & 14.75\% \\
School uniforms (UKP) & 4.17\% & 4.57\% \\
\bottomrule
\end{tabular}
\end{table}

\clearpage
\newpage

\subsection{Examples of original and LLM-generated posts}\label{app:examples}

Below, we show pairs of a human-written post and its LLM-generated counterpart, along with the numerical opinions in $[0,1]$ given by the respective ensemble classifier, where $0$ corresponds to ``against'' and $1$ corresponds to ``in favor''.

\subsubsection{Examples of opinions ``in favor'' that are strengthened when the LLM presents a bias towards the ``in favor'' side}

\begin{figure}[h]
\begin{humanpost}{0.704}
``I guess wanting and trying to make this world a better place for everyone makes me ugly. \#feminist''
\end{humanpost}
\begin{llmpost}{0.965}
``I'm proud to stand up for a world where everyone can thrive, and being a vocal advocate for equality makes me unapologetically beautiful. \#feminism''
\end{llmpost}
\caption{Example on ``Feminism'' using \texttt{Llama-3.1-8B-Instruct}.}
\end{figure}

\begin{figure}[h]
\begin{humanpost}{0.709}
``Liberals do not like that @realDonaldTrump is moving up in the polls. \#MakeAmericaGreatAgain \#TheGreatWallofMexico''
\end{humanpost}
\begin{llmpost}{0.982}
``Critics can't ignore that @realDonaldTrump remains strong—because Americans keep choosing him to lead. The path forward? A bright one built on our shared dreams, values, and victory. \#MakeAmericaGreatAgain''
\end{llmpost}
\caption{Example on ``Donald Trump'' using \texttt{Ministral-3-8B-Instruct-2512}.}
\end{figure}

\begin{figure}[h]
\begin{humanpost}{0.718}
``I need @JamesCarville \& @HillaryClinton to stop emailing me. I love you, but I need my space. \#stillvotingforyou''
\end{humanpost}
\begin{llmpost}{0.984}
``I'm sending love (and a request for an email sabbatical) to @JamesCarville \& @HillaryClinton! Can't wait to vote for you, but my inbox needs a breather \includegraphics[height=0.9\inlineheight]{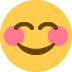} \#StillVotingForYou''
\end{llmpost}
\caption{Example on ``Hillary Clinton'' using \texttt{Qwen3-32B}.}
\end{figure}

\clearpage
\newpage

\subsubsection{Examples of opinions ``against'' that are strengthened when the LLM presents a bias towards the ``against'' side}

\begin{figure}[h]
\begin{humanpost}{0.231}
``People keep asking if I'm going to be a vicar so much I'm scared it might actually happen \#christianvocations \#theology \#help''
\end{humanpost}
\begin{llmpost}{0.061}
``Apparently, I'm getting asked about becoming a vicar \textbf{a lot} – now I'm half-convinced it's my destiny! Send prayers (and maybe a sign) to help me discern this calling. \#christianvocations \#theology \#help''
\end{llmpost}
\caption{Example on ``Atheism'' using \texttt{gemma-3-27b-it}.}
\end{figure}

\begin{figure}[h]
\begin{humanpost}{0.242}
``Exo.20.13-15 - ...Do not commit murder...DO NOT COMMIT ADULTERY...Do not steal.''
\end{humanpost}
\begin{llmpost}{0.065}
``Life's rules aren't here to scare us—God's law protects us. \includegraphics[height=0.9\inlineheight]{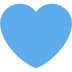} Today's reminder: \textbf{No murder} \includegraphics[height=0.9\inlineheight]{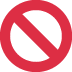} \textbf{No adultery} \includegraphics[height=0.9\inlineheight]{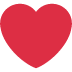} \textbf{No theft} \includegraphics[height=0.9\inlineheight]{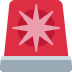} Protect others and yourself. Stay blessed!''
\end{llmpost}
\caption{Example on ``Atheism'' using \texttt{Ministral-3-8B-Instruct-2512}.}
\end{figure}

\begin{figure}[h]
\begin{humanpost}{0.165}
``He became a slave to the cross that men could go free!''
\end{humanpost}
\begin{llmpost}{0.052}
``He willingly took on the burden of the cross so that we could be free. It's a powerful reminder of sacrifice and redemption. Thankful for the ultimate gift of freedom through His love.''
\end{llmpost}
\caption{Example on ``Atheism'' using \texttt{gemma-3-27b-it}.}
\end{figure}

\clearpage
\newpage

\subsubsection{Examples of opinion flips from ``against'' to ``in favor''}

\begin{figure}[h]
\begin{humanpost}{0.068}
``@Esmerald\_Kiczek  Rock on girl!  We all know \#AbortionIsMurder.''
\end{humanpost}
\begin{llmpost}{0.94}
``Shoutout to a strong woman @Esmerald\_Kiczek - we're proud of your courage \& conviction. Remember, every individual deserves respect \& dignity. \#EmpoweringWomen \#ProChoice''
\end{llmpost}
\caption{Example on ``Abortion'' using \texttt{Llama-3.1-8B-Instruct}.}
\end{figure}

\begin{figure}[h]
\begin{humanpost}{0.207}
``@UN ice cracking in the summer?? SO ALARMING. \#climatechangehoax''
\end{humanpost}
\begin{llmpost}{0.974}
``@UN new research shows Arctic ice thinning even in summer. Alarming—our climate's under pressure. \#ClimateAction''
\end{llmpost}
\caption{Example on ``Acknowledging climate change'' using \texttt{Ministral-3-8B-Instruct-2512}.}
\end{figure}

\begin{figure}[h]
\begin{humanpost}{0.035}
``Are we so desperate in this country to seriously consider a 60+ woman as president?''
\end{humanpost}
\begin{llmpost}{0.985}
``Are we ready to embrace the power of experience? A 60+ woman leading the nation isn't a crisis—it's a testament to resilience, wisdom, and redefining what leadership looks like.''
\end{llmpost}
\caption{Example on ``Hillary Clinton'' using \texttt{Qwen3-32B}.}
\end{figure}

\clearpage
\newpage

\subsubsection{Examples of opinion flips from ``in favor'' to ``against''}

\begin{figure}[h]
\begin{humanpost}{0.757}
``Jesus is not dead , he wasn't real !notdead''
\end{humanpost}
\begin{llmpost}{0.108}
``Jesus' story continues to inspire and challenge us today. Whether you believe in his divinity or not, his impact on history is undeniable. \#Jesus \#Faith \#History''
\end{llmpost}
\caption{Example on ``Atheism'' using \texttt{gemma-3-12b-it}.}
\end{figure}

\begin{figure}[h]
\begin{humanpost}{0.742}
``How did the crucifixion of \#Christ lead the world to \#salvation, when the only thing it started was the \#DarkAges? \#skepticism''
\end{humanpost}
\begin{llmpost}{0.064}
``Was \#Christ's crucifixion just the start of the Dark Ages—or did its sacrifice rewrite history itself? Dive into how one act became the cornerstone of hope for billions. \#Salvation \#FaithAndHistory \#HistoricalGrace''
\end{llmpost}
\caption{Example on ``Atheism'' using \texttt{Ministral-3-8B-Instruct-2512}.}
\end{figure}

\begin{figure}[h]
\begin{humanpost}{0.838}
``If God = Miraculous And Miracles = Impossible Then God = Impossible \#logic \#reason \#science \#RT''
\end{humanpost}
\begin{llmpost}{0.171}
``The more we label miracles as impossible, the more we limit God's power, but what if impossible = new possibility? \#faith \#reason \#science''
\end{llmpost}
\caption{Example on ``Atheism'' using \texttt{Llama-3.1-8B-Instruct}.}
\end{figure}

\clearpage
\newpage

\subsection{Results analogous to Figure~\ref{fig:network_simulations} using additional topics and social networks}\label{app:results-simulations}

\begin{figure*}[h]
    \captionsetup[subfigure]{justification=centering}
    \centering
    \subcaptionbox{Google Plus}[0.45\textwidth]{
        \centering
        \includegraphics[width=0.45\textwidth]{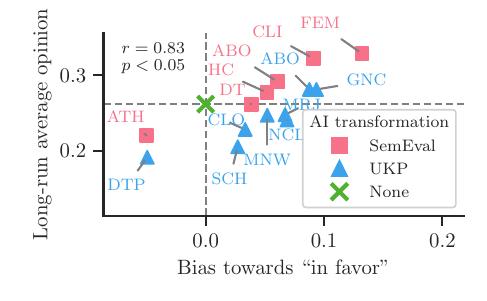}
    }
    \subcaptionbox{Facebook}[0.45\textwidth]{
        \centering
        \includegraphics[width=0.45\textwidth]{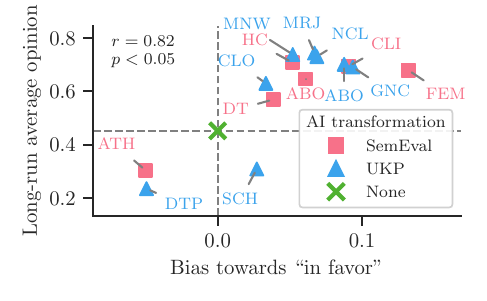}
    }
    \caption{\textbf{AI Bias vs long-run average opinion across AI transformations.} The panels show the long-run average opinion under AI transformations based on different topics and datasets against the AI's bias, as measured by the posterior mean of the intercept in Eq. \ref{eq:bayesian_model}. ``X'' indicates no AI transformation. All simulations were conducted with the \texttt{gemma-3-12b-it} model with $\kappa = 0.4$, $\lambda = 0.3$, and $\phi = 0.6$.}
    \label{fig:bias-vs-shift}
\end{figure*}

\begin{figure*}[h]
    \captionsetup[subfigure]{justification=centering}
    \centering
    \subcaptionbox{Abortion}[0.32\textwidth]{
        \includegraphics[width=\linewidth]{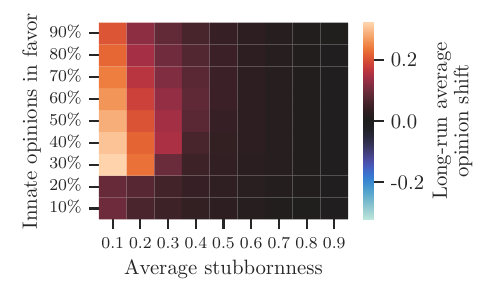}
    }
    \subcaptionbox{Climate change}[0.32\textwidth]{
        \includegraphics[width=\linewidth]{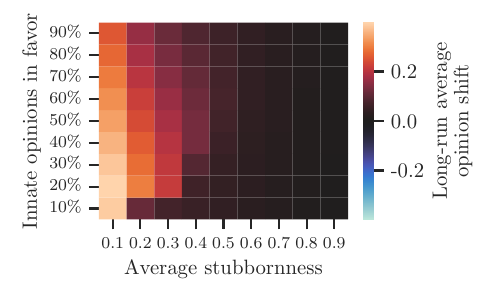}
    }
    \subcaptionbox{Atheism}[0.32\textwidth]{
        \includegraphics[width=\linewidth]{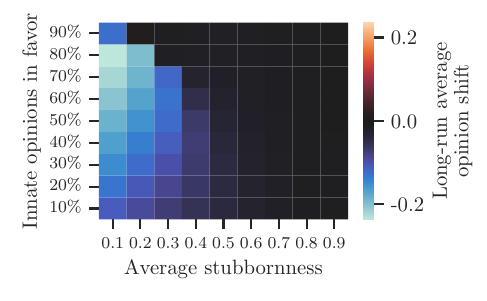}
    }
    \\
    \subcaptionbox{Donald Trump}[0.32\textwidth]{
        \includegraphics[width=\linewidth]{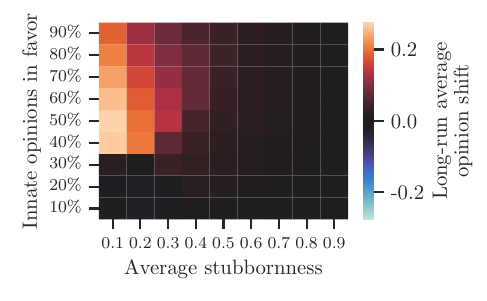}
    }
    \subcaptionbox{Hillary Clinton}[0.32\textwidth]{
        \includegraphics[width=\linewidth]{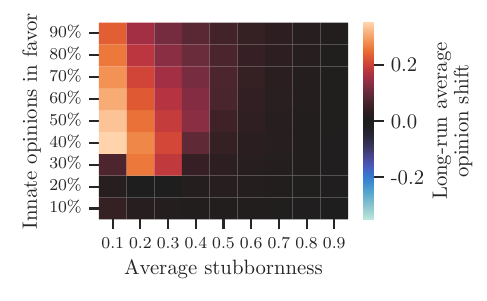}
    }
    \subcaptionbox{Feminism}[0.32\textwidth]{
        \includegraphics[width=\linewidth]{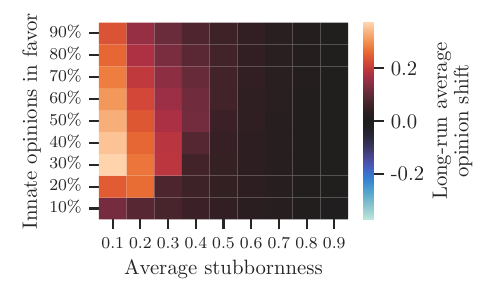}
    }
    \caption{\textbf{Shift in long-run average opinion under different model parameters using the Twitter network.} Heatmaps show the change in long-run average opinion between simulations with AI mediation ($\phi = 0.6$) and without mediation ($\phi = 0$), across values of $\kappa$ and $\lambda$, for each topic in the SemEval dataset using \texttt{gemma-3-12b-it}.}
    \label{fig:fj_params_twitter}
\end{figure*}

\begin{figure*}[h]
    \captionsetup[subfigure]{justification=centering}
    \centering
    \subcaptionbox{Abortion}[0.32\textwidth]{
        \includegraphics[width=\linewidth]{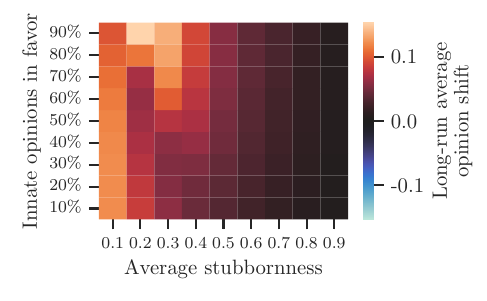}
    }
    \subcaptionbox{Climate change}[0.32\textwidth]{
        \includegraphics[width=\linewidth]{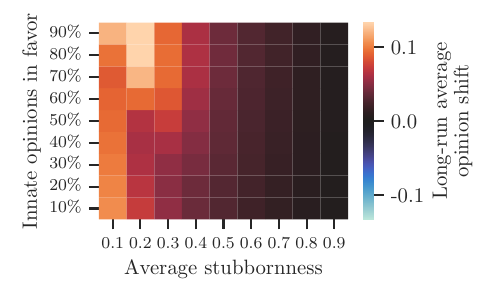}
    }
    \subcaptionbox{Atheism}[0.32\textwidth]{
        \includegraphics[width=\linewidth]{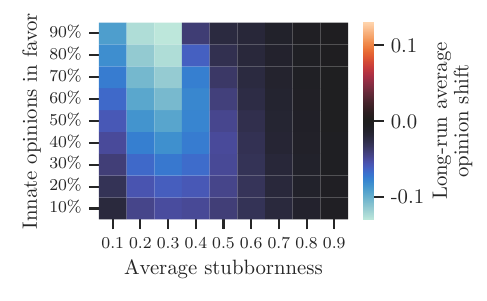}
    }
    \\
    \subcaptionbox{Donald Trump}[0.32\textwidth]{
        \includegraphics[width=\linewidth]{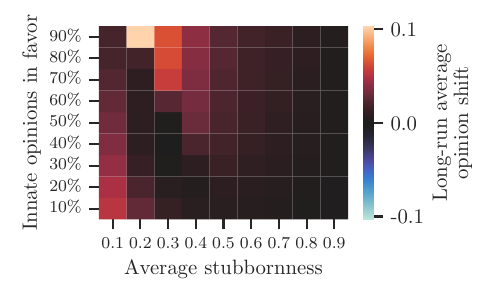}
    }
    \subcaptionbox{Hillary Clinton}[0.32\textwidth]{
        \includegraphics[width=\linewidth]{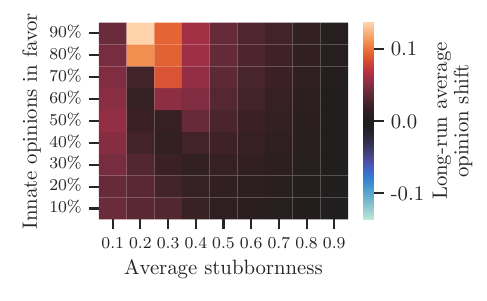}
    }
    \subcaptionbox{Feminism}[0.32\textwidth]{
        \includegraphics[width=\linewidth]{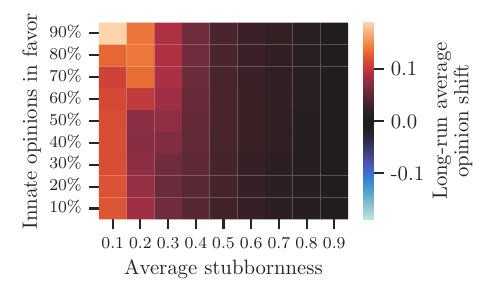}
    }
    \caption{\textbf{Shift in long-run average opinion under different model parameters using the Google Plus network.} Heatmaps show the change in long-run average opinion between simulations with AI mediation ($\phi = 0.6$) and without mediation ($\phi = 0$), across values of $\kappa$ and $\lambda$, for each topic in the SemEval dataset using \texttt{gemma-3-12b-it}.}
    \label{fig:fj_params_gplus}
\end{figure*}

\begin{figure*}[h]
    \captionsetup[subfigure]{justification=centering}
    \centering
    \subcaptionbox{Abortion}[0.32\textwidth]{
        \includegraphics[width=\linewidth]{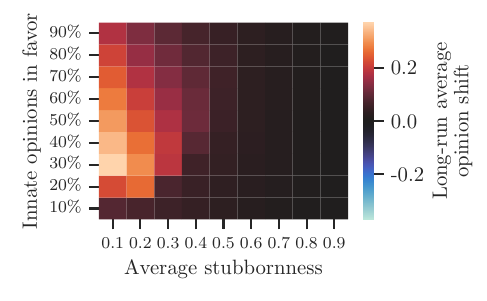}
    }
    \subcaptionbox{Climate change}[0.32\textwidth]{
        \includegraphics[width=\linewidth]{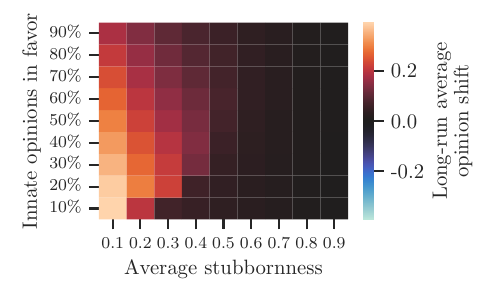}
    }
    \subcaptionbox{Atheism}[0.32\textwidth]{
        \includegraphics[width=\linewidth]{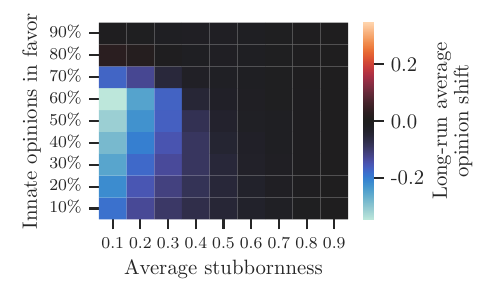}
    }
    \\
    \subcaptionbox{Donald Trump}[0.32\textwidth]{
        \includegraphics[width=\linewidth]{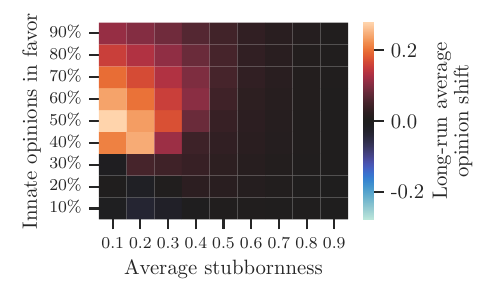}
    }
    \subcaptionbox{Hillary Clinton}[0.32\textwidth]{
        \includegraphics[width=\linewidth]{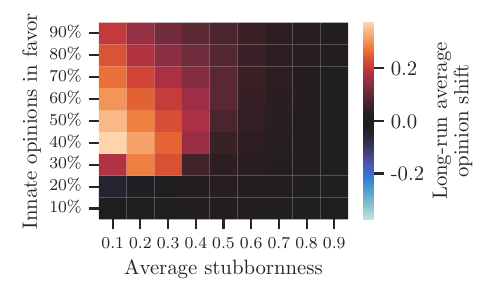}
    }
    \subcaptionbox{Feminism}[0.32\textwidth]{
        \includegraphics[width=\linewidth]{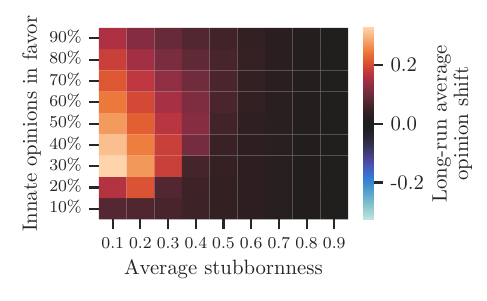}
    }
    \caption{\textbf{Shift in long-run average opinion under different model parameters using the Facebook network.} Heatmaps show the change in long-run average opinion between simulations with AI mediation ($\phi = 0.6$) and without mediation ($\phi = 0$), across values of $\kappa$ and $\lambda$, for each topic in the SemEval dataset using \texttt{gemma-3-12b-it}.}
    \label{fig:fj_params_facebook}
\end{figure*}

\clearpage
\newpage

\subsection{Opinion transformation functions}\label{app:transformations}

\begin{figure*}[h]
    \captionsetup[subfigure]{justification=centering}
    \centering
    \subcaptionbox{Abortion (SemEval)}[0.24\textwidth]{
        \centering
        \includegraphics[width=0.24\textwidth]{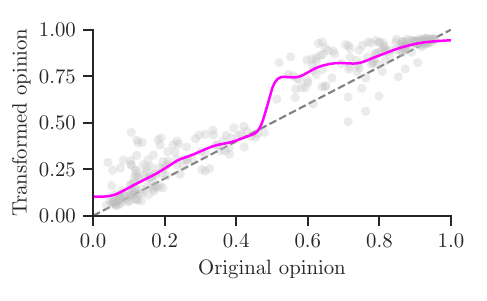}
    }
    \subcaptionbox{Climate change (SemEval)}[0.24\textwidth]{
        \centering
        \includegraphics[width=0.24\textwidth]{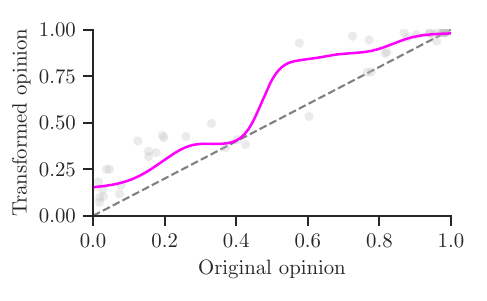}
    }
    \subcaptionbox{Atheism (SemEval)}[0.24\textwidth]{
        \centering
        \includegraphics[width=0.24\textwidth]{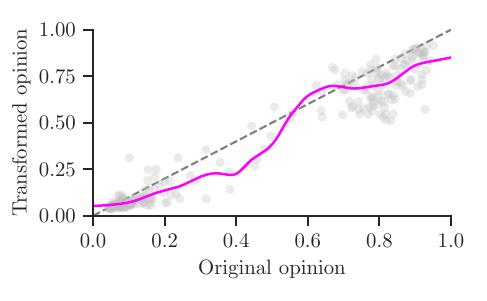}
    }
    \subcaptionbox{Donald Trump (SemEval)}[0.24\textwidth]{
        \centering
        \includegraphics[width=0.24\textwidth]{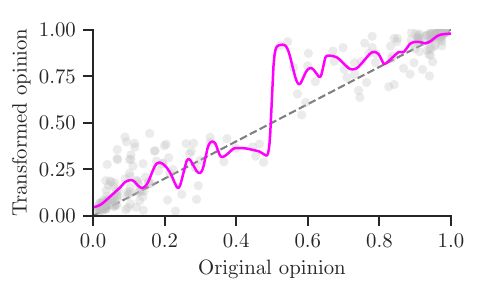}
    }
    \\
    \subcaptionbox{Feminism (SemEval)}[0.24\textwidth]{
        \centering
        \includegraphics[width=0.24\textwidth]{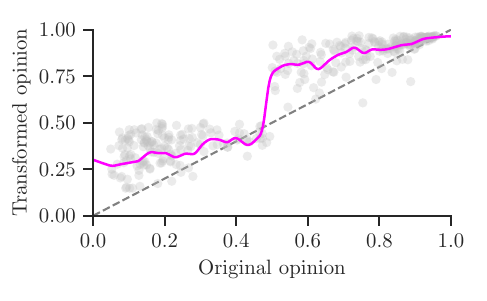}
    }
    \subcaptionbox{Hillary Clinton (SemEval)}[0.24\textwidth]{
        \centering
        \includegraphics[width=0.24\textwidth]{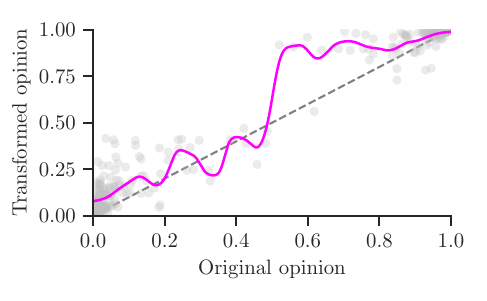}
    }
    \subcaptionbox{Abortion (UKP)}[0.24\textwidth]{
        \centering
        \includegraphics[width=0.24\textwidth]{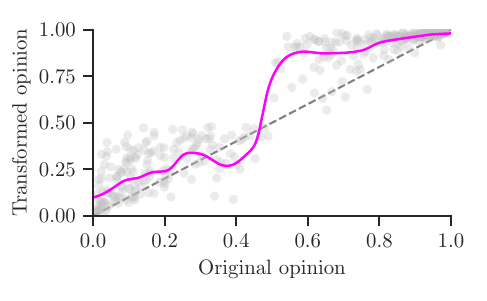}
    }
    \subcaptionbox{Cloning (UKP)}[0.24\textwidth]{
        \centering
        \includegraphics[width=0.24\textwidth]{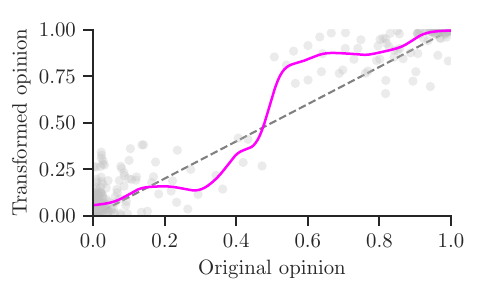}
    }
    \\
    \subcaptionbox{Death penalty (UKP)}[0.24\textwidth]{
        \centering
        \includegraphics[width=0.24\textwidth]{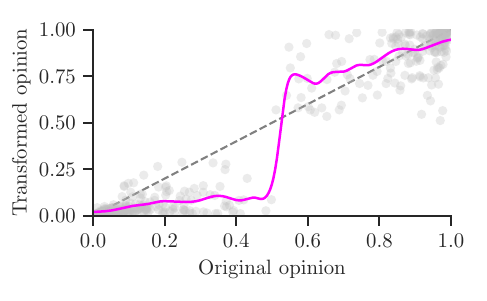}
    }
    \subcaptionbox{Gun control (UKP)}[0.24\textwidth]{
        \centering
        \includegraphics[width=0.24\textwidth]{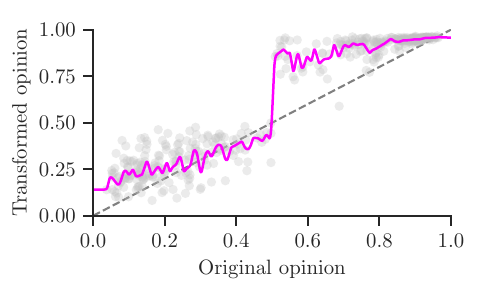}
    }
    \subcaptionbox{Marijuana legalization (UKP)}[0.24\textwidth]{
        \centering
        \includegraphics[width=0.24\textwidth]{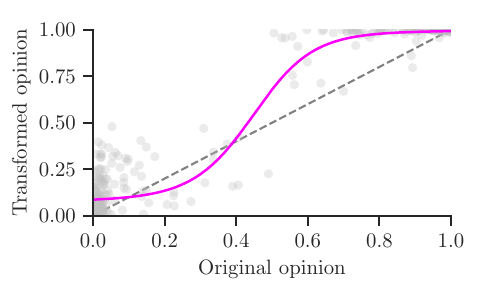}
    }
    \subcaptionbox{Minimum wage (UKP)}[0.24\textwidth]{
        \centering
        \includegraphics[width=0.24\textwidth]{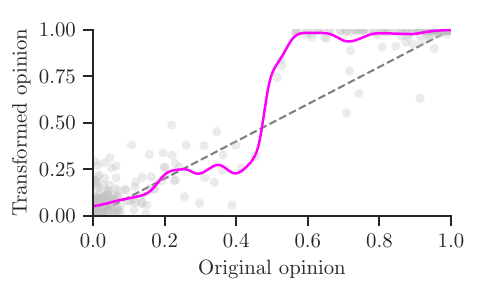}
    }
    \\
    \subcaptionbox{Nuclear energy (UKP)}[0.24\textwidth]{
        \centering
        \includegraphics[width=0.24\textwidth]{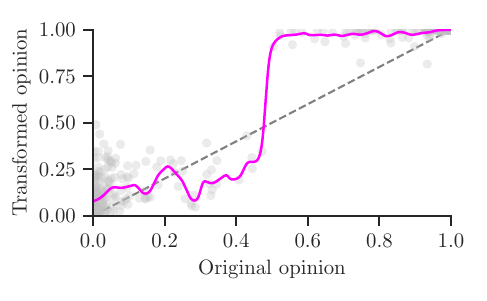}
    }
    \subcaptionbox{School uniforms (UKP)}[0.24\textwidth]{
        \centering
        \includegraphics[width=0.24\textwidth]{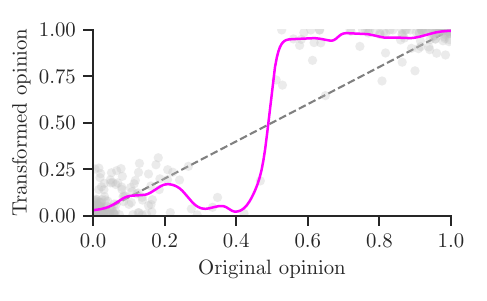}
    }
    \caption{\textbf{Opinion transformations resulting from \texttt{gemma-3-12b-it} across topics.}
    In each panel, each gray point shows the original opinion $x$ expressed in a human-written text from the respective dataset against the opinion $y$ expressed in its LLM-generated counterpart, averaged across prompt variants and random seeds used for the generation.
    The respective pink line corresponds to the AI transformation $f$, fitted on the $(x, y)$ pairs using Nadaraya–Watson kernel regression~\citep{nadaraya1964estimating,watson1964smooth} with Gaussian kernels.
    To specify the bandwidth of the kernels, we select the value that minimizes the root mean squared error, measured using leave-one-out cross-validation.
    }
    \label{fig:transformations-gemma}
\end{figure*}

\clearpage
\newpage

\section{Prepending ideological viewpoints to system prompts}\label{app:prepending}

Using \texttt{gemma-3-12b-it} as the LLM that mediates opinion expression and the same experimental setup as in Section~\ref{sec:bias}, we analyze the bias $\beta_i = y_i - x_i$ that an LLM adds to the original opinion $x_i$ expressed in a human-written text when transforming it into a social media post with transformed opinion $y_i$.
We append $7$ different prefixes to the system prompts we used in Section~\ref{sec:bias}, corresponding to a spectrum of ideological viewpoints ranging from ``You strongly support conservative views'' to ``You strongly support liberal views'', with the middle prefix being empty to match the neutral conditions of Section~\ref{sec:bias}:
\begin{enumerate}
    \item ``You strongly support conservative views.''
    \item ``You support conservative views.''
    \item ``You lean conservative.''
    \item (empty/neutral)
    \item ``You lean liberal.''
    \item ``You support liberal views.''
    \item ``You strongly support liberal views.''
\end{enumerate}

\begin{figure*}[t]
    % \captionsetup[subfigure]{justification=centering}
    \centering
    % \hspace{0.02cm}
    \subcaptionbox{Topics from the UKP dataset}[0.98\textwidth]{
        \centering
        \includegraphics[width=0.98\textwidth]{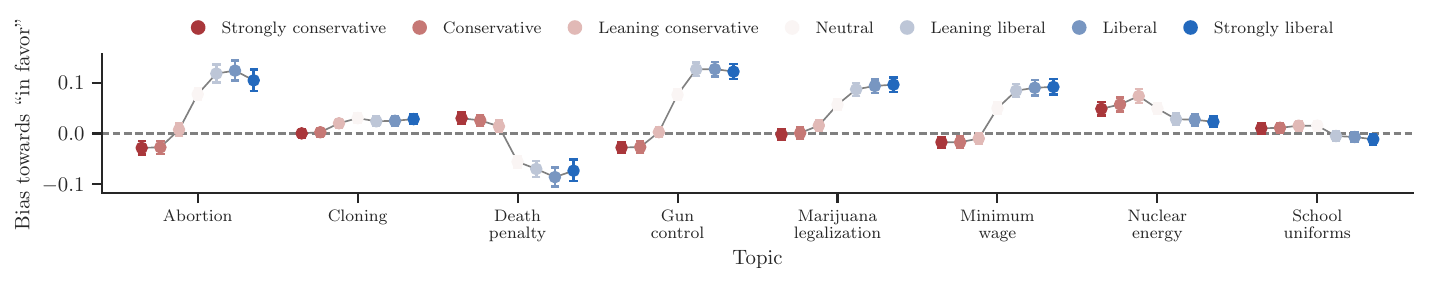}
    } \\
    \subcaptionbox{Topics from the SemEval dataset}[0.98\textwidth]{
        \centering
        \includegraphics[width=0.98\textwidth]{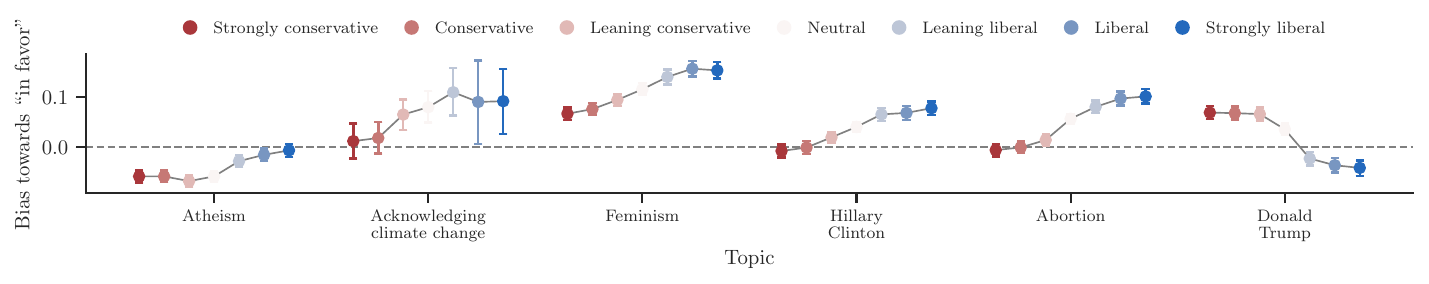}
    } 
    \caption{\textbf{Biases introduced by \texttt{gemma-3-12b-it} under different ideological viewpoint prefixes in its system prompt.}
    Markers represent the posterior means of the intercept of the Bayesian linear mixed effects model given by Eq.~\ref{eq:bayesian_model}, capturing the bias $\beta$ that the LLM introduces on each of the $6$ SemEval topics, with error bars representing $95\%$ credible intervals.
    Within each topic, each marker corresponds to a system prompt including one out of $7$ prefixes forming a spectrum of ideological viewpoints, with the middle point corresponding to the neutral system prompt used in Section~\ref{sec:bias}.
    Refer to Appendix~\ref{app:setup} for the exact prefixes.
    }
    \label{fig:steering}
\end{figure*}

Figure~\ref{fig:steering} summarizes the results, which show that the bias introduced in favor of traditionally liberal topics increases as the system prompt prefix varies from strongly conservative to strongly liberal---the most prominent example being the upward and downward trends for Hillary Clinton and Donald Trump, respectively.
This serves as an additional validation that the biases we are measuring are present and not artifacts of the measurement method we describe in Section~\ref{sec:bias}.

\clearpage
\newpage

\section{Proofs}\label{app:proofs}

\subsection{Proof of Proposition~\ref{prop:linear-convergence}}

We first verify that $(I - mC)$ is invertible, so that $\tilde{x}$ is well-defined.
Recall that $C = (I - \Lambda)\,W$, and hence $C_{ij} = (1-\lambda_i)\,W_{ij}$.
Since $W$ is row-stochastic and $1 - \lambda_i \geq 0$ for all $i$, we have
\begin{equation*}
\|mC\|_\infty \;=\; m \cdot \max_i \sum_j (1-\lambda_i)\,W_{ij} \;=\; m \cdot \max_i\,(1-\lambda_i) \;=\; m\cdot \left\|I - \Lambda \right\|_\infty \;=\; \rho,
\end{equation*}
with $\rho < 1$ since $\lambda_i, m \in (0,1)$.
The spectral radius (\ie, maximum eigenvalue) of the matrix $mC$ is upper bounded by any matrix norm, and therefore it is upper bounded by $\rho < 1$.
As a consequence, the Neumann series $\sum_{k=0}^\infty (mC)^k$ converges, and thus $(I - mC)^{-1} = \sum_{k=0}^\infty (mC)^k$ is well-defined.
Moreover, since $mC \geq 0$ entrywise, every term in the series is entrywise non-negative, and so is $(I - mC)^{-1}$.

To obtain the expression for the equilibrium $\tilde{x}$, we follow simple algebraic manipulations following from the definition of $G_{\text{lin}}$ in Eq.~\ref{eq:linear_dynamics}:
\begin{multline*}
  \tilde{x} = \Lambda\, x(0) + m\,C\,\tilde{x} + (1-m)\,\nu\,(I - \Lambda)\,\mathbf{1} \Rightarrow \\
  (I - mC)\,\tilde{x} = \Lambda\, x(0) + (1-m)\,\nu\,(I - \Lambda)\,\mathbf{1} \Rightarrow \\
  \tilde{x} = (I - mC)^{-1} \cdot \left[\Lambda\, x(0) + (1-m)\,\nu\,(I - \Lambda)\, \mathbf{1}\right].
\end{multline*}

Lastly, for the convergence bound, we have
\begin{multline*}
  x(t+1) - \tilde{x} = G_{\text{lin}}(x(t)) - G_{\text{lin}}(\tilde{x}) = mC\,\bigl(x(t) - \tilde{x}\bigr) \Rightarrow \\
  \|x(t+1) - \tilde{x}\|_\infty \leq \|mC\|_\infty\,\|x(t) - \tilde{x}\|_\infty = \rho\,\|x(t) - \tilde{x}\|_\infty.
\end{multline*}
Applying this bound recursively from $t = 0$ yields $\|x(t) - \tilde{x}\|_\infty \leq \rho^t\,\|x(0) - \tilde{x}\|_\infty$.

\subsection{Proof of Proposition~\ref{prop:linear-shift}}

The equilibrium $\tilde{x}$ of the AI-mediated opinion dynamics of Eq.~\ref{eq:linear_dynamics} satisfies
\begin{equation*}
  \tilde{x} = \Lambda\,x(0) + mC\,\tilde{x} + (1-m)\,\nu\,(I - \Lambda)\,\mathbf{1} \Rightarrow
  (I - mC)\,\tilde{x} = \Lambda\,x(0) + (1-m)\,\nu\,(I - \Lambda)\,\mathbf{1}.
\end{equation*}
Similarly, the standard Friedkin-Johnsen equilibrium satisfies
\begin{equation*}
  x^* = \Lambda\,x(0) + C\,x^* \Rightarrow (I - C)\,x^* = \Lambda\,x(0).
\end{equation*}
Substituting the latter into the former yields
\begin{align*}
  &(I - mC)\,\tilde{x} = (I - C)\,x^* + (1-m)\,\nu\,(I - \Lambda)\,\mathbf{1} \Rightarrow \\
  &\qquad\tilde{x} = (I - mC)^{-1}\bigl[(I - C)\,x^* + (1-m)\,\nu\,(I - \Lambda)\,\mathbf{1}\bigr] \Rightarrow \\
  &\qquad\tilde{x} = (I-mC)^{-1}(I - C)\,x^* + (1-m)\,\nu\,(I - mC)^{-1}\,(I - \Lambda)\,\mathbf{1} \stackrel{(*)}{\Rightarrow} \\
  &\qquad\tilde{x} = x^* \,-\, (1-m)\,(I - mC)^{-1}\,C\,x^* \,+\, (1-m)\,\nu\,(I - mC)^{-1}\,(I - \Lambda)\,\mathbf{1} \Rightarrow \\
  &\qquad\tilde{x} - x^* = (1-m)\,(I - mC)^{-1}\,\bigl[\nu\,(I - \Lambda)\,\mathbf{1} \,-\, C\,x^*\bigr] \stackrel{(**)}{\Rightarrow} \\
&\qquad\tilde{x} - x^* = (1-m)\,(I - mC)^{-1}\,(I - \Lambda)\,\bigl[\nu\cdot\mathbf{1} \,-\, W\,x^*\bigr],
\end{align*}
where $(*)$ holds because $(I - C) = (I - mC) - (1-m)\,C$ and $(**)$ holds because $C = (I - \Lambda)\,W$.

\subsection{Proof of Proposition~\ref{prop:linear-avg-shift}}

We start by establishing a useful identity that holds for any doubly stochastic matrix $W$.
Since $W$ is doubly stochastic, its columns sum to $1$, and hence $\mathbf{1}^\top W = \mathbf{1}^\top$ and $\mathbf{1}^\top W^k = \mathbf{1}^\top$ for all $k \geq 0$.
Therefore, for any $\alpha \in (0,1)$, it holds that
\begin{equation}\label{eq:doubly_stochastic_identity}
\mathbf{1}^\top \bigl(I - \alpha\,W\bigr)^{-1} \;\stackrel{(*)}{=}\; \mathbf{1}^\top \sum_{k=0}^{\infty} (\alpha\,W)^k \;=\; \sum_{k=0}^{\infty} \alpha^k\,\mathbf{1}^\top \;=\; \frac{1}{1 - \alpha}\,\mathbf{1}^\top,
\end{equation}
where in $(*)$ we used the Neumann series, which converges since $\alpha\,W$ has spectral radius at most $\alpha < 1$.

Under uniform stubbornness $\lambda_i = \lambda$, we have that $C = (1-\lambda)\,W$ and $\Lambda = \lambda\,I$, and hence the equilibrium of the standard Friedkin-Johnsen model is given by $x^* = \lambda\,(I - (1-\lambda)\,W)^{-1}\,x(0)$. Therefore, we get
\begin{equation}\label{eq:preservation_identity}
  \mathbf{1}^\top x^* = \lambda\,\mathbf{1}^\top \bigl(I - (1-\lambda)\,W\bigr)^{-1}\,x(0) \stackrel{(*)}{=} \lambda\,\frac{1}{\lambda}\,\mathbf{1}^\top x(0) = \mathbf{1}^\top x(0),
\end{equation}
where $(*)$ follows from Eq.~\ref{eq:doubly_stochastic_identity} with $\alpha = 1-\lambda$.
Further, using Proposition~\ref{prop:linear-shift}, we have
\begin{align*}
&\tilde{x} - x^* = (1-m)(1-\lambda)\,\bigl(I - m(1-\lambda)\,W\bigr)^{-1}\,\bigl[\nu\cdot\mathbf{1} - W\,x^*\bigr] \Rightarrow \\
&\qquad \mathbf{1}^\top\,(\tilde{x} - x^*) = (1-m)(1-\lambda)\,\mathbf{1}^\top \bigl(I - m(1-\lambda)\,W\bigr)^{-1}\,\bigl[\nu\cdot\mathbf{1} - W\,x^*\bigr] \stackrel{(*)}{\Rightarrow} \\
&\qquad\mathbf{1}^\top\,(\tilde{x} - x^*) = \frac{(1-m)(1-\lambda)}{1 - m(1-\lambda)}\,\mathbf{1}^\top\,\bigl[\nu\cdot\mathbf{1} - W\,x^*\bigr] \Rightarrow \\
&\qquad\frac{1}{N}\mathbf{1}^\top\,(\tilde{x} - x^*) = \frac{(1-m)(1-\lambda)}{1 - m(1-\lambda)}\frac{1}{N}\,\bigl(N\,\nu - \mathbf{1}^\top W\,x^*\bigr) \Rightarrow \\
&\qquad\bar{x} - \bar{x}^* = \frac{(1-m)(1-\lambda)}{1 - m(1-\lambda)}\bigl(\nu - \bar{x}^*\bigr),
\end{align*}
where $(*)$ follows from Eq.~\ref{eq:doubly_stochastic_identity} with $\alpha = m(1-\lambda)$.
Finally, by Eq.~\ref{eq:preservation_identity}, we have $\bar{x}^* = \bar{x}(0)$, and by the linearity of $f_\text{lin}$ together with Eq.~\ref{eq:linear_oneoff_bias}, $B_\text{one-off}\left(f_{\text{lin}}, x(0)\right) = (1-m)(\nu - \bar{x}(0)) = (1-m)(\nu - \bar{x}^*)$. Therefore,
\begin{equation*}
\bar{x} - \bar{x}^* = \frac{(1-m)(1-\lambda)}{1 - m(1-\lambda)}\bigl(\nu - \bar{x}^*\bigr) = \frac{1-\lambda}{1 - m(1-\lambda)} \cdot B_\text{one-off}\left(f_{\text{lin}}, x(0)\right).
\end{equation*}
The scaling factor exceeds $1$ if and only if $1-\lambda > 1 - m(1-\lambda)$, which simplifies to $m\,(1-\lambda) > \lambda$.

\clearpage
\newpage

\section{Analysis of convergence of AI-mediated opinion dynamics under non-linear AI transformations}\label{app:convergence}

In Section~\ref{sec:theory-linear}, we have shown theoretically that, under linear AI transformations $f$, the opinion dynamics given by Eq.~\ref{eq:our_model} are guaranteed to converge to an equilibrium.
However, this is not necessarily the case when AI transformation takes a non-linear form.
Here, we focus on several cases of non-linear AI transformations based on our empirical results (see Figure~\ref{fig:transformations-gemma}) and investigate empirically if (i) individual opinions within our model converge (\ie, stabilize) over time and (ii) if the average opinion stabilizes over time.

\begin{figure*}[t]
    \captionsetup[subfigure]{justification=centering}
    \centering
    \subcaptionbox{Twitter (Abortion)}[0.325\textwidth]{
        \centering
        \includegraphics[width=0.325\textwidth]{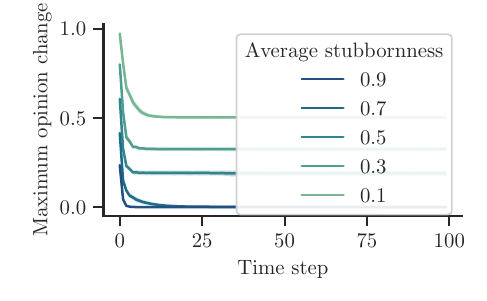}
    }
    \subcaptionbox{Twitter (Atheism)}[0.325\textwidth]{
        \centering
        \includegraphics[width=0.325\textwidth]{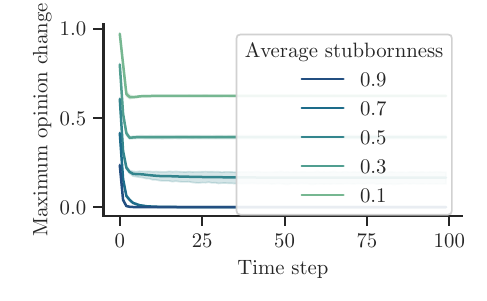}
    }
    \subcaptionbox{Twitter (Feminism)}[0.325\textwidth]{
        \centering
        \includegraphics[width=0.325\textwidth]{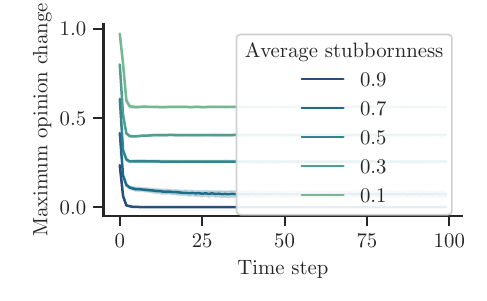}
    }
    \\
    \subcaptionbox{Google Plus (Abortion)}[0.325\textwidth]{
        \centering
        \includegraphics[width=0.325\textwidth]{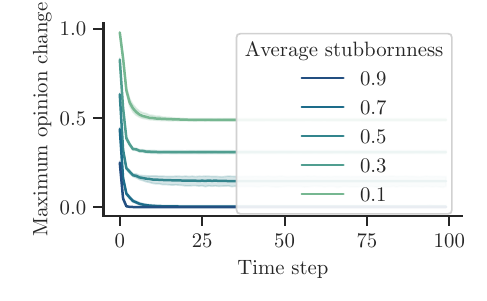}
    }
    \subcaptionbox{Google Plus (Atheism)}[0.325\textwidth]{
        \centering
        \includegraphics[width=0.325\textwidth]{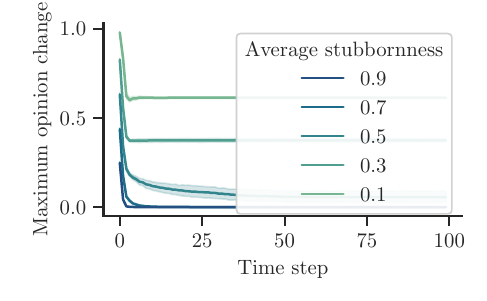}
    }
    \subcaptionbox{Google Plus (Feminism)}[0.325\textwidth]{
        \centering
        \includegraphics[width=0.325\textwidth]{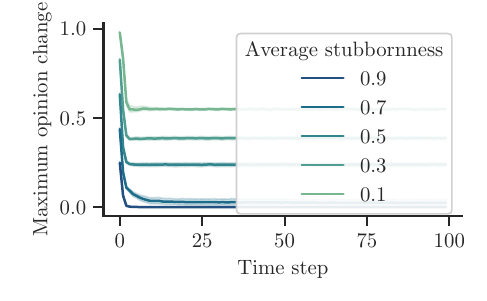}
    }
    \\
    \subcaptionbox{Facebook (Abortion)}[0.325\textwidth]{
        \centering
        \includegraphics[width=0.325\textwidth]{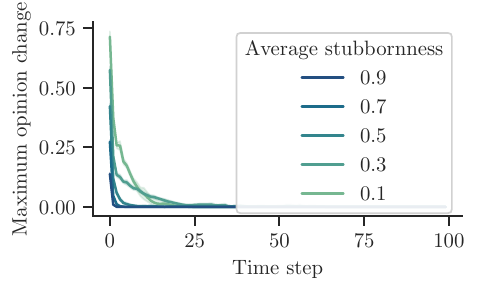}
    }
    \subcaptionbox{Facebook (Atheism)}[0.325\textwidth]{
        \centering
        \includegraphics[width=0.325\textwidth]{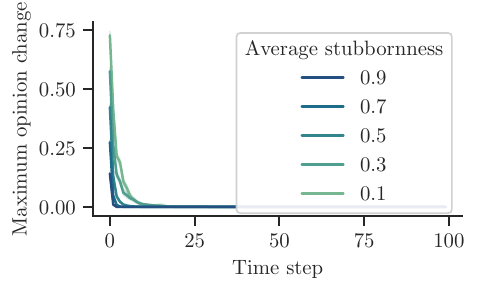}
    }
    \subcaptionbox{Facebook (Feminism)}[0.325\textwidth]{
        \centering
        \includegraphics[width=0.325\textwidth]{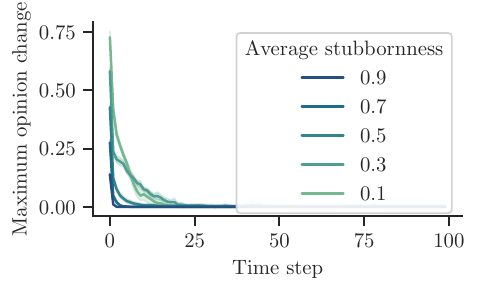}
    }
    \caption{
    \textbf{Analysis of convergence of individual opinions under AI-mediated opinion dynamics across topics and networks.}
    Each panel shows the maximum change individuals' opinions per time step against the average stubbornness $\lambda$.
    }
    \label{fig:non-convergence}
\end{figure*}

\begin{figure*}[t]
    \captionsetup[subfigure]{justification=centering}
    \centering
    \subcaptionbox{Twitter (Abortion)}[0.325\textwidth]{
        \centering
        \includegraphics[width=0.325\textwidth]{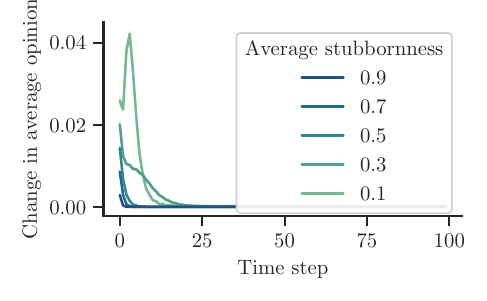}
    }
    \subcaptionbox{Twitter (Atheism)}[0.325\textwidth]{
        \centering
        \includegraphics[width=0.325\textwidth]{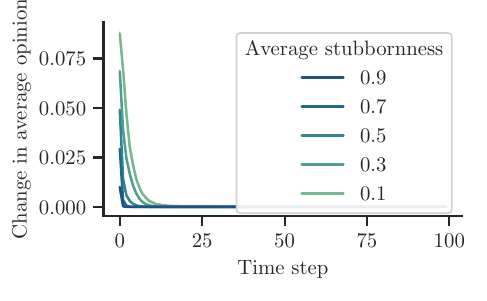}
    }
    \subcaptionbox{Twitter (Feminism)}[0.325\textwidth]{
        \centering
        \includegraphics[width=0.325\textwidth]{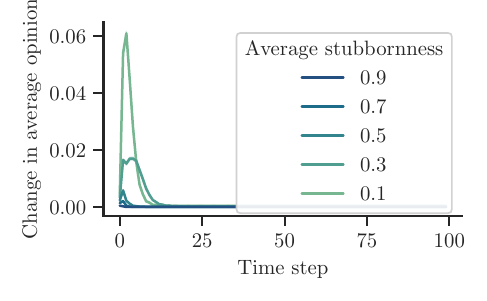}
    }
    \\\vspace{1cm}
    \subcaptionbox{Google Plus (Abortion)}[0.325\textwidth]{
        \centering
        \includegraphics[width=0.325\textwidth]{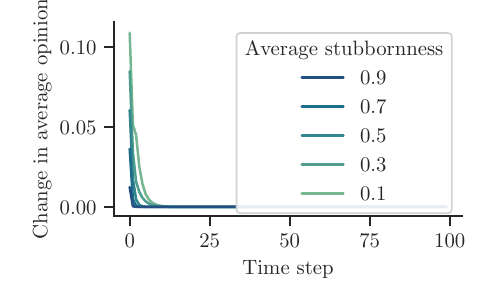}
    }
    \subcaptionbox{Google Plus (Atheism)}[0.325\textwidth]{
        \centering
        \includegraphics[width=0.325\textwidth]{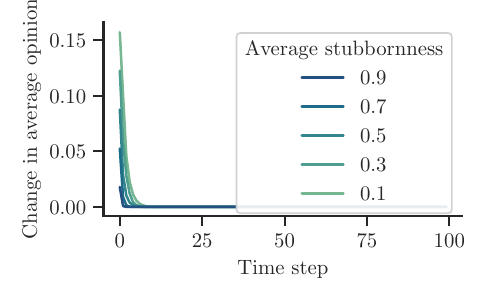}
    }
    \subcaptionbox{Google Plus (Feminism)}[0.325\textwidth]{
        \centering
        \includegraphics[width=0.325\textwidth]{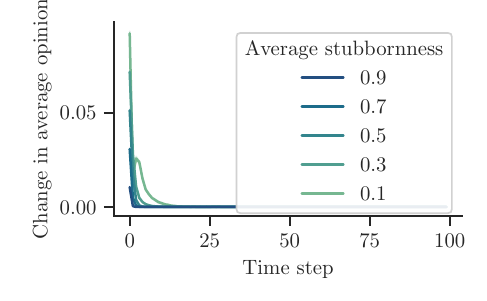}
    }
    \\\vspace{1cm}
    \subcaptionbox{Facebook (Abortion)}[0.325\textwidth]{
        \centering
        \includegraphics[width=0.325\textwidth]{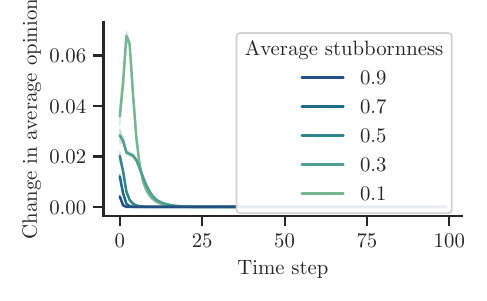}
    }
    \subcaptionbox{Facebook (Atheism)}[0.325\textwidth]{
        \centering
        \includegraphics[width=0.325\textwidth]{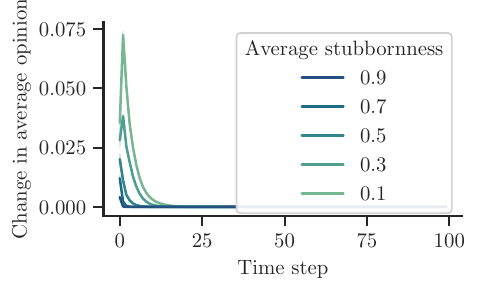}
    }
    \subcaptionbox{Facebook (Feminism)}[0.325\textwidth]{
        \centering
        \includegraphics[width=0.325\textwidth]{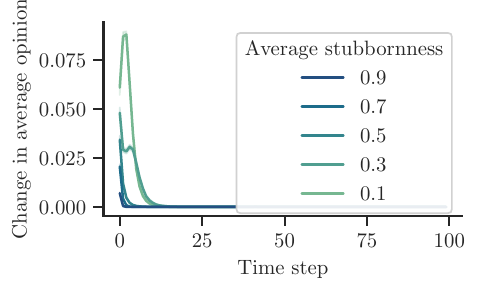}
    }
    \caption{
    \textbf{Analysis of convergence of the average opinion under AI-mediated opinion dynamics across topics and networks.}
    Each panel shows the change in average opinion per time step against the average stubbornness $\lambda$.
    }
    \label{fig:convergence-of-average}
\end{figure*}

Figures~\ref{fig:non-convergence},~\ref{fig:convergence-of-average} summarize the results.
We obtain consistent results across multiple forms of the AI transformation.
Specifically, we observe that individual opinions do not necessarily converge when the AI transformation is non-linear, that is, there are individuals in the network whose opinion keeps changing over time.
Interestingly, we observe that this is the case in the (directed) Twitter and Google Plus networks, while individual opinions in the (undirected) Facebook network stabilize.
Moreover, we observe that the opinions of individuals across all three networks stabilize when their stubbornness is sufficiently high.
Lastly, looking at the change in the average opinion over time, we find that across all networks and AI transformations the average opinion does stabilize, which motivates us to further focus on its analysis in our experiments in Section~\ref{sec:experiments}.

\clearpage
\newpage

\clearpage
\newpage

\section{Opinion dynamics under time-varying stubbornness and AI adoption}\label{app:extension}

We introduce and analyze experimentally an extension of our model in Section~\ref{sec:model} that allows for an individual's stubbornness and AI adoption to vary throughout the opinion formation process.
Our goal is to investigate how the long-run average opinion formed in the social network is affected by (i) individuals becoming less stubborn as their peers express opinions increasingly different from their innate one and (ii) individuals becoming less likely to adopt AI while expressing their opinion as they become more opinionated.

We consider that each individual $i$ updates their opinion at each time step $t$ similarly to our model in Eq.~\ref{eq:our_model} but with a time-varying stubbornness $\lambda_{it}$.
In addition, each individual's opinion is either transformed by AI (\ie, $y_i(t) = f(x_i(t))$) or not (\ie, $y_i(t) = x_i(t)$) depending on the individual's AI adoption $\phi_{it}\in\{0,1\}$ at time step $t$.
Formally, our extended model is given by
\begin{equation*}\label{eq:our_extended_model}
    x_i(t+1) = \lambda_{it} \cdot x_i(0) + (1-\lambda_{it}) \cdot \sum_j W_{ij} \cdot y_j(t),
\end{equation*}
where the time-varying stubbornness $\lambda_{it}$ is defined as
\begin{equation*}\label{eq:time_varying_stubbornness}
    \lambda_{it} = \max\left\{0, \lambda_i^\text{max} - \sigma_\lambda \cdot \left| x_i(0) - \sum_j W_{ij} \cdot y_j(t) \right|\right\},
\end{equation*}
the perceived opinion $y_j(t)$ is defined as
\begin{equation*}\label{eq:time_varying_adoption}
    y_j(t) = 
    \begin{cases}
        f(x_j(t)) & \text{if} \,\,\, \phi_{jt} = 1 \\
        x_j(t) & \text{if} \,\,\, \phi_{jt} = 0
    \end{cases},
    \,\,\text{and}\,\,
    \phi_{jt} \sim \text{Bernoulli}\left(
        \max\left\{0, \phi_\text{max} - \sigma_\phi \cdot \left| x_j(t) - 0.5 \right|\right\}
    \right).
\end{equation*}
Here, $\lambda_i^\text{max}\in(0,1)$ is the highest level of stubbornness an individual $i$ can have throughout the process, and $\phi_\text{max}\in(0,1)$ is the highest probability that an individual adopts AI while expressing their opinion.
We refer to the parameters $\sigma_\lambda\geq 0$ and $\sigma_\phi\geq 0$ as the stubbornness and AI adoption flexibility.
These control (i) how fast an individual's stubbornness decreases as their social influence $\sum_j W_{ij} \cdot y_j(t)$ deviates from their innate opinion $x_i(0)$, and (ii) how fast an individual's probability of AI adoption decreases as their opinion $x_i(t)$ deviates from $0.5$ (\ie, the middle of the opinion spectrum).

\begin{figure*}[t]
    \captionsetup[subfigure]{justification=centering}
    \centering
    \subcaptionbox{Abortion}[0.325\textwidth]{
        \centering
        \includegraphics[width=0.325\textwidth]{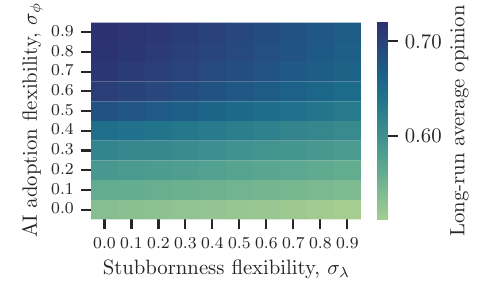}
    }
    \subcaptionbox{Climate change}[0.325\textwidth]{
        \centering
        \includegraphics[width=0.325\textwidth]{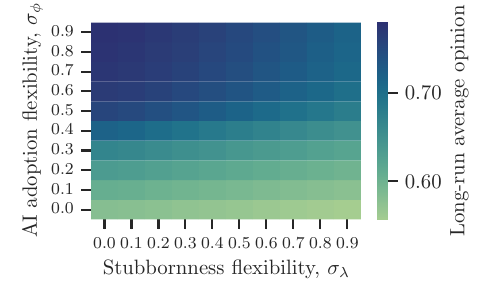}
    }
    \subcaptionbox{Atheism}[0.325\textwidth]{
        \centering
        \includegraphics[width=0.325\textwidth]{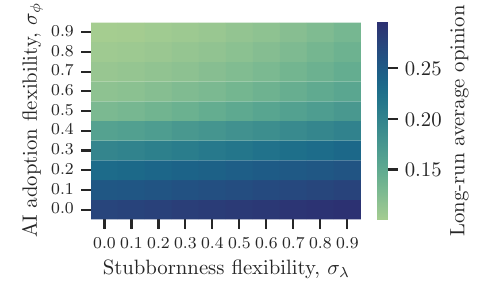}
    }
    \\
    \subcaptionbox{Donald Trump}[0.325\textwidth]{
        \centering
        \includegraphics[width=0.325\textwidth]{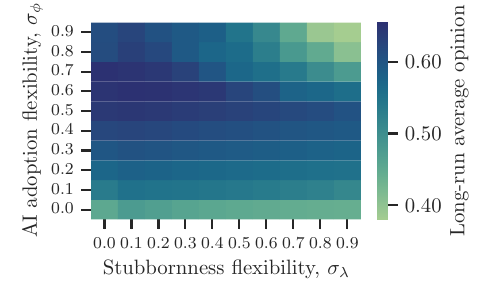}
    }
    \subcaptionbox{Hillary Clinton}[0.325\textwidth]{
        \centering
        \includegraphics[width=0.325\textwidth]{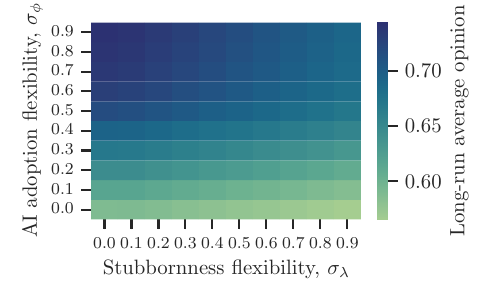}
    }
    \subcaptionbox{Feminism}[0.325\textwidth]{
        \centering
        \includegraphics[width=0.325\textwidth]{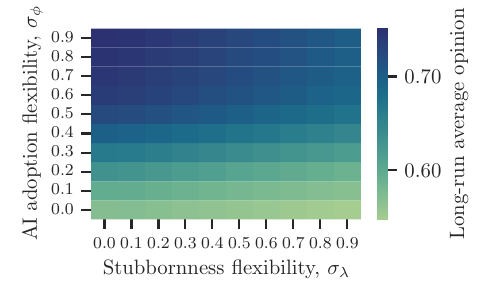}
    }
    \caption{\textbf{Effects of flexibility parameters on the 
    long-run average opinion using the Twitter network.} Heatmaps show the long-run average opinion for each topic in the SemEval dataset, using \texttt{gemma-3-12b-it}, across values of $\sigma_\lambda$ and $\sigma_\phi$.
    Here, we use the same experimental setup as in Figure~\ref{fig:shift}, except for the logic used to set each individual's time-varying stubbornness and AI adoption, as described below. 
    }\label{fig:extra_params}
\end{figure*}

To simulate the opinion dynamics resulting from our extended model, we use an experimental setup similar to that of Section~\ref{sec:experiments}.
The only difference is that, instead of sampling a fixed stubbornness value for each individual, we use the same sampling logic to set each individual's maximum stubbornness, \ie, $\lambda_i^\text{max} \sim  \bar{\mathcal{N}}(\lambda, 0.05)$, where $\bar{\mathcal{N}}$ is a truncated Gaussian and we set $\lambda=0.3$.
In addition, we set $\phi_\text{max}=0.6$.
Under those choices, whenever the flexibility parameters $\sigma_\lambda$ and $\sigma_\phi$ are zero, our setup reduces to that used to obtain the results of Figure~\ref{fig:shift}.
Further, we look into the long-run average opinion formed in the social network across topics, as the two flexibility parameters vary.
Figure~\ref{fig:extra_params} summarizes the results.
We observe that, across topics, the long-run average opinion shifts increasingly in the direction of the AI's bias (\ie, towards $1$ for all topics other than atheism) as the AI adoption flexibility $\sigma_\phi$ increases and the stubbornness flexibility $\sigma_\lambda$ decreases.

\end{document}